\newif\ifXHTML\XHTMLfalse 

\ifXHTML
       \documentclass[12pt]{article}
       \usepackage[authoryear]{natbib}
\else
\documentclass[rmp,twocolumn,nofootinbib]{revtex4}
\fi
\usepackage{graphics}
\usepackage{epsfig}
\usepackage{graphicx}
\usepackage{amsmath}
\usepackage{amssymb}
\usepackage{dsfont}
\usepackage{dcolumn}
\usepackage{bm}
\usepackage{makeidx}
\usepackage[usenames]{color}

\begin{document}
\title{Spin-dependent phenomena and device concepts explored in (Ga,Mn)As}

\ifXHTML
\author{%
T.~Jungwirth\\
{\small Institute of Physics  ASCR, Cukrovarnick\'a~10,}\\
{\small 162~53 Praha~6, Czech Republic}\\
{\small School of Physics and Astronomy, University of Nottingham,}\\
{\small Nottingham NG7~2RD, UK}\\[2em]
}

\maketitle

\else
\author{T. Jungwirth}
\affiliation{Institute of Physics ASCR, v.v.i., Cukrovarnick\'a 10, 162 53 Praha 6, Czech Republic}
\affiliation{School of Physics and Astronomy, University of Nottingham,
Nottingham NG7 2RD, UK}
\author{J.~Wunderlich}
\affiliation{Institute of Physics ASCR, v.v.i., Cukrovarnick\'a 10, 162 53 Praha 6, Czech Republic}
\affiliation{Hitachi Cambridge Laboratory, Cambridge CB3 0HE, United Kingdom}
\author{V. Nov\'ak}
\author{K.~Olejn\'{\i}k}
\affiliation{Institute of Physics ASCR, v.v.i., Cukrovarnick\'a 10, 162 53 Praha 6, Czech Republic}
\author{B.~L.~Gallagher}
\author{R.~P.~Campion}
\author{K.~W.~Edmonds}
\author{A.~W. Rushforth}
\affiliation{School of Physics and Astronomy, University of Nottingham, Nottingham NG7 2RD, United Kingdom}
\author{A.~J.~Ferguson}
\affiliation{Microelectronics Group, Cavendish Laboratory, University of Cambridge, CB3 0HE, UK}
\author{P.~N\v{e}mec}
\affiliation{Faculty of Mathematics and Physics, Charles University in Prague, Ke Karlovu 3, 121 16 Prague 2, Czech Republic}
\fi

\begin{abstract}
Over the past two decades, the research of (Ga,Mn)As has led to a deeper understanding of relativistic spin-dependent phenomena in magnetic systems. It has also led to discoveries of new effects and demonstrations of unprecedented functionalities of experimental spintronic devices with general applicability to a wide range of materials. This is a review of the basic material properties that make (Ga,Mn)As a favorable test-bed system for spintronics research and a discussion of contributions of (Ga,Mn)As studies in the general context of the spin-dependent phenomena and device concepts. Special focus is on the spin-orbit coupling induced effects and the reviewed topics include the interaction of spin with electrical current, light, and heat.
\end{abstract}

\ifXHTML\cleardoublepage\else\maketitle\fi \tableofcontents

\section{Introduction}
\label{intro}
Under equilibrium growth  conditions the incorporation of magnetic Mn ions into III-As semiconductor crystals is limited to approximately  0.1\%.  To circumvent  the solubility problem  a non-equilibrium, low-temperature molecular-beam-epitaxy (LT-MBE) technique was employed which led to first successful growths of (In,Mn)As and (Ga,Mn)As ternary alloys with more than 1\% Mn and to the discovery of ferromagnetism in these materials \cite{Ohno:1992_a,Munekata:1993_a,Ohno:1996_a,Shen:1997_a,Hayashi:1997_a,VanEsch:1997_a,Ohno:1998_a,Shimizu:1999_a,Hayashi:2001_a}. 

The compounds qualify as ferromagnetic semiconductors to the extent that their magnetic properties can be altered by the usual semiconductor electronics engineering variables, such as doping, electric fields, or light. The achievement of ferromagnetism in an ordinary III-V semiconductor with Mn concentrations exceeding 1\% demonstrates on its own the sensitivity of magnetic properties to doping. Several experiments have verified that changes in the carrier density and distribution in thin (III,Mn)As films due to an applied gate voltage can induce reversible changes of the Curie temperature $T_c$ and other magnetic and magneto-transport properties \cite{Ohno:2000_a,Chiba:2003_a,Chiba:2006_b,Wunderlich:2007_a,Chiba:2008_a,Olejnik:2008_a,Owen:2008_a,Stolichnov:2008_a,Riester:2009_a,Sawicki:2009_a,Mikheev:2012_a,Niazi:2013_a,Chiba:2013_a}. Experiments in which ferromagnetism in a (III,Mn)As system is turned on and off optically or in which recombination of spin-polarized carriers injected from the ferromagnetic semiconductor yields emission of circularly polarized light clearly demonstrated  the interaction of spin and light  in these materials \cite{Munekata:1997_a,Koshihara:1997_a,Ohno:1999_b}.

(Ga,Mn)As has become a test-bed material for the research of phenomena in which charge carriers respond to spin and {\em vice versa}. By exploiting the large spin polarization of carriers in (Ga,Mn)As and building on the well established heterostructure growth and microfabrication techniques in semiconductors, high quality magnetic tunnel junctions have been demonstrated showing  large tunneling magnetoresistances  (TMRs) \cite{Tanaka:2001_a,Chiba:2004_a,Saito:2005_a,Mattana:2005_a}.  In the studies of the inverse magneto-transport effects, namely spin-transfer torques (STTs) in tunnel junctions \cite{Chiba:2004_b} and domain walls, \cite{Yamanouchi:2004_a,Yamanouchi:2006_a,Wunderlich:2007_c,Adam:2009_a,Wang:2010_a,Curiale:2012_a,Ranieri:2012_a} the dilute-moment p-type (Ga,Mn)As is unique for its low saturation magnetization and strongly spin-orbit coupled valence band \cite{Sinova:2004_b,Garate:2008_d,Hals:2008_a}. Compared to common transition-metal ferromagnets this implies a more significant role of the field-like (non-adiabatic) STT complementing the antidamping-like (adiabatic) STT and lower currents required to excite magnetization dynamics. Moreover, the leading role of magnetocrystalline anisotropies over the dipolar shape anisotropy fields allows for the control of the direct and inverse magneto-transport phenomena by tuning the lattice strains {\em ex situ} by microfabrication \cite{Wunderlich:2007_c,Wenisch:2007_a} or {\em in situ} by piezo-electric transducers \cite{Rushforth:2008_a,Overby:2008_a,Goennenwein:2008_a,Ranieri:2012_a}. 

In general, TMR \cite{Julliere:1975_a,Moodera:1995_a,Myiazaki:1995_a} and STT \cite{Slonczewski:1996_a,Berger:1996_a,Zhang:2004_c} are examples of  spin-dependent phenomena which can be understood within the basically non-relativistic two-channel  model of conduction in ferromagnets \cite{Mott:1964_a}, and in which spins are transported between at least two non-collinear parts of a non-uniform magnetic structure with the magnetization in one part serving as a reference to the other one. Besides these more commonly considered spintronic effects, (Ga,Mn)As studies have extensively focused on relativistic phenomena which in principle can be observed in  uniform magnetic structures and where the spin-dependence of the transport stems from the internal spin-orbit coupling in carrier bands.  An archetypical example among these effects is the anisotropic magnetoresistance (AMR) discovered by Kelvin more than 150 years ago in wires of Ni and Fe \cite{Thomson:1857_a}. Research in (Ga,Mn)As led to the observation of a tunneling anisotropic magnetoresistance (TAMR)  \cite{Brey:2004_b,Gould:2004_a}. Unlike the TMR which corresponds to the different resistances of the parallel and antiparallel magnetizations in two magnetic electrodes separated by the tunnel barrier, the TAMR relies on the rotation of the magnetization in a single magnetic electrode while the other electrode can be non-magnetic. Huge and electrically tuneable relativistic anisotropic magneto-transport phenomena were observed in the Coulomb blockade (CB) devices  in which (Ga,Mn)As formed the island or the gate electrode of a single electron transistor (SET) \cite{Wunderlich:2006_a,Schlapps:2009_a,Ciccarelli:2012_a}. The TAMR and CB-AMR were subsequently  reported in other systems including common transition-metal ferromagnets and antiferromagnets \cite{Moser:2006_a,Gao:2007_a,Park:2008_a,Bernand-Mantel:2009_a,Park:2010_a}.  

For the inverse magneto-transport effects, the relativistic counterpart of the STT is the  current induced spin-orbit torque (SOT) \cite{Bernevig:2005_c,Manchon:2008_b}. Similar to the TAMR/CB-AMR, the SOT can be observed in uniform magnets, the seminal experiment was performed in (Ga,Mn)As \cite{Chernyshov:2009_a}, and subsequently the phenomenon was reported in other systems including transition metal ferromagnets \cite{Miron:2010_a}.  For the SOT, the above mentioned favorable characteristics of (Ga,Mn)As, namely the strong spin-orbit coupling in the carrier bands and exchange coupling of carrier spins with the dilute local moments, combines with the broken space-inversion symmetry in the host zinc-blende lattice. The broken space-inversion symmetry  is a necessary condition for observing the relativistic SOT \cite{Bernevig:2005_c,Manchon:2008_b}. 

Theoretical studies of the intrinsic nature of the anomalous Hall effect (AHE) \cite{Luttinger:1958_a,Onoda:2002_a,Jungwirth:2002_a} and experiments in (Ga,Mn)As interpreted by this theory  \cite{Jungwirth:2002_a,Nagaosa:2010_a} have inspired a renewed interest in the AHE in a broad class of ferromagnets \cite{Nagaosa:2010_a}. Simultaneously they led to predictions of a directly related intrinsic spin Hall effect (SHE) \cite{Murakami:2003_a,Sinova:2004_a} in which the spin-dependent transverse deflection of electrons originating from the relativistic band structure occurs in a non-magnetic conductor. The intrinsic SHE proposal triggered an intense theoretical debate and prompted the experimental discovery of the phenomenon \cite{Kato:2004_d,Wunderlich:2004_a}. The SHE has become a common tool to electrically detect or generate spin currents \cite{Jungwirth:2012_a} and the intrinsic SHE combined with the STT can allow for an in-plane current induced switching of  the free magnetic electrode in a TMR magnetic tunnel junction \cite{Liu:2012_a}. An intense discussion has ensued on  the alternative, SHE-STT based or SOT based interpretations  of these  in-plane current induced spin reorientation effects \cite{Miron:2011_b,Liu:2012_a,Garello:2013_a}. Research in (Ga,Mn)As continues to contribute to this research area in a distinct way; experimental and theoretical studies in (Ga,Mn)As have uncovered that the intrinsic SHE and SOT can be linked by a common  microscopic  origin \cite{Kurebayashi:2013_a}, the same one that was originally proposed for interpreting  the AHE data in (Ga,Mn)As \cite{Jungwirth:2002_a}.  

The SHE, STT, and SOT phenomena are at the forefront of the research field of electrically controlled spin manipulation  and play an important role in the development of a new generation of magnetic random access memories (MRAMs), tunable oscillators, and other spintronic devices \cite{Ralph:2007_a,Chappert:2007_a}. Optical excitations of magnetic systems by laser pulses have traditionally represented a complementary research field whose aim is to explore magnetization dynamics at short time scales and enable ultrafast spintronic devices \cite{Kirilyuk:2010_a}. The optical counterparts of the STT and SOT, in which current carriers are replaced by photo-carriers and which have been identified in laser induced spin dynamics studies in (Ga,Mn)As \cite{Rossier:2003_a,Nunez:2004_b,Nemec:2012_a,Tesarova:2012_b}, build a bridge between these two important fields of spintronics research.  The direct-gap GaAs host allowing for the generation of a high density of photo-carriers, optical selection rules linking light and carrier-spin polarizations, and the carrier spins interacting with magnetic moments on Mn via exchange coupling make (Ga,Mn)As a unique ferromagnetic system for exploring the interplay of photonics and spintronics. 

Thermopower, also known as the Seebeck effect, is the ability of conductors to generate electric voltages from thermal gradients. A subfield of spintronics, termed spin-caloritronics, explores the possibility of controlling charge and spin by heat and {\em vice versa} \cite{Bauer:2012_a}. In (Ga,Mn)As, experiments on the anomalous Nernst effect (ANE) \cite{Pu:2008_a}, which is the spin-caloritronics counterpart to the AHE, confirmed the validity of the Mott relation between the off-diagonal electrical and thermal transport coefficients in a ferromagnet \cite{Wang:2001_b}.  The experiments also firmly established the intrinsic nature of both the AHE and ANE in metallic (Ga,Mn)As. The anisotropic magneto-thermopower (AMT)  \cite{Ky:1966_a}  is a phenomenon in which the Seebeck coefficient of a uniform magnetic conductor depends on the angle between the applied temperature gradient and magnetization. Measurements of this counterpart to the AMR electrical-transport effect in (Ga,Mn)As \cite{Pu:2006_a} initiated a renewed interest in the phenomenon in a broad class of magnetic materials \cite{Wisniewski:2007_a,Tang:2011_a,Mitdank:2012_a,Anwar:2012_a}. The spin-caloritronic counterpart of the TMR effect in magnetic tunnel junctions is observed when the voltage gradient across the junction is replaced with a temperature gradient. The resulting tunneling magneto-thermopower (TMT) represents the difference between the Seebeck coefficients for the parallel and antiparallel magnetizations of the tunnel junction electrodes \cite{Walter:2011_a,Liebing:2011_a}. The relativistic analogue in a tunnel junction with only one magnetic electrode is the tunneling anisotropic magneto-thermopower (TAMT)  whose  observation was reported in (Ga,Mn)As \cite{Naydenova:2011_a}, reminiscent of the discovery of the TAMR \cite{Gould:2004_a}. Another spin-caloritronics effect which is distinct from the magneto-thermopower (magneto-Seebeck) phenomena is the spin-Seebeck effect \cite{Uchida:2008_a,Uchida:2010_a,Jaworski:2010_a,Sinova:2010_b}. Here the thermal gradient in a ferromagnet induces a spin-current which is then converted into electrical voltage via, e.g., the SHE in an attached non-magnetic electrode \cite{Uchida:2008_a,Uchida:2010_a,Jaworski:2010_a,Sinova:2010_b}. Experiments in (Ga,Mn)As \cite{Jaworski:2010_a} provided a direct evidence that, unlike the Seebeck effect in normal conductors, the spin-Seebeck effect does not originate from charge flow. The intriguing origin of the spin-Seebeck effect has been extensively debated \cite{Bauer:2012_a,Tikhonov:2013_a} since these seminal experiments.

In Section~\ref{material} we provide an overview of the material properties of (Ga,Mn)As with the emphasis on characteristics that make (Ga,Mn)As a favorable model system for spintronics research. For more detailed discussions of the materials aspects of the research of (Ga,Mn)As  in the context of  the family of (III,Mn)V and other magnetic materials we refer to other comprehensive review articles \cite{Matsukura:2002_a,Dietl:2003_a,Jungwirth:2006_a,Sato:2010_a,Dietl:2013_a}. The focus of this review are the spin-dependent phenomena and devices concepts explored in (Ga,Mn)As, and their relevance within the broad spintronics research field. These are discussed in Section~\ref{spintronics}. Our aim is to find conceptual links between the seemingly diverse areas of spintronic studies in (Ga,Mn)As. Simultaneously, we attempt  to provide intuitive physical pictures of the spin-dependent phenomena and functionalities for not only describing the specific observations in the ferromagnetic semiconductor (Ga,Mn)As but also for highlighting their applicability to other materials including the common transition metal ferromagnets, and other types of magnetic-order such as   antiferromagnets. While (Ga,Mn)As and the related ferromagnetic semiconductors have so far failed to allow for practical spintronic functionalities at room temperature, transition metal ferromagnets are commonly used in commercial spintronic devices \cite{Chappert:2007_a} and antiferromagnets can readily combine room temperature operation with not only metal but also semiconductor electronic structure \cite{Jungwirth:2010_a}. In Section~\ref{sum} we provide a brief summary of the spintronics research directions inspired by (Ga,Mn)As.

\section{Test-bed material for spintronics research} 
\label{material}

\subsection{Electronic structure and magnetism in (Ga,Mn)As}
\label{electronic_magnetism}
The elements in the (Ga,Mn)As compound have
nominal atomic structures
[Ar]$3d^{10}4s^2p^1$ for Ga, [Ar]$3d^{5}4s^2$ for Mn, and
[Ar]$3d^{10}4s^2p^3$ for As.  This circumstance correctly suggests
that the most stable position of Mn in the GaAs host lattice, at least up to a certain level of Mn doping, is
on the Ga site  where its two $4s$-electrons can participate in
crystal bonding in much the same way as the two Ga $4s$-electrons.
Because of the missing valence $4p$-electron,
the substitutional Mn$_{\rm Ga}$ impurity acts as an acceptor.
In the electrically neutral state, the isolated Mn$_{\rm Ga}$ has the character of
a local moment with zero angular momentum and spin $S=5/2$ (Land\'e g-factor $g=2$) due to the five $3d$ electrons and a moderately bound hole. GaAs is an intermediate band-gap III-V semiconductor, with $E_g=1.5$~eV at low temperatures. The experimental acceptor binding energy of an isolated Mn impurity substituting for Ga is of an intermediate strength,
$E_a^0\approx0.1$~eV \cite{Chapman:1967_a,Blakemore:1973_a,Bhattacharjee:2000_a,Yakunin:2004_b,Madelung:2003_a}.

The perturbation of the crystal potential of GaAs due to a single Mn impurity has three main components \cite{Masek:2010_a}. (i) The first is the long-range hydrogenic-like potential
of a single acceptor in GaAs which alone would produces a bound state at about 30 meV above the valence band \cite{Marder:1999_a}.
(ii) The  second contribution is a short-range central-cell potential. It is specific to a given impurity and reflects the difference in the electro-negativity of the impurity and the host atom \cite{Harrison:1980_a}. For a conventional non-magnetic acceptor Zn$_{\rm Ga}$, which is the 1st nearest neighbor of Ga in the periodic table, the atomic {\em p}-levels are shifted  by $\sim 0.25$~eV which increases the binding energy by $\sim 5$~meV. For Mn, the 6th nearest neighbor of Ga, the {\em p}-level shift is $\sim 1.5$~eV which when compared to Zn$_{\rm Ga}$ implies
the  central-cell contribution   to the acceptor level of Mn$_{\rm Ga}$  $\sim 30$~meV \cite{Bhattacharjee:2000_a}.
(iii) The remaining part of the Mn$_{\rm Ga}$ binding energy is due to the spin-dependent hybridization of Mn $d$-states  with neighboring As $p$-states. Its contribution, which has been directly inferred from spectroscopic measurements of uncoupled Mn$_{\rm Ga}$ impurities \cite{Schneider:1987_a,Linnarsson:1997_a,Bhattacharjee:2000_a}, is again comparable to the binding energy of the hydrogenic single-acceptor potential. Combining  (i)-(iii) accounts for the experimental binding energy of the Mn$_{\rm Ga}$ acceptor of 0.1~eV. An important caveat to these elementary considerations is that the short-range potentials alone of strengths inferred in (ii) and (iii)  would not produce a bound-state above the top of the valence band but only a broad region of scattering states inside the valence band.

\begin{figure}[h!]
\hspace*{-.3cm}\includegraphics[width=1\columnwidth,angle=0]{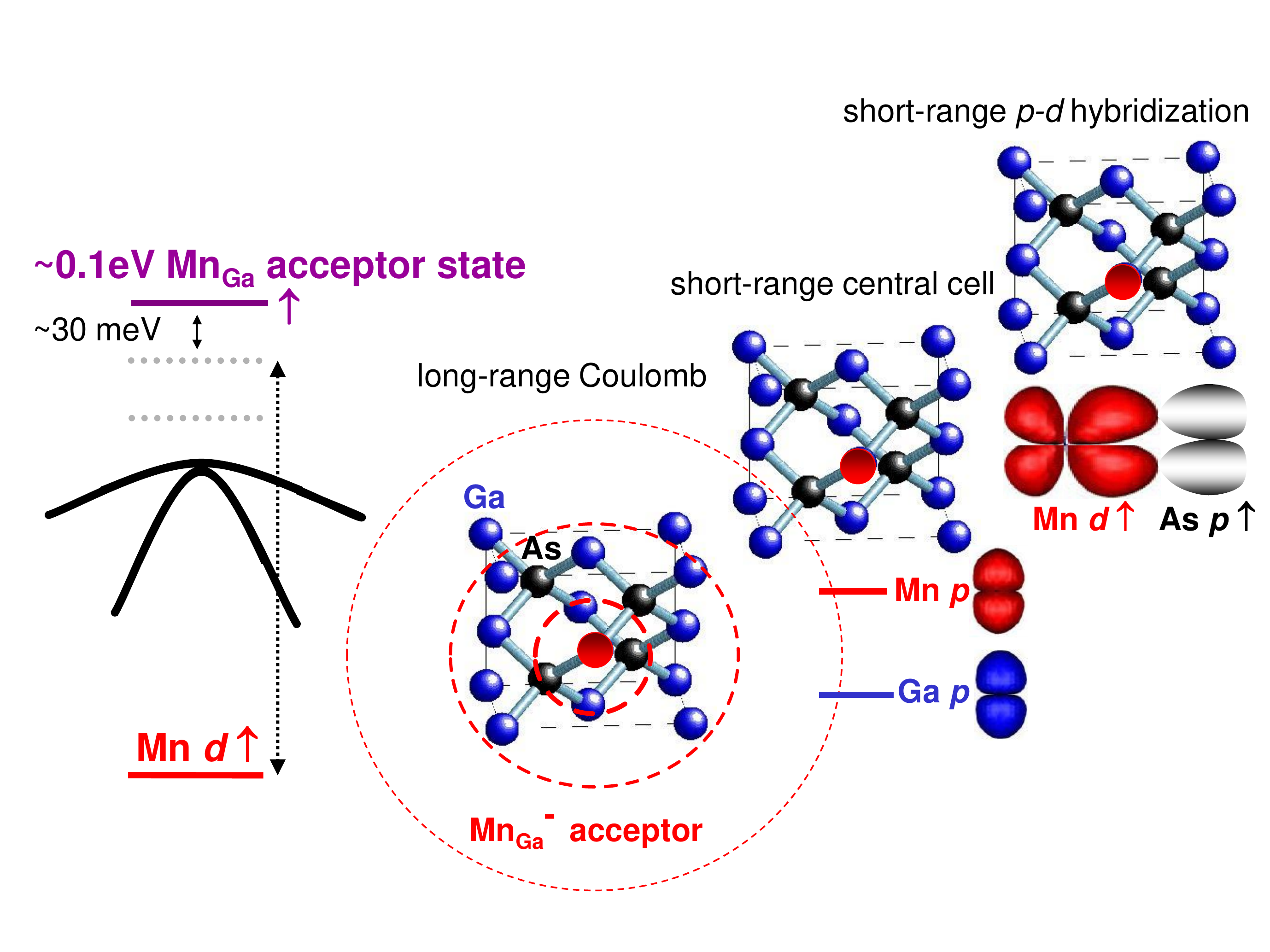}

\caption{(Color online) Schematic illustration of the long-range Coulomb and the two short-range potentials each contributing $\sim 30$~meV to the binding energy of the Mn$_{\rm Ga}$ acceptor. From Supplemental Material of \cite{Masek:2010_a}.
}
\label{Mn_cartoons}
\end{figure}

The low-energy
degrees of freedom in (Ga,Mn)As materials are the orientations of
Mn local moments and the occupation numbers of acceptor levels
near the top of the valence band.  The number of local moments  and the number
of holes
may differ from the number of
Mn$_{\rm Ga}$ impurities in the GaAs host due to the presence of charge and moment compensating defects.
Hybridization between
Mn $d$-orbitals and valence As/Ga $sp$-orbitals, mainly the As $p$-orbitals on the neighboring
sites, leads to an antiferromagnetic exchange interaction between the
spins that they carry \cite{Schneider:1987_a,Linnarsson:1997_a,Bhattacharjee:2000_a,Okabayashi:1998_a}.

At concentrations $\ll 1$\% of substitutional Mn,
the average distance between Mn impurities (or between holes
bound to Mn ions) is much larger than the size of the bound hole
characterized approximately by the impurity effective Bohr radius. These very dilute (Ga,Mn)As
systems are insulating, with the holes occupying a narrow impurity band, and paramagnetic.
Experimentally, ferromagnetism in (Ga,Mn)As is
observed when Mn doping reaches approximately 1\% and the system is still below but near the insulator-to-metal transition \cite{Ohno:1999_a,Campion:2003_b,Potashnik:2002_a,Jungwirth:2007_a}.
 ($x=1$\% Mn-doping corresponds
to Mn density $c=4 x/a^3=2.2\times 10^{20}$~cm$^{-3}$ where $a$ is the lattice constant
in Ga$_{1-x}$Mn$_x$As.)

At these Mn concentrations,
the localization length of the holes is extended to a degree that allows them to mediate, via the $p-d$
hybridization,
ferromagnetic exchange interaction between  Mn local moments,
even though the moments are dilute. 

Beyond a critical Mn doping, which in experiments is about 1.5\%, Mn
doped GaAs exhibits a  transition to a state in which the Mn
impurity levels overlap sufficiently strongly that the ground state
is metallic, {\em i.e.}, that states at the Fermi level are not bound
to a single or a group of Mn atoms but are delocalized across the
system \cite{Matsukura:2002_a,Jungwirth:2006_a,Jungwirth:2007_a}. In the metallic regime
Mn can, like a shallow acceptor (C, Be, Mg, Zn, e.g.),
provide delocalized holes with a low-temperature density
comparable to Mn density \cite{Ruzmetov:2004_a,MacDonald:2005_a,Jungwirth:2005_b}.
The transition to the metallic state
occurs at Mn density which is about two orders of magnitude larger than in GaAs doped with shallow acceptors \cite{Silva:2004_a}.  This is because of the central cell and $p-d$~hybridization contributions to the binding energy which make Mn acceptors  more
localized than the shallow acceptors.  
A crude estimate of the critical metal-insulator transition density can be obtained with a
short-range potential model, using the experimental binding energy
and assuming an effective mass of valence band holes, $m^{\ast}=0.5m_e$.  This model implies an isolated
acceptor level with effective Bohr radius $a_0=(\hbar^2/2m^{\ast}E_a^0)^{1/2}=10\,\AA$.
The radius $a_0$ then equals the Mn impurity spacing scale $c^{-1/3}$
at $c\approx 10^{21}$~cm$^{-3}$. This explains qualitatively the higher
metal-insulator-transition critical density in Mn doped GaAs compared to
the case of systems doped with shallow, more hydrogenic-like acceptors
which have binding energies $E_a^0\approx 30$~meV \cite{Madelung:2003_a,Silva:2004_a}.

Unlike the metal-insulator phase transition, which is sharply defined
in terms of the temperature $T=0$ limit of the conductivity, the crossover
in the character of states near the Fermi level in semiconductors with increased doping is
gradual \cite{Shcklovskii:1984_a,Lee:1985_a,Paalanen:1991_a,Jungwirth:2006_a,Dietl:2007_b,Dietl:2007_d}.
At very weak doping, the Fermi level resides inside
a narrow  impurity band (assuming some compensation) separated from the valence band by an energy gap of a magnitude
close to the impurity binding energy.
In this regime strong electronic correlations are
an essential element of the physics and a single-particle picture has limited
utility.  Well into the metallic state, on the other hand, the impurities are sufficiently
close together, and the long-range Coulomb potentials which contribute to the binding
energy of an isolated impurity are sufficiently screened, that
the system can be viewed as an imperfect crystal with disorder-broadened and
shifted host bands.
In this regime, electronic correlations are usually less strong
and a single-particle picture often suffices. The short-range components of the
Mn binding energy in GaAs, which are not screened by the carriers, move the crossover to higher
dopings and contribute significantly to carrier scattering in the metallic state. The picture of disorder-broadened and
shifted Bloch bands has to be applied, therefore, with care even in the most metallic (Ga,Mn)As materials. While for some properties it may provide even a semiquantitatively reliable description for other properties it may fail, as we discuss in more detail below. 

Although neither picture is very helpful for describing the physics in the crossover regime
which spans some finite
range of dopings, the notion of the impurity band on the lower doping side from the crossover
and of the disordered exchange-split host band on the higher doping side from
the crossover  still have a clear qualitative meaning. The former implies that
 there is a deep minimum in the density-of-states
between separate impurity and host band states.
In the latter case the impurity band and the host band merge into one inseparable
band whose tail may still contain localized states depending on the carrier concentration and disorder. In metallic ferromagnetic (Ga,Mn)As materials, hard X-ray
angle-resolved photoemission \cite{Gray:2012_a} and the differential off- and on-resonance photoemission \cite{Marco:2013_a} data do not show  a separation or intensity drop near the Fermi energy that would indicate the presence of a gap between the valence band and a Mn impurity band. The host and impurity bands are merged in ferromagnetic (Ga,Mn)As according to these spectroscopic measurements. Note that terms overlapping and merging impurity and valence bands describe the same
basic physics in (Ga,Mn)As. This is because
the Mn-acceptor states span several unit cells even in the very dilute
limit and  many unit cells as the impurity band broadens with increasing doping.
The localized and the delocalized
states then have a similarly
mixed As-Ga-Mn {\em spd}-character. This applies to
systems on either side of the metal-insulator transition. 
By recognizing that the bands are merged, that is, overlapped and mixed, in ferromagnetic (Ga,Mn)As materials, the distinction between ÔvalenceÕ and ÔimpurityÕ states becomes mere semantics which can lead to seemingly controversial statements on the material's electronic structure  but has no fundamental physics relevance.

\begin{figure}[h!]
\hspace{0cm}\includegraphics[width=.7\columnwidth,angle=-90]{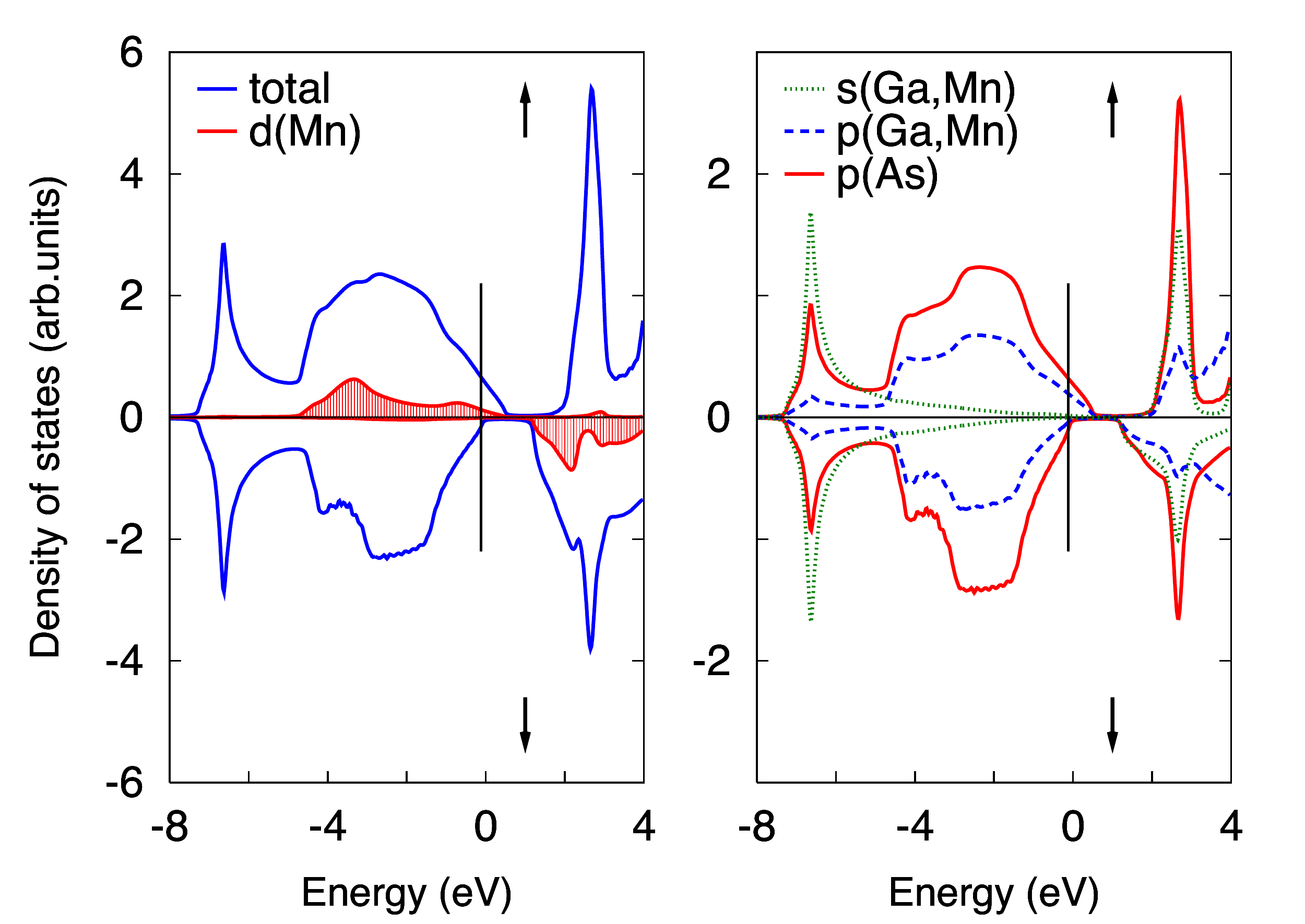}
\caption{(Color online) TBA-Anderson density of states of
Ga$_{0.9}$Mn$_{0.1}$As and its orbital composition. The position of
the Fermi energy is indicated by a vertical line. From Supplemental Material of \cite{Masek:2010_a}.
}
\label{TBA_10percent}
\end{figure}

A microscopic theory directly linked to the above qualitative considerations is based on the $spd$ tight-binding approximation (TBA) Hamiltonian of (Ga,Mn)As in which electronic correlations on the localized Mn $d$-orbitals are treated using the Anderson model of the magnetic impurity \cite{Masek:2010_a}. In  Fig.~\ref{TBA_10percent} we plot an examples of the total and orbital resolved densities of states (DOSs) for 10\% of Mn$_{\rm Ga}$ impurities. The Mn-{\em d} spectral weight is peaked at several eV's below the top of the valence band, in agreement with photoemission data \cite{Okabayashi:1998_a,Gray:2012_a,Marco:2013_a}, and is significantly smaller near the Fermi energy $E_F$. The Fermi level states at the top of the valence band have a dominant As(Ga) $p$-orbital character. The {\em p-d} coupling strength, $N_0\beta\equiv N_0J_{ex}=\Delta/(Sx)$ ($N_0=1/\Omega_{u.c.}$ where $\Omega_{u.c.}$ is the unit cell volume) \cite{Jungwirth:2006_a}, determined from the calculated valence band exchange splitting $\Delta$ (and  taking $S=5/2$) is close to the upper bound of the reported experimental range of $N_0\beta\sim 1-3$~eV \cite{Matsukura:1998_a,Okabayashi:1998_a,Szczytko:1999_a,Bhattacharjee:2000_a,Omiya:2000_a}. This is regarded as a moderately weak {\em p-d} coupling because the corresponding Fermi level states of the (Ga,Mn)As have a similar orbital character to the states in the host GaAs valence band. These spectral features  are among the key characteristics of the hole mediated ferromagnetism in (Ga,Mn)As. 

The effective Hamiltonian theory of (Ga,Mn)As, based on the kinetic-exchange (Zener) model \cite{Dietl:1997_a,Jungwirth:1999_a,Dietl:2000_a,Jungwirth:2006_a}, assumes also a value of $N_0\beta$ within the above experimental range, namely $N_0\beta=1.2$~eV ($J_{ex}=55$~meV~nm$^3$) which is closer to the lower experimental bound \cite{Jungwirth:2006_a}. It is this moderate {\em p-d} hybridization that allows it to be treated perturbatively and to perform the Schrieffer-Wolff transformation from the microscopic TBA-Anderson Hamiltonian to the effective model in which valence band states experience a spin-dependent kinetic-exchange field \cite{Jungwirth:2006_a}. Hence, the effective kinetic-exchange model and the microscopic TBA-Anderson theory provide a consistent physical picture of ferromagnetic (Ga,Mn)As. These two models of the electronic structure of (Ga,Mn)As have represented the most extensively used basis for analyzing the spin-dependent phenomena and device functionalities in (Ga,Mn)As. 

\begin{figure}[h!]
\hspace{0cm}\includegraphics[width=.7\columnwidth,angle=-90]{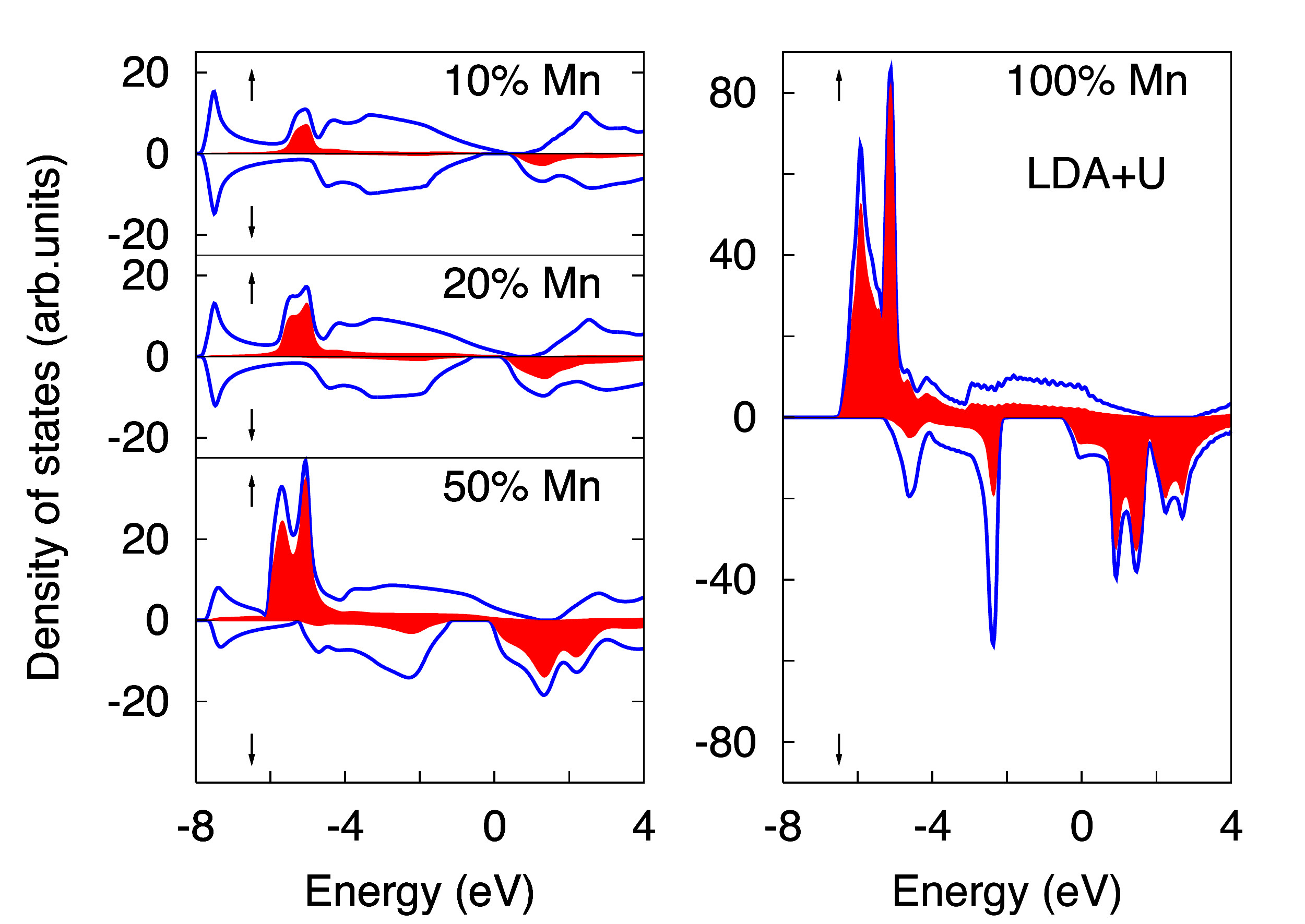}
\caption{(Color online) Density of states for (Ga,Mn)As mixed crystals with various content of Mn obtained in the GGA+U theory. Blue line represents the total DOS while the red  area shows the partial density of Mn $d$-states. From Supplemental Material of \cite{Masek:2010_a}.
}
\label{LDA+U}
\end{figure}

In Fig.~\ref{LDA+U} we show DOSs over the entire Mn$_{\rm Ga}$ doping range obtained from the GGA+U density functional calculations \cite{Masek:2010_a,Sato:2010_a}. The GGA+U, the TBA-Anderson, and the kinetic-exchange Zener theories all provide a consistent picture of the band structure of ferromagnetic (Ga,Mn)As. Simultaneously, it is important to keep in mind that the moderate acceptor binding energy of Mn$_{\rm Ga}$ shifts the insulator-to-metal transition to orders of magnitude higher doping densities than in the case of common shallow non-magnetic acceptors, as mentioned above \cite{Jungwirth:2007_a,Masek:2010_a}. Disorder and correlation effects, therefore, play a comparatively more significant role in (Ga,Mn)As than in  degenerate semiconductors with common shallow dopants and any simplified one-particle band picture of ferromagnetic (Ga,Mn)As can only represent a proxy to the electronic structure of the material.

As seen in Fig.~\ref{LDA+U}, the bands evolve continuously from the intrinsic non-magnetic semiconductor GaAs, via the degenerate ferromagnetic semiconductor (Ga,Mn)As to the ferromagnetic metal MnAs. From this it can by expected that $T_c$ of MnAs, with the value close to room temperature (350~K for cubic MnAs inclusions in (Ga,Mn)As \cite{Yokoyama:2005_a,Kovacs:2011_a}), sets the upper theoretical bound of achievable $T_c$'s in (Ga,Mn)As across the entire doping range. In experiment, as we discuss in Section~\ref{trends}, the Mn$_{\rm Ga}$ doping  is limited to approximately 10\% with corresponding $T_c$ reaching 190~K in uniform thin-film crystals prepared by optimized LT-MBE synthesis and post-growth annealing. In these samples the hole density is in the $\sim 10^{20}-10^{21}$~cm$^{-3}$ range, i.e., several orders of magnitude higher then densities in commonly used non-magnetic semiconductors but also 1-2 orders of magnitude lower than is typical for metals.

\subsubsection{Curie point singularities}
\label{Curie}

Ferromagnetic (Ga,Mn)As with Mn doping ranging from $\sim 1$ to $\sim 10$\% is a very heavily doped compound semiconductor or can be also regarded at these high Mn concentrations as a random alloy. Quantities like the residual resistivity are then inevitably affected by strong disorder effects.  Even in the most metallic (Ga,Mn)As materials
the hole mean free path is comparable to the separation of the Mn impurities so the diffusivity is low. Typically, the product of the Fermi wavevector and the mean free path is, $k_F \Lambda=\hbar\mu k_F^2/e \sim 1-10$, estimated from the  experimental mobilities $\mu$ and hole densities \cite{Jungwirth:2007_a}. For thermodynamic properties, as well as for the spintronics effects discussed in Section~\ref{spintronics}, the disordered nature of (Ga,Mn)As can, however, play a less significant role. This makes the spin-dependent phenomena and device functionalities discovered and explored in (Ga,Mn)As  applicable to a broad class of materials  beyond the dilute moment ferromagnetic semiconductor compounds. 

\begin{figure}
\hspace*{-0.cm}\includegraphics[width=1\columnwidth,angle=0]{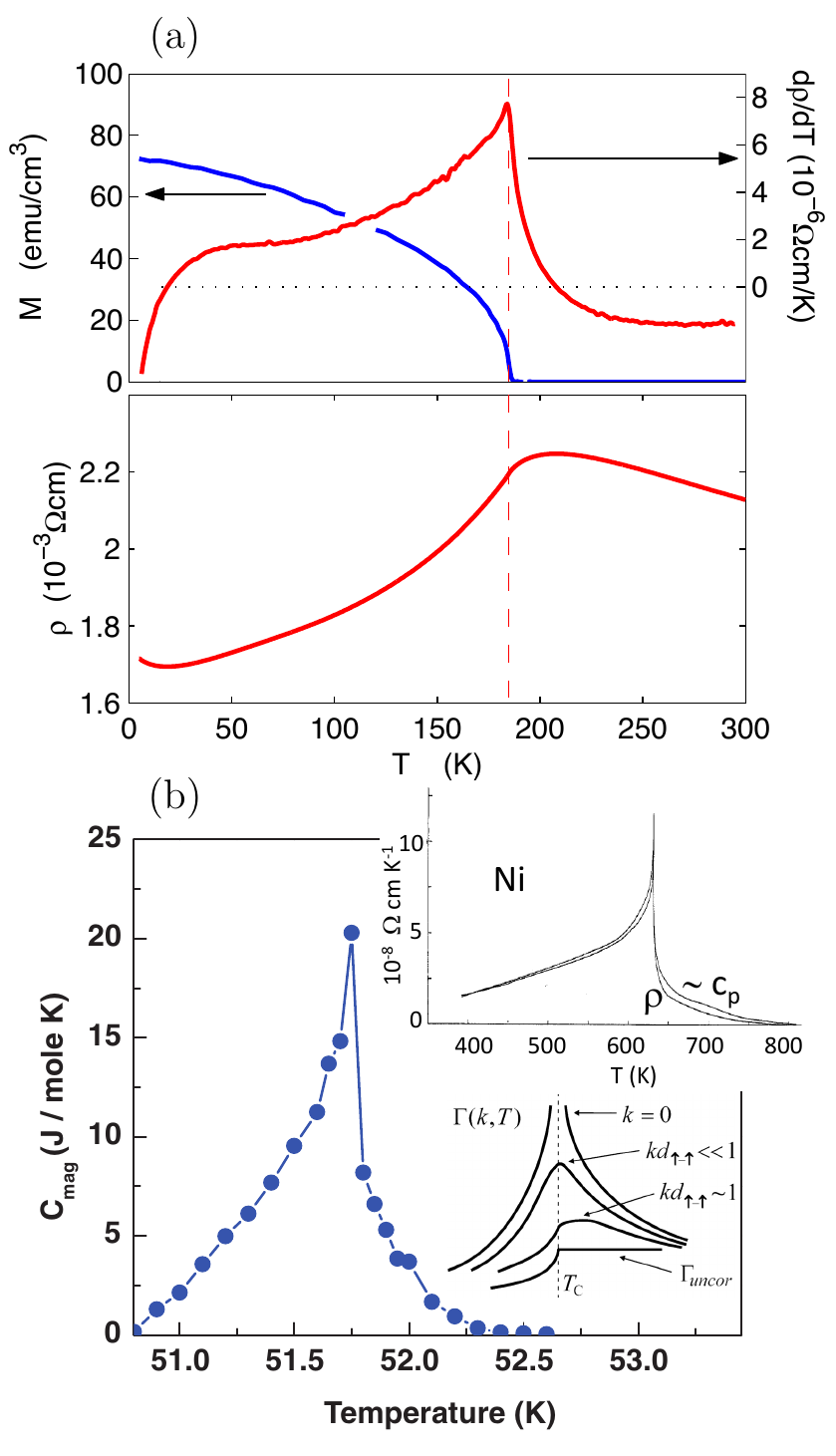}
%
%
%
%
%
%
%
%
%
\caption{(Color online)
(a) Temperature dependent remanent magnetization, resistivity and temperature derivative of the resistivity  of a nominally   11\% Mn-doped (Ga,Mn)As. From Supplementary information of  \cite{Jungwirth:2010_b}.
(b) Magnetic contribution to the specific heat of a 2.6\% Mn-doped (Ga,Mn)As.  Adapted from \cite{Yuldashev:2010_a}. Upper inset: Temperature derivative of resistance and a multiple of the specific heat plotted against temperature for Ni. From \cite{Joynt:1984_a}. Lower inset: Schematic diagram of the spin-spin correlation function in low and large $k$-vector limits \cite{Fisher:1968_a}.
}
\label{drdtfig1}
\end{figure}

An example of the seemingly surprising similarity between the basic magnetic characteristics of (Ga,Mn)As and the common transition metal ferromagnets such as Ni is shown in Fig.~\ref{drdtfig1}. Here we illustrate that (Ga,Mn)As can have Curie point singularities \cite{Novak:2008_a,Yuldashev:2010_a} which are typical of uniform itinerant ferromagnets \cite{Joynt:1984_a,Shacklette:1974_a}. Fig.~\ref{drdtfig1}(a) shows remanent magnetization $M(T)$ which vanishes sharply at $T\rightarrow T_c^-$. For the same 11\% Mn-doped sample, Fig.~\ref{drdtfig1}(a) also shows the resistivity $\rho(T)$ and its temperature derivative, $d\rho/dT$. While $\rho(T)$ has a broad shoulder near $T_c$, $d\rho/dT$ has a singularity at  $T_c$ which precisely coincides with $T_c$ inferred from the remanence measurement in the same (Ga,Mn)As material \cite{Novak:2008_a,Jungwirth:2010_b,Nemec:2012_b}. We explain below that the Curie point singularity in $d\rho/dT$ is related to the singularity in the specific heat which was also detected in (Ga,Mn)As \cite{Yuldashev:2010_a} and is shown in Fig.~\ref{drdtfig1}(b). The  specific heat measurements were performed in lower Mn-doped samples (Ga,Mn)As (2.6\% Mn-doping in Fig.~\ref{drdtfig1}(b)) and therefore the singularity occurs in these samples at a correspondingly lower $T_c$.

Since seminal works of de~Gennes and Friedel \cite{DeGennes:1958_a} and Fisher and Langer \cite{Fisher:1968_a}, critical behavior of resistivity has been one of the central problems in the physics of itinerant ferromagnets. Theories of coherent scattering from long wavelength spin fluctuations, based on the original paper by de~Gennes and Friedel, have been used to explain the large peak in the resistivity $\rho(T)$ at $T_c$ observed in Eu-chalcogenide dense-moment magnetic semiconductors \cite{Haas:1970_a}. The emphasis on the long wavelength limit of the spin-spin correlation function, reflecting critical behavior of the magnetic susceptibility,  is justified in these systems by the small density of carriers relative to the density of magnetic moments, and corresponding small Fermi wavevectors of carriers.

As pointed out by Fisher and Langer \cite{Fisher:1968_a}, the resistivity anomaly in high carrier density transition metal ferromagnets is qualitatively different and associated with the critical behavior of correlations between nearby moments. When approaching $T_c$ from above, thermal fluctuations between nearby moments are partially suppressed by short-range magnetic order.  Their singular behavior is like that of the internal energy and unlike that of the magnetic susceptibility. The singularity at $T_c$ occurs in $d\rho/dT$ and is closely related to the critical behavior of the specific heat. While Fisher and Langer expected this behavior for $T\rightarrow T_c^+$ and a dominant role of uncorrelated spin fluctuations at $T\rightarrow T_c^-$, later studies of elemental transition metals found a proportionality between $d\rho/dT$ and specific heat on both sides of the Curie point, as shown in the upper inset of Fig.~\ref{drdtfig1}(b) \cite{Joynt:1984_a,Shacklette:1974_a}.

The character of the transport anomaly in (Ga,Mn)As is distinct from the critical contribution to transport in the dense-moment magnetic semiconductors \cite{Haas:1970_a} and is reminiscent of the $d\rho/dT$ singularity in transition metal ferromagnets \cite{Joynt:1984_a,Shacklette:1974_a}. Ferromagnetism in (Ga,Mn)As originates from
spin-spin coupling between local Mn-moments and valence band holes, $J\sum_i\delta({\bf r}-{\bf R}_i){\boldsymbol\sigma}\cdot{\bf S}_i$ \cite{Dietl:1997_a,Jungwirth:1999_a,Dietl:2000_a,Jungwirth:2006_a}. Here ${\bf S}_i$ represents the local spin  and ${\boldsymbol\sigma}$ the hole spin operator. This local-itinerant exchange  interaction plays a central role in theories of the critical transport anomaly. When treated in the Born approximation, the interaction yields a carrier scattering rate from magnetic fluctuations, and the corresponding contribution to $\rho(T)$, which is proportional to the static spin-spin correlation function, $\Gamma({\bf R}_i,T)\sim J^2[\langle{\bf S}_i\cdot{\bf S}_0\rangle - \langle{\bf S}_i\rangle\cdot \langle{\bf S}_0\rangle]$ \cite{DeGennes:1958_a}. Typical temperature dependences of the uncorrelated part, $\Gamma_{uncor}({\bf R}_i,T)\sim \delta_{i,0}J^2[S(S+1) - \langle{\bf S}_i\rangle^2]$, and of the Fourier components of the correlation function, $\Gamma({\bf k},T)=\sum_{i\neq 0}\Gamma({\bf R}_i,T)\exp({\bf k\cdot R}_i)$, are illustrated in the lower inset of Fig.~\ref{drdtfig1}(b) \cite{Fisher:1968_a}. At small wavevectors, $\Gamma({\bf k},T)$ and correspondingly $\rho(T)$ have a peak at $T_c$. At $k$ similar to the inverse separation of the local moments ($kd_{\uparrow - \uparrow}\sim 1$) the peak broadens into a shoulder while the singular behavior at $T_c$ is in the temperature derivative of the spin-spin correlator and, therefore, in $d\rho/dT$.

$M^2$ expansion providing a good fit to the magnetic contribution to the resistivity at $T<T_c$ \cite{Novak:2008_a} corresponds to the dominant contribution from  $\Gamma_{uncor}$ on the ferromagnetic side of the transition. The shoulder in $\rho(T)$ on the paramagnetic side and the presence of the singularity in $d\rho/dT$  suggest that  large wavevector components of $\Gamma({\bf k},T)$ dominate the temperature dependence of the scattering in the  $T\rightarrow T_c^+$ critical region \cite{Novak:2008_a}. The large $k$-vector limit is consistent with the ratio between hole and Mn local-moment densities approaching unity in high quality (Ga,Mn)As materials with low charge compensation by unintentional impurities \cite{Nemec:2012_b}. 

\subsubsection{Localization effects in transport}
\label{localization}

While the sharp transport Curie point singularities highlight that ferromagnetic (Ga,Mn)As epilayers can have a high degree of uniformity \cite{Kodzuka:2009_a} and can behave similarly to common, weakly-disordered itinerant ferromagnets, the magnitude of the resistivity at zero and finite frequencies and over the broad temperature range is significantly affected by the vicinity of the metal-insulator transition in (Ga,Mn)As. The valence band calculations treating disorder in the first-order Born approximation overestimate the experimental conductivities of metallic (Ga,Mn)As by up to a factor of 10 \cite{Jungwirth:2002_c,Sinova:2002_a}. This discrepancy is  removed by accounting for strong disorder and localization effects using, e.g., exact-diagonalization calculations  \cite{Yang:2003_b,Jungwirth:2007_a}. Even the most metallic (Ga,Mn)As materials with delocalized carriers at the Fermi level may contain localized states in the valence band tail which modify the finite-frequency absorption spectra \cite{Burch:2006_a,Chapler:2011_a,Jungwirth:2007_a,Jungwirth:2010_b}.  

The low diffusivity of carriers implies that quantum interference and electron-electron interactions can produce sizable effects in (Ga,Mn)As. Weak localization (WL)
quantum corrections are due to constructive interference between partial waves undergoing multiple scattering from a state with wavevector $k$ to a state $-k$ and partial waves traversing the time reversed trajectory. The effect is also referred to as coherent backscattering and it leads to a reduction of the conductivity. A  distinct, electron-electron interaction quantum correction to the conductivity \cite{Lee:1985_a} can arise in disordered conductors which often has a similar magnitude to the WL correction. This arises because electron-electron interactions cannot be treated independently of the disorder scattering for strong disorder.

Explicit expressions for the WL corrections can be obtained for $L_{\Phi}\gg \Lambda\gg\lambda_F$, where $L_{\Phi}$, $\Lambda$, and $\Lambda_F$ are the carrier  phase coherence length, mean free path, and  Fermi wavelength. The second condition can be  rewritten as $k_F \Lambda\gg 1$ where $k_F$ is the Fermi wavevector. The corrections are of order $(k_F \Lambda)^{-1}$ and so become important for small $k_F \Lambda$. It has been argued  that higher order corrections are small and that the condition $k_F \Lambda\gg 1$ can be relaxed to $k_F \Lambda > 1$. Application of a magnetic field can suppress the resistance enhancement due to WL as it removes time-reversal invariance leading to negative magnetoresistance. The magnetic field begins to have a significant effect when $\ell_B\sim L_{\Phi}$, where $\ell_B=(\hbar/eB)^{1/2}$ is the magnetic length, and the magnetic field completely suppresses WL when $\ell_B\sim\Lambda$. Since WL quantum corrections are suppressed by  sufficiently large magnetic fields one expects a similar suppression by the internal magnetization. For dense moment ferromagnets like Fe, Ni, etc., $\mu_0M\sim 2$~T and the mean free path is usually quite large so WL is strongly suppressed. However WL is observed for example in highly disordered Ni films \cite{Aprili:1997_a}. For the dilute moment ferromagnet (Ga,Mn)As, $\mu_0M\sim 50$~mT while the field needed to suppresses WL, {\em i.e.} when $\ell_B\sim\Lambda\sim 1$~nm, is $\sim 1000$~T. So one expects WL effects to be present, and since typically $k_F \Lambda\sim 1-10$, they may be large.

The identification of WL contributions to the temperature dependence of resistance is difficult as they generally co-exist with other temperature dependent contributions and because the expected functional form can be very different for the different possible phase breaking mechanisms. In disordered ferromagnets like (Ga,Mn)As, spin disorder scattering can, e.g.,  produce large magnetoresistance, particularly close to the  localization boundary \cite{Kramer:1993_a,Nagaev:1998_a,Omiya:2000_a}.
\begin{figure}[h]
\vspace*{-0cm}

\hspace*{-0cm}\includegraphics[width=1\columnwidth,angle=-0]{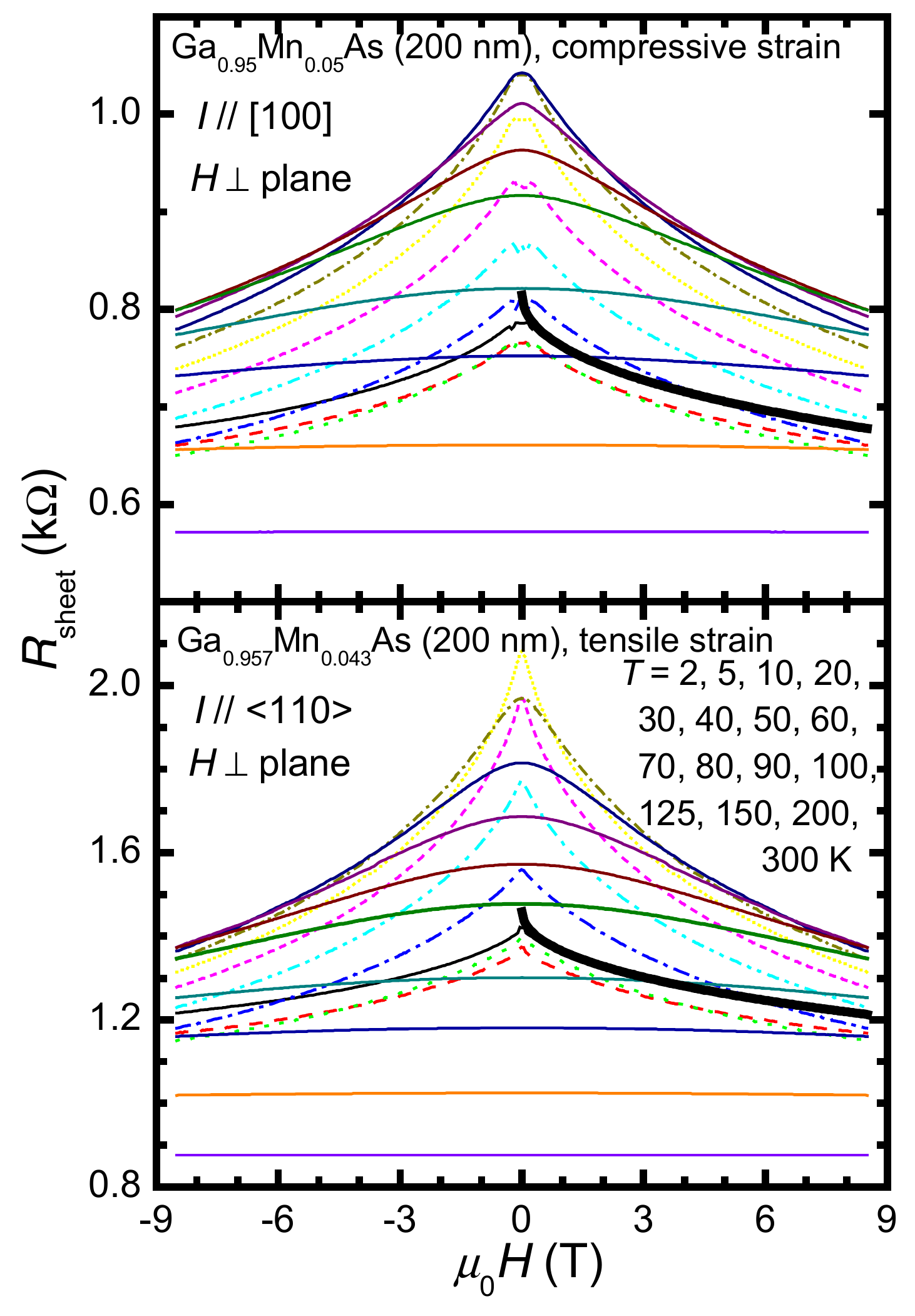}
\vspace*{-0.5cm} 
\caption{(Color online) Field and temperature dependencies of resistance in
Ga$_{0.95}$Mn$_{0.05}$As/GaAs (compressive strain, upper panel) and in
tensile strained Ga$_{0.957}$Mn$_{0.043}$As/(In,Ga)As (lower panel) for
magnetic field perpendicular to the film plane. Starting from up,
subsequent curves at $B = 0$ correspond to temperatures in K: 70,
60, 80, 50, 90, 40, 100, 30, 125, 20, 2, 5, 10, 150, 200, 300 (upper
panel) and to 50, 60, 40, 70, 30, 80, 90, 20, 100, 2, 10, 5,
125, 150, 200, 300 (lower panel).
From \cite{Matsukura:2004_a}.}
\label{fig_quant4}
\end{figure}

For external magnetic fields less than the coercive fields the magnetoresistance
response is usually dominated by
AMR (see Sections~\ref{Mott-Dirac},\ref{AHE-AMR}). At larger fields a negative isotropic magnetoresistance
is observed which can be very large for low conductivity material \cite{Matsukura:2004_a}. This could be due to the suppression of spin
disorder \cite{Lee:1987_a}.
However, as shown in Fig.~\ref{fig_quant4} \cite{Matsukura:2004_a},
the negative magnetoresistance does not seem to saturate, even in extremely strong magnetic fields.
It has been argued \cite{Matsukura:2004_a} that the negative magnetoresistance arises from WL and gives a correction consistent with the predicted form proportional to $-B^{1/2}$ \cite{Kawabata:1980_a}, which assumes a complete suppression of spin-disorder and spin-orbit scattering (see Fig.~\ref{fig_quant4}). 

The role of spin-orbit coupling in WL phenomena in (Ga,Mn)As has been extensively discussed  \cite{Neumaier:2007_a,Rokhinson:2007_a,Garate:2008_c}.
In the context of the spintronic phenomena and functionalities in (Ga,Mn)As and their applicability to other materials, discussed in Section~\ref{spintronics}, an important conclusion is drawn from numerical studies of WL in  (Ga,Mn)As \cite{Garate:2008_c}. They showed that while WL corrections can significantly contribute to the absolute residual resistivity, the relative changes in resistivity associated with magnetization reorientations, namely the AMR ratios, are nearly independent on whether the WL corrections are included or not \cite{Garate:2008_c}. These results, which agree qualitatively with analytical considerations on simpler models \cite{Bhatt:1985_a}, illustrate that the intrinsically strong disorder in (Ga,Mn)As can qualitatively play a minor role in not only the thermodynamic properties but also in the spintronic phenomena reflecting the interactions of carrier spins with electrical current, light, or heat. What determines these phenomena is primarily the magnetic exchange and spin-orbit fields acting on the carrier states. Disorder can mix the carrier states but as long as this mixing does not significantly alter the effects of the exchange field and spin-orbit coupling on the carriers the spintronic phenomena remain robust against disorder. This explains the qualitative and often semi-quantitative success, and justifies the applicability, of microscopic theories of spintronic phenomena in (Ga,Mn)As starting from  a Bloch-band description of the material's electronic structure. Simultaneously it should be noted that due to strong disorder and the vicinity of the metal-insulator transition a full quantitative description is unlikely to be achievable within any of the existing theoretical models of (Ga,Mn)As.

We conclude this section by discussing the universal conductance fluctuations (UCFs) in (Ga,Mn)As. These result from the interference between partial waves from scattering centers within a conductor. In the usual semi-classical theory of electron conduction this is neglected since it is assumed that such effects will be averaged away. However, for conductors of size comparable with $L_{\Phi}$ the interference effects are intrinsically non-self-averaging. This leads to corrections to the conductivity of order $e^2/h$. Application of a magnetic field modifies the interference effects, giving reproducible but aperiodic UCFs  \cite{Lee:1987_a}
of amplitude $\sim e^2/h$. One can think of a conductor with dimensions $>L_{\Phi}$
as made up of a number of independent phase coherent sub-units leading to averaging. UCFs are then diminished for dimensions $\gg L_{\Phi}$ and only WL due to the coherent backscattering may still contribute in macroscopic samples.

At temperatures which are a significant fraction of the Curie temperature one expects spin-disorder and spin-orbit scattering to lead to the phase coherence length $L_{\Phi}\sim\Lambda$,
strongly suppressing quantum corrections. However, in high quality metallic (Ga,Mn)As it has been argued
\cite{Matsukura:2004_a} that $L_{\Phi}$ need not be very small at low temperatures because virtually all spins contribute to the ferromagnetic ordering and the large splitting of the valence band makes both spin-disorder and spin-orbit scattering relatively inefficient. The strong magneto-crystalline anisotropies also tend to suppress magnon scattering at low temperatures.
\begin{figure}[h]
\vspace*{-0cm}

\hspace*{-0.5cm}\includegraphics[width=1\columnwidth,angle=-0]{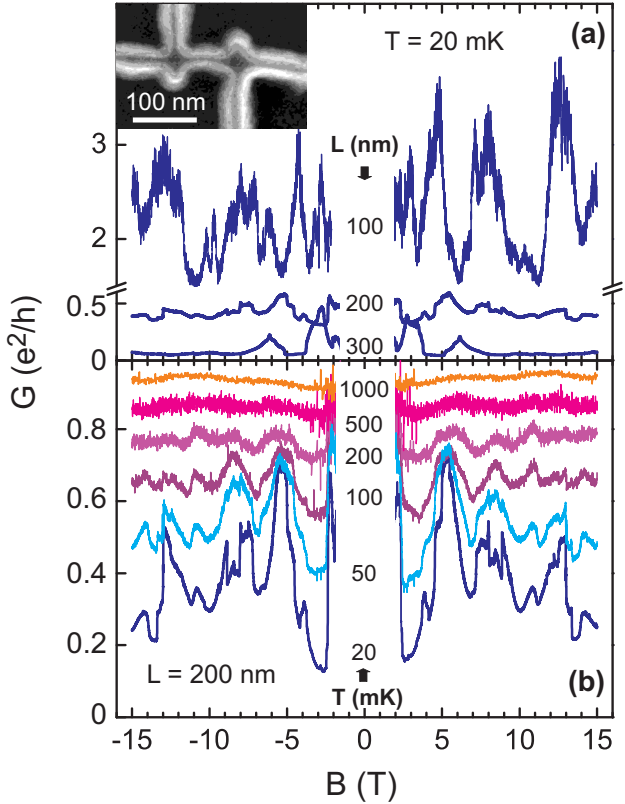}
\vspace*{-0cm} \caption{(Color online) (a) Conductance fluctuations for three
wires of different length $L$. For the shortest wire the amplitude
of the conductance fluctuations is about $e^2/h$, expected
for conductors with all spatial dimensions smaller or comparable
to $L_{\Phi}$. The inset shows an electron micrograph of a 20~nm
wide wire with a potential probe separation of $\sim 100$~nm.
(b) Conductance  vs. magnetic field of the 200~nm wire for different
temperatures between 20~mK and 1~K. From \cite{Wagner:2006_a}.}
\label{ucf}
\end{figure}

Recent observations \cite{Wagner:2006_a,Vila:2006_a}
of large UCFs in (Ga,Mn)As microdevices, and the evidence for the closely related Aharonov-Bohm effect (ABE) in (Ga,Mn)As microrings,
confirm that $L_{\Phi}$ can be large at low temperatures.
Fig.~\ref{ucf} shows UCFs measured \cite{Wagner:2006_a}
in (Ga,Mn)As wires of approximate width 20~nm and thickness 50~nm.
Panel (a) shows that the UCF amplitude is $\sim e^2/h$ in a 100~nm long wire at 20~mK.
This directly demonstrates that $L_{\Phi}\sim 100$~nm.
Similar measurements in higher conductivity (Ga,Mn)As give
$L_{\Phi}\sim 100$~nm at 100mK.
These are large values corresponding to a phase relaxation time that
is orders of magnitude larger than the elastic scattering time.

\begin{figure}[h]
\vspace*{-0cm}

\hspace*{-0.5cm}\includegraphics[width=1\columnwidth,angle=0]{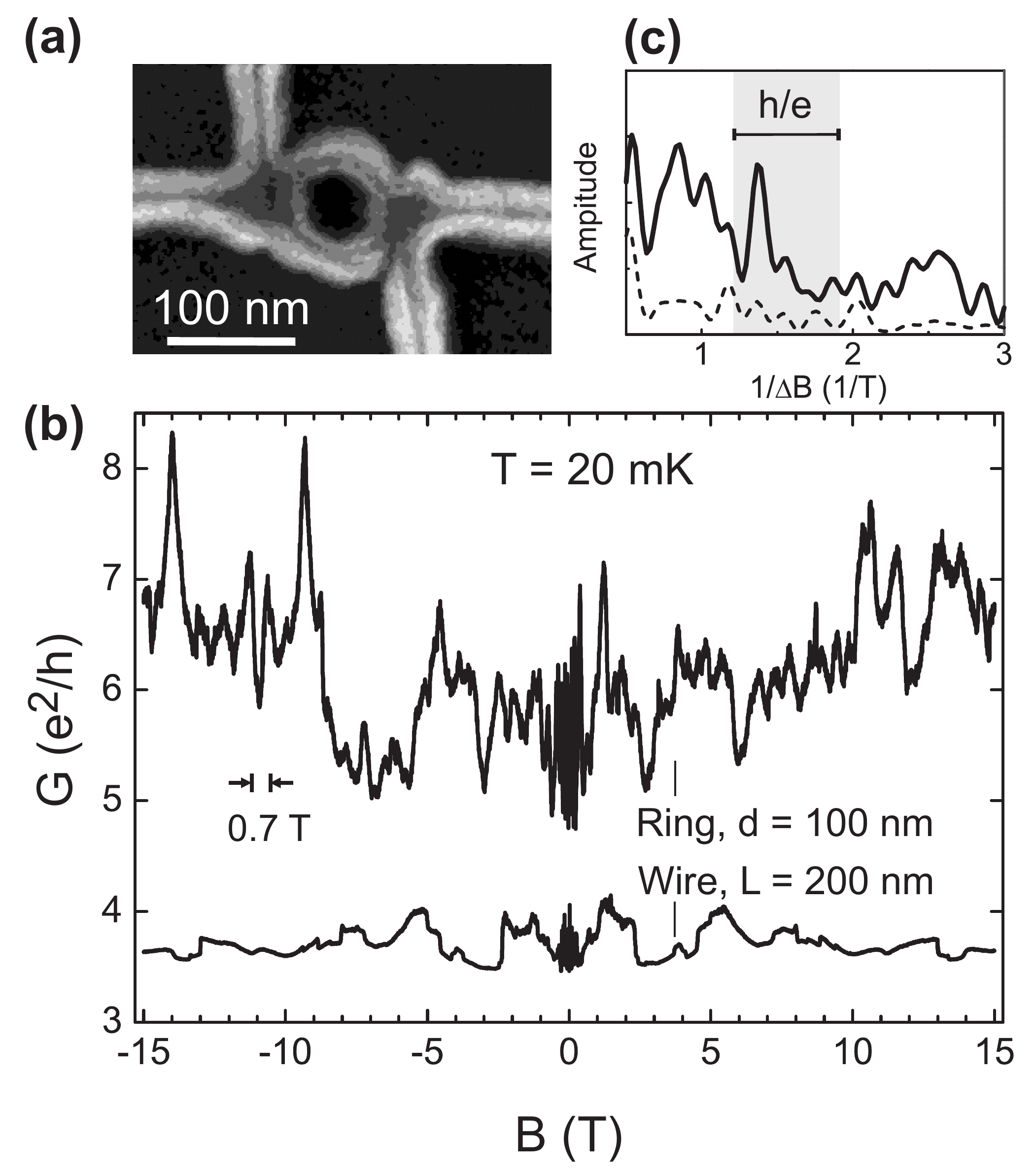}
\vspace*{0cm} \caption{(a) Electron micrograph of a (Ga,Mn)As ring sample
with a diameter of $\sim 100$~nm. (b) Comparison of the magnetoconductance
trace of the ring sample with the conductance
of a wire of comparable length and 20~nm width. (c) Corresponding
FFT taken from the conductance of ring and wire.
The region where  ABE oscillations are expected is highlighted.
From \cite{Wagner:2006_a}.}
\label{ABeffect}
\end{figure}

Fig.~\ref{ABeffect} shows measurements \cite{Wagner:2006_a}
of the magnetic field dependence of the conductivity of a lithographically defined
100~nm diameter (Ga,Mn)As ring compared to that of a 200~nm long (Ga,Mn)As wire.
Additional small period oscillations are observed for the ring which
the Fourier transform shows to be consistent with the expected ABE period.
This confirms the long  $L_{\Phi}$ indicated by the large amplitude UCFs and confirms that almost all spins are participating in the magnetic order with strong suppression of spin scattering.

\subsection{Doping trends in basic magnetic and transport properties of (Ga,Mn)As}
\label{trends}
\subsubsection{Low Mn-doped bulk materials}
\label{bulk}
Narrow impurity bands have been clearly observed in Mn doped GaAs samples with
carrier densities much lower than the metal-insulator transition density,
for example in equilibrium grown bulk
materials with Mn density $c=10^{17}-10^{19}$~cm$^{-3}$ \cite{Brown:1972_a,Woodbury:1973_a,Blakemore:1973_a}.
The energy gap between the impurity band and the valence band, $E_a$, can be measured
by studying the temperature dependence of longitudinal and Hall conductivities, which show activated
behavior because of thermal excitation of holes from the
impurity band to the much more conductive valence band \cite{Blakemore:1973_a,Woodbury:1973_a,Marder:1999_a}.

The activation energy decreases with increasing
Mn density 
\cite{Blakemore:1973_a}.
The lowering of impurity binding energies at larger $c$, which is expected to
scale with the mean impurity
separation, 
is apparent already in the equilibrium grown
bulk materials with $c=10^{17}-10^{19}$~cm$^{-3}$.   The degenerate semiconductor regime was, however, not reached in the bulk materials. 

\subsubsection{Synthesis of high Mn-doped epilayers}
\label{epilayers}
\begin{figure}[ht]
\hspace{0cm}\includegraphics[width=1\columnwidth,angle=0]{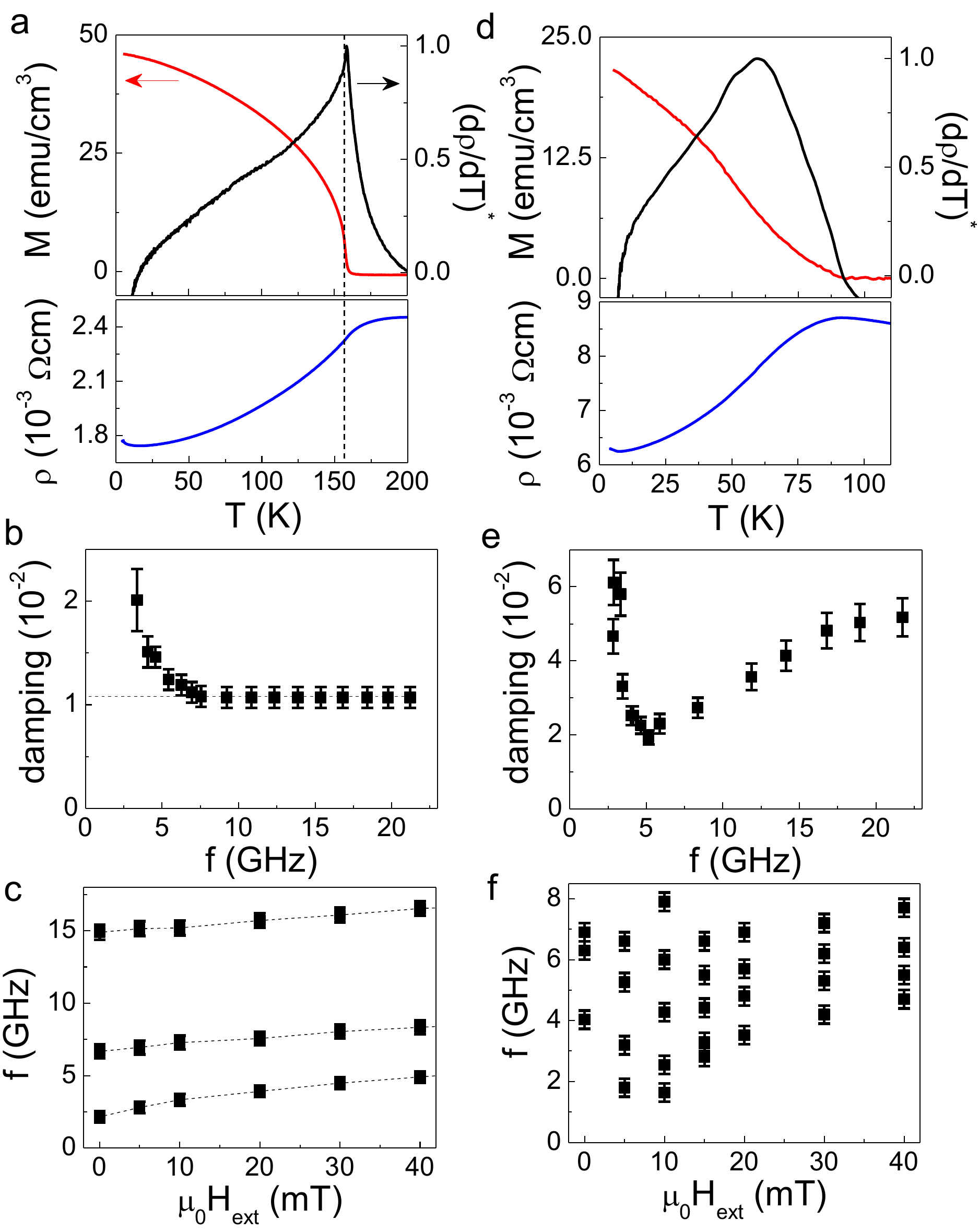}
\caption{
(Color online) (a) Magnetization $M$, temperature derivative of the resistivity normalized to the peak value $(d\rho/dT)^\ast$, and resistivity $\rho(T)$ of an optimized 20~nm thick epilayer with 7\% nominal Mn-doping. (b) and (c) Frequency dependence of the damping factor and field dependence of the SWR frequencies of the same sample. (d) -- (f) Same as (a) -- (c) for a material differing by having only one of the synthesis parameters not optimized (epilayer thickness of 500~nm being too large). From \cite{Nemec:2012_b}.}
\label{synth1}
\end{figure}

A comprehensive experimental assessment of basic doping trends  including the regimes near and above the insulator-to-metal transition became possible since late 1990's with the development of LT-MBE (Ga,Mn)As films  \cite{Ohno:1998_a}. The epilayers can be doped well beyond the equilibrium Mn solubility limit while avoiding phase segregation and maintaining a high degree of uniformity \cite{Kodzuka:2009_a}. Because of the highly non-equilibrium nature of the heavily-doped ferromagnetic (Ga,Mn)As, the growth and post-growth annealing procedures have to be individually optimized for each Mn-doping level in order to obtain films which are as close as possible to idealized uniform (Ga,Mn)As mixed crystals with the minimal density of compensating and other unintentional defects. This is illustrated in Fig.~\ref{synth1} showing, side by side, basic electrical and magnetic characteristics of two medium, 7\% Mn-doped epilayers \cite{Nemec:2012_b}. The left column shows data measured on a material which was prepared under optimized conditions for the given nominal Mn-doping. The sample has sharp Curie point singularities in magnetization and $d\rho/dT$ (Fig.~\ref{synth1}a). Magnetization precession damping factor and spin-wave resonances (SWRs) obtained from magneto-optical measurements (Figs.~\ref{synth1}b,c) confirm the high magnetic quality of the material. The initial decrease of the damping factor with frequency followed by a frequency independent part (Fig.~1b) is typical of uniform ferromagnets \cite{Walowski:2008_a}. It allows  to accurately separate the intrinsic Gilbert damping constant $\alpha$, corresponding to the frequency independent part, from effects that lead to inhomogeneous broadening of  FMR linewidths. Similarly, the observed Kittel SWR modes of a uniform ferromagnet (Fig.~1c) allows to measure accurately the magnetic anisotropy and spin stiffness parameters of (Ga,Mn)As. 

The right column data (Figs.~\ref{synth1}d-f) were measured on a 7\% Mn-doped epilayer differing from the sample of the left column in only one of the synthesis parameters not being optimized. The stoichiometry, substrate growth temperature, postgrowth annealing temperature and time, and epilayer thickness are among the key synthesis parameters. All these parameters were equally optimized in the two samples except for the epilayer thickness. In the medium and high Mn-doped samples, full material optimization  is possible only for film thicknesses $\lesssim 50$~nm. The epilayer whose measurements are shown in the right panels of Fig.~\ref{synth1} is 500~nm thick. Its magnetization and transport Curie point singularities are largely smeared out, the damping factor is strongly frequency dependent, and alternating number of SWRs is observed with increasing applied field whose spacings are inconsistent with Kittel modes. The material is non-uniform, the magnetization and transport data indicate strong moment and charge compensation by extrinsic impurities, and for this material it is impossible to reliably extract any of the intrinsic micromagnetic parameters of (Ga,Mn)As. 

\begin{figure}[ht]
\hspace{0cm}\includegraphics[width=1\columnwidth,angle=0]{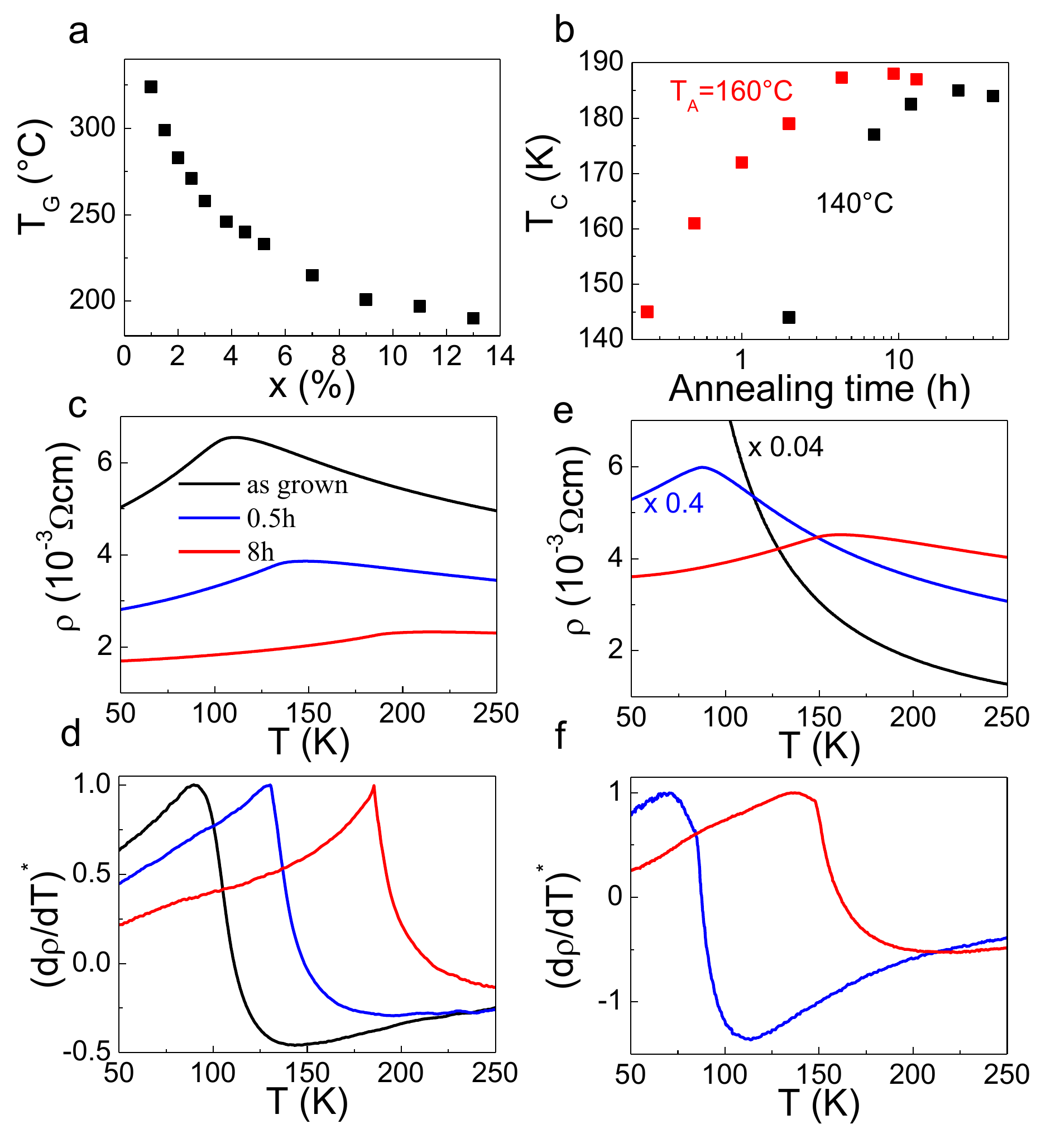}
\caption{
(Color online) (a) Optimal growth temperature $T_G$ as a function of the nominal Mn-doping $x$. %
(b) Dependence of the Curie temperature $T_c$ on the annealing time for two different annealing temperatures $T_A$ in a 15~nm thick (Ga,Mn)As epilayer with 13\% nominal Mn doping grown at optimal $T_G$. 
(c), (d) $\rho(T)$ and $(d\rho/dT)^\ast$ in the $x=13\%$ epilayer grown at optimal $T_G$ in the as-grown state, for optimal $T_A$ and annealing time 0.5h, and for optimal $T_A$ and optimal annealing time of 8h. (e), (f), Same as (c), (d) for a $x=13\%$ epilayer grown at 10$^\circ$ below the optimal $T_G$; $(d\rho/dT)^\ast$ is not plotted for the as-grown insulating and paramagnetic sample. From \cite{Nemec:2012_b}.}
\label{synth2}
\end{figure}

In Fig.~\ref{synth2} we illustrate that even in films thinner than 50~nm, apparently small changes in the remaining key synthesis parameters can significantly affect the material quality \cite{Nemec:2012_b}. Staying  near the 1:1 stoichiometric As:(Ga+Mn) ratio is favorable for the LT-MBE growth  of (Ga,Mn)As \cite{Myers:2006_a,Wang:2008_e}. Fig.~\ref{synth2}a shows the optimal growth temperature $T_G$ for the stoichiometric growth as a function of the nominal Mn-doping $x$.  The optimal $T_G$ remains near (from the lower temperature side) the 2D/3D growth-mode boundary which implies its strong dependence on $x$. Fig.~\ref{synth2}b shows $T_c$ as a function of the annealing time for the optimal $T_G=190^\circ$C for the 13\% Mn doped sample and for two annealing temperatures. One is the optimal annealing temperature $T_A=160^\circ$C and the other one is 20$^\circ$ lower. The maximun $T_c=188$~K sample is obtained by optimizing simultaneously the annealing time and $T_A$. Figs.~\ref{synth2}c,d illustrate how the increasing $T_c$ is accompanied by the improving material quality (reduction of extrinsic compensation and sample inhomogeneity) over the annealing time for optimal $T_G$ and $T_A$. The importance of  the optimal $T_G$ during the growth is highlighted in Figs.~\ref{synth2}e,f showing the same annealing sequence measurements as in Figs.~\ref{synth2}c,d on a 13\% doped sample grown at a temperature of only $10^\circ$ below the optimal $T_G$. In contrast to the material grown at the optimal $T_G$, the sample is insulating and paramagnetic in the  as-grown state. Ferromagnetism and metallic conduction can be recovered by annealing, however, the compensation and inhomogeneity cannot be removed and the ferromagnetic transition temperature remains tens of degrees below the $T_c$ of the sample grown at the optimal $T_G$. Similarly lower quality samples are obtained by growing at higher than optimal $T_G$. 

Figs.~\ref{synth1} and \ref{synth2} illustrate the following general conclusions drawn from  extensive material optimization studies \cite{Nemec:2012_b}. Inferring doping trends in basic material properties of (Ga,Mn)As from sample series mixing as-grown and annealed materials is unsuitable as the quality of the samples may strongly vary in  such a series. Choosing one {\em a priori} fixed $T_G$, $T_A$, and annealing time for a range of Mn-dopings is unlikely to produce a high-quality, uniform and uncompensated (Ga,Mn)As material even for one of the considered  dopings and is bound to produce low-quality samples for most of the studied Mn-dopings. Finally, optimized (Ga,Mn)As samples require exceedingly long annealing times for film thicknesses $\gtrsim50$~nm  and are impossible to achieve  in $\sim100$~nm and thicker films by the known (Ga,Mn)As synthesis approaches. 

\begin{figure}[ht]
\hspace{0cm}\includegraphics[width=1\columnwidth,angle=0]{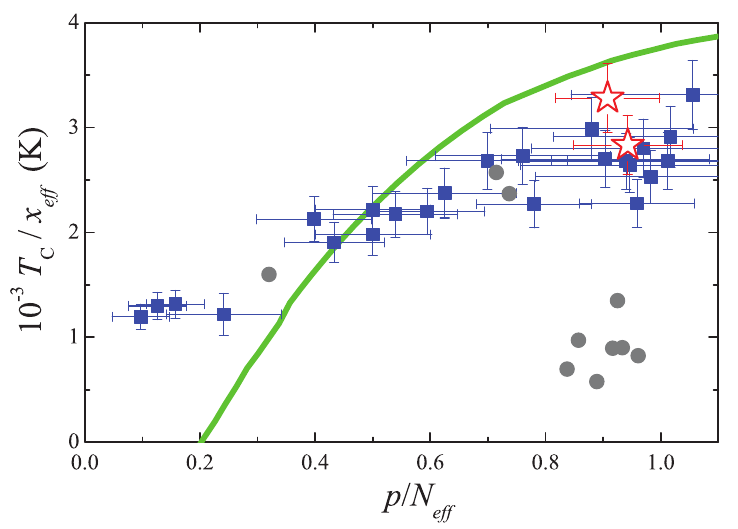}
\caption{(Color online) Curie temperature $T_c$versus hole density $p$ normalized to $x_{eff}$ ($N_{eff} = 4 x_{eff}/a^3$) representing the concentration of Mn magnetic moments which contribute to the magnetic order. Blue squares correspond
to samples from Ref.~\cite{Edmonds:2012_a} prepared under optimized growth conditions where hole density $p$ is obtained from high-field
Hall measurements. Gray circles correspond to samples from Ref.~\cite{Dobrowolska:2012_a},
where $p$ is obtained from ion channeling measurements. Red stars
correspond to samples from Ref.~\cite{Rushforth:2008_c} prepared under optimized growth conditions, where $p$ is obtained from ion
channeling measurements. The green line is the prediction of the
microscopic calculation of Ref.~\cite{Jungwirth:2005_b}.}
\label{T_c-p}
\end{figure}

When limited attention is paid to the details of the synthesis of the highly non-equilibrium (Ga,Mn)As alloy, seemingly contradictory experimental results can be found in these materials \cite{Burch:2006_a,Tang:2008_a,Dobrowolska:2012_a,Dobrowolska:2012_b} as compared to measurements on samples prepared under the above optimized growth conditions \cite{Jungwirth:2010_b,Edmonds:2012_a}. As an example we show in Fig.~\ref{T_c-p} measurements of $T_c$ versus hole density $p$ \cite{Dobrowolska:2012_a,Edmonds:2012_a}. The data are normalized to $x_{eff}$ ($N_{eff} = 4 x_{eff}/a^3$) representing the concentration of Mn magnetic moments which contribute to the magnetic order. The
results obtained in Ref.~\cite{Dobrowolska:2012_a} indicated a strong suppression of $T_c$
in (Ga,Mn)As layers with close to one hole per substitutional
Mn. It was thus suggested that $T_c$ in ferromagnetic (Ga,Mn)As is determined by the location of the Fermi level within a narrow
impurity band, separated from the valence band. On the other hand, experiments on epilayers prepared under the optimized growth conditions found that $T_c$  takes its largest values in weakly compensated samples when $p$ is comparable to the concentration of substitutional Mn acceptors. This
is inconsistent with models in which the Fermi level is located within a narrow isolated impurity band and corroborates predictions for $T_c$ of the above discussed microscopic theories (see Fig.~\ref{T_c-p}) in which valence and impurity bands are merged in ferromagnetic (Ga,Mn)As.

Reliable measurements of systematic doping trends in intrinsic semiconducting and magnetic properties  of materials which represent as close as possible idealized uniform (Ga,Mn)As mixed crystals with the minimal density of compensating and other unintentional defects require the careful optimization of the synthesis. Many studies of the spintronics phenomena in (Ga,Mn)As, discussed below in Section~\ref{spintronics}, have also benefited from the high quality optimized epilayers. This applies in particular to experiments sensitive to small tilts of carrier spins from the equilibrium direction which is the  case, e.g., of the magneto-optical phenomena observed in the pump-and-probe experiments discussed in Section~\ref{LIT}. While for the detailed analysis the optimally synthesized and thoroughly characterized (Ga,Mn)As epilayers are always favorable, many of the spintronics effects and functionalities have been demonstrated in materials with extrinsic disorder not fully removed from the film. As shown in Figs.~\ref{synth1} and \ref{synth2} these materials can still be ferromagnetic and conductive   and as discussed in Sections~\ref{Curie} and \ref{localization} the spintronics phenomena can be, at least on a qualitative level, relatively robust against strong disorder, whether intrinsic or extrinsic. 

\subsubsection{Curie temperature and conductivity}
\label{Tc_cond}
\begin{figure}[h!]
\hspace{0cm}\includegraphics[width=1\columnwidth,angle=0]{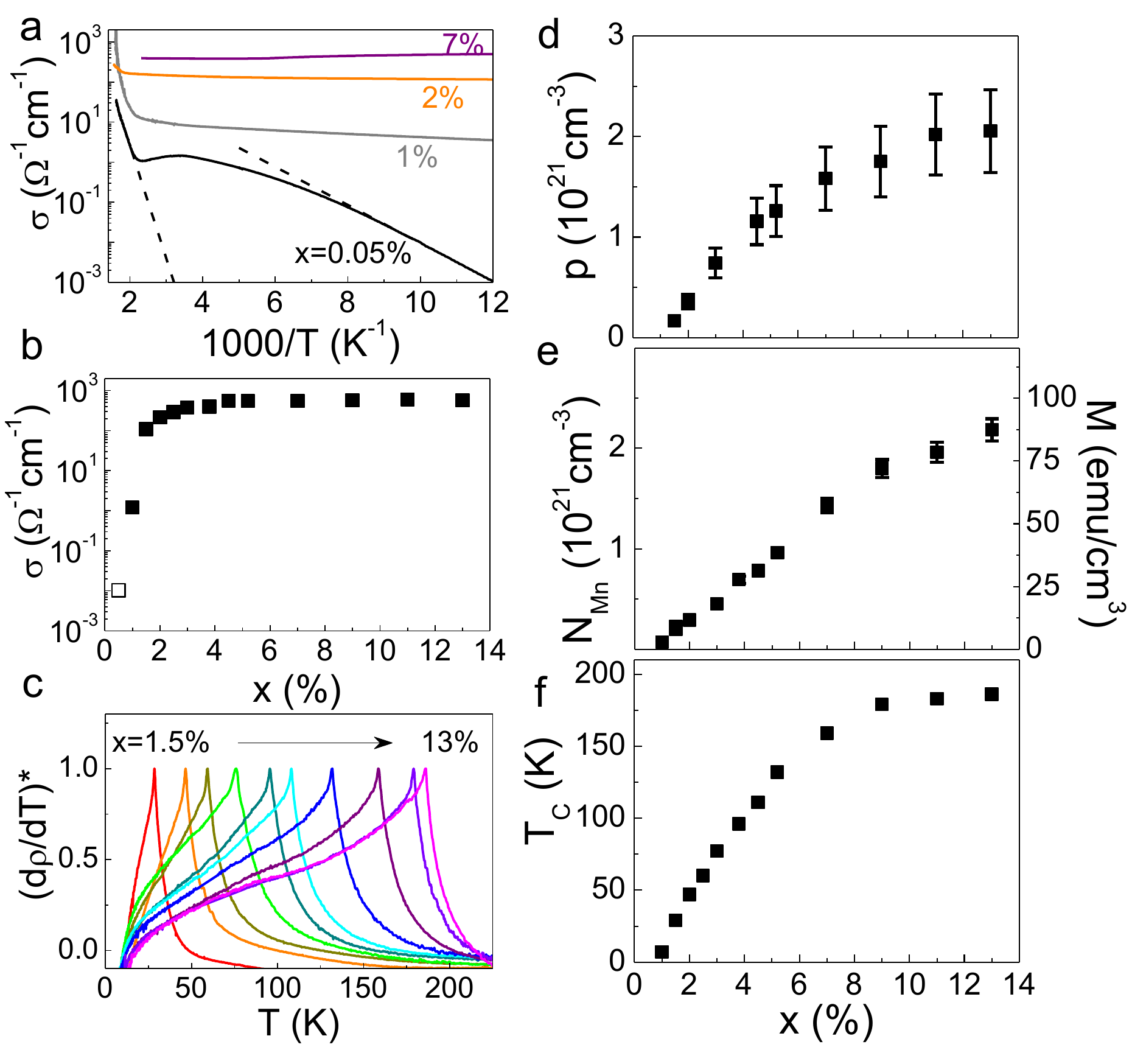}
\caption{(Color online) (a) Temperature dependence of the conductivity $\sigma(T)$ of optimized (Ga,Mn)As epilayers with depicted nominal Mn doping. Dashed lines indicate the activated parts of $\sigma(T)$ of the insulating paramagnetic (Ga,Mn)As with 0.05\% Mn doping, corresponding to the Mn acceptor level and the band gap, respectively. (b) Conductivity at 4~K as a function of the nominal Mn doping. Open symbol corresponds to a paramagnetic sample. (c) Sharp Curie point singularities in the temperature derivative of the resistivity in the series of optimized ferromagnetic (Ga,Mn)As epilayers with metallic conduction.
(d-f) hole density $p$, magnetization $M$ and corresponding Mn moment density $N_{Mn}$, and Curie temperature $T_c$ as a function of the nominal Mn doping in the series of optimized (Ga,Mn)As epilayers. 
From \cite{Nemec:2012_b}.
}
\label{doping_trends}
\end{figure}

Uniform (Ga,Mn)As materials with minimized extrinsic disorder can be divided into the following groups: at nominal dopings below $\sim0.1\%$ the (Ga,Mn)As materials are paramagnetic,  strongly insulating, showing signatures of the activated transport corresponding to valence band -- impurity band transitions at intermediate temperatures, and valence band -- conduction band transitions at high temperatures (see Fig.~\ref{doping_trends}(a)) \cite{Jungwirth:2007_a,Nemec:2012_b}. For higher nominal dopings, $0.5 \lesssim x \lesssim 1.5\%$, no clear signatures of activation from the valence band to the impurity band are seen in the dc transport, indicating that the bands start to overlap and mix, yet the materials remain insulating. At $x\approx1.5\%$, the low-temperature conductivity of the film increases abruptly by several orders of magnitude (see Fig.~\ref{doping_trends}(b)), and the system turns into a degenerate semiconductor. The onset of ferromagnetism occurs already on the insulating side of the transition at $x\approx 1\%$. All ferromagnetic samples over a broad nominal Mn-doping range can have sharp Curie point singularities when synthesized under individually optimized growth and post-growth annealing conditions (see Fig.~\ref{doping_trends}(c)).

The hole concentration $p$ can be measured by the slope of the Hall curve at high fields with an error bar due to the multi-band nature estimated to $\sim 20\%$.\cite{Jungwirth:2005_b} Within this uncertainty, the overall trend shows increasing $p$ with increasing doping in the optimized materials, as shown in Fig.~\ref{doping_trends}(d). Similarly, the saturation moment and $T_c$ steadily increase with increasing nominal doping up to $x\approx 13\%$, as shown in Figs.~\ref{doping_trends}(e),(f). Assuming 4.5$\mu_B$ per Mn atom \cite{Jungwirth:2005_a} the density $c\equiv N_{Mn}$ of uncompesated Mn$_{\rm Ga}$ moments can be inferred from the magnetization data (see left y-axis in Fig.~\ref{doping_trends}(e)).  Since there is no apparent deficit of $p$ compared to $N_{Mn}$, and since the interstitial Mn impurity \cite{Maca:2002_a,Yu:2002_a,Edmonds:2002_b} compensates one local moment but two holes it can be concluded that interstitial Mn, which is the key contributor to extrinsic disorder, is  removed in the optimally grown and annealed epilayers. Hence, a broad series of optimized (Ga,Mn)As materials can be prepared with reproducible characteristics, showing an overall trend of increasing saturation moment with increasing $x$ , increasing $T_c$ (reaching 188~K), and increasing hole density. The materials have no measurable
charge or moment compensation of the substitutional Mn$_{\rm Ga}$ impurities and have a large degree of uniformity.

Fig.~\ref{stiffness_etc} demonstrates that the intrinsic micromagnetic parameters of (Ga,Mn)As measured on the optimized materials show also a smooth monotonic  doping dependence \cite{Nemec:2012_b}. As detailed below, their values are characteristic of common band ferromagnets and all the semiconducting and magnetic properties summarized in Figs.~\ref{doping_trends} -- \ref{stiffness_etc} are consistent with the microscopically established electronic structure of (Ga,Mn)As. The control and reproducibility of material properties of (Ga,Mn)As have been confirmed in the optimized films by multiple material synthesis and characterization experiments in  different MBE chambers \cite{Nemec:2012_b,Edmonds:2012_a}.

\subsubsection{Micromagnetic parameters}
\label{micromag}
Micromagnetic parameters of (Ga,Mn)As and related (III,Mn)V ferromagnetic semiconductors were studied by magnetization, magneto-transport, magneto-optical, or ferromagnetic/spin-wave resonance (FMR/SWR) measurements \cite{Munekata:1993_a,Ohno:1998_a,Dietl:2001_b,Abolfath:2001_a,Sinova:2004_b,Rappoport:2004_a,Sawicki:2004_a,Zhou:2007_a,Liu:2007_e,Wenisch:2007_a,Humpfner:2006_a,Pappert:2007_a,Wunderlich:2007_c,Rushforth:2008_a,Rushforth:2008_b,Goennenwein:2008_a,Khazen:2008_a,Gould:2008_a,Stolichnov:2008_a,Overby:2008_a,Owen:2008_a,Chiba:2008_a,Bihler:2009_a,Potashnik:2002_a,Gourdon:2007_a,Wang:2007_f,Zemen:2009_a,Werpachowska:2010_a,Cubukcu:2010_a,Cubukcu:2010_b,Haghgoo:2010_a,Nemec:2012_b,Ranieri:2012_a}. A large experimental scatter  of the measured micromagnetic parameters can be found in the literature which reflects partly the issues related to the control of extrinsic disorder in the synthesis of (Ga,Mn)As. The experimental scatter also reflects, however, the favorable intrinsic tuneability of  (Ga,Mn)As properties by varying the temperature, hole and Mn-moment density, III-V substrate on which the (Ga,Mn)As film is deposited, or by alloying the magnetic film with other III or V elements, by device microfabrication, by applying electrostatic or piezoelectric fields on the film, etc.  

\begin{figure}[h!]
\hspace{0cm}\includegraphics[width=1\columnwidth,angle=0]{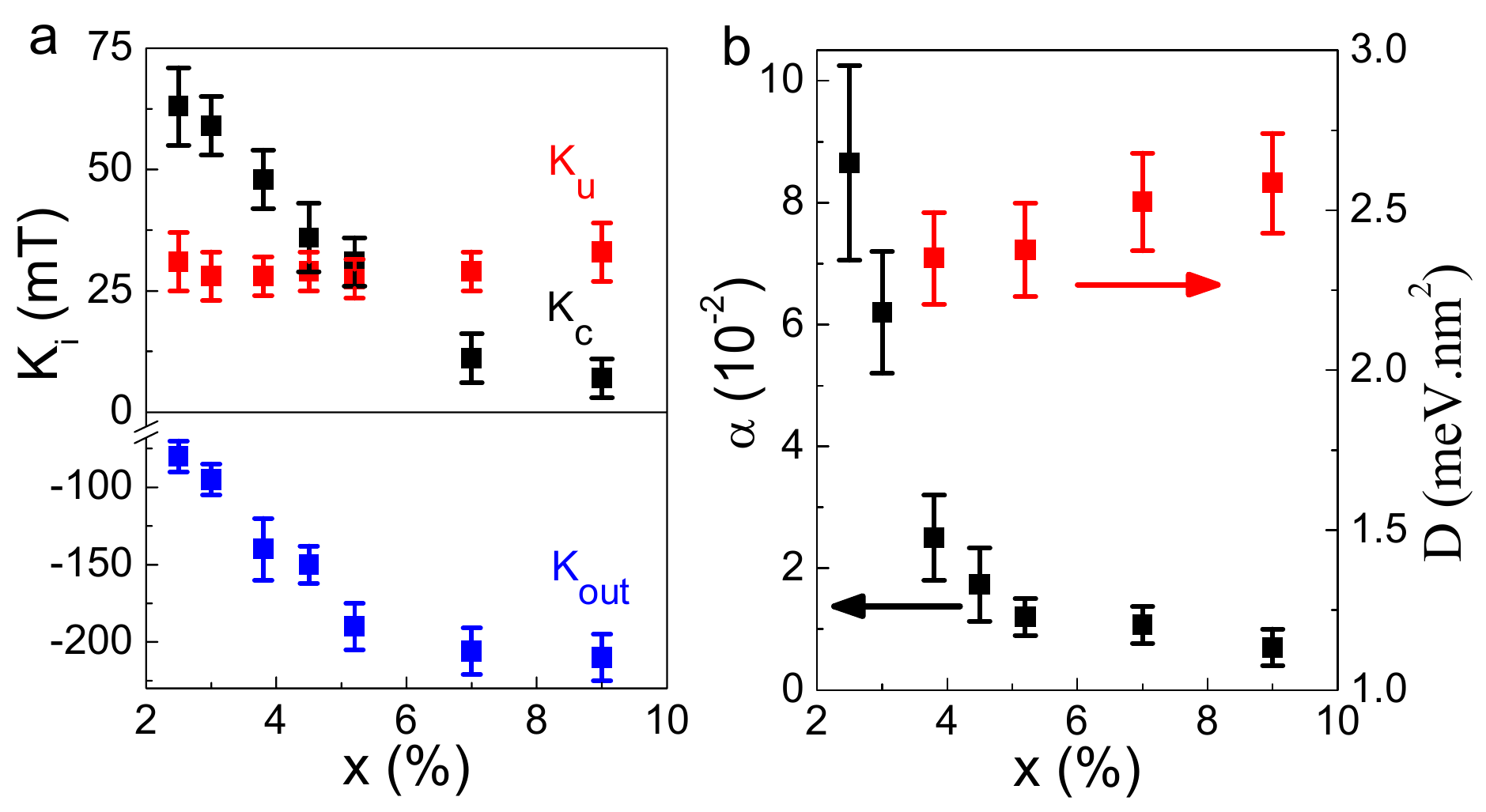}
\caption{
(Color online) (a) Dependence of magnetic anisotropy constants on nominal Mn doping. (b) Dependence of the Gilbert damping constant $\alpha$ and the spin stiffness constant $D$ on nominal Mn doping. From \cite{Nemec:2012_b}. Measurements were performed at 15 K.}
\label{stiffness_etc}
\end{figure}

When measuring the micromagnetic parameters on the optimally and consistently synthesized series of bare (Ga,Mn)As epilayers on a GaAs substrate, fully reproducible and systematic trends can be inferred when simultaneously determining  the magnetic anisotropy $K_i$, Gilbert damping $\alpha$, and spin stiffness $D$ constants from one set of measurements. This has been demonstrated, e.g., on a series of (Ga,Mn)As/GaAs epilayers over a broad range of Mn-dopings by employing the magneto-optical pump-and-probe technique, as shown in Fig.~\ref{stiffness_etc}  \cite{Nemec:2012_b}. 

The magnetic anisotropy fields are dominated by three components. The out-of-plane component $K_{out}$ is a sum of the thin-film shape anisotropy and the magnetocrystalline anisotropy due to the substrate lattice-matching growth strain. In (Ga,Mn)As grown on GaAs the strain in the (Ga,Mn)As epilayer is compressive and $K_{out}$ favors for most Mn-dopings in-plane magnetization (see Fig.~\ref{stiffness_etc}(a)). However, when using an InGaAs substrate or adding P into the magnetic film, the growth strain can change from compressive to tensile, $K_{out}$ flips sign and the film turns into an out-of-plane ferromagnet \cite{Dietl:2001_b,Abolfath:2001_a,Yamanouchi:2004_a,Rushforth:2008_b,Cubukcu:2010_b}. This transition from an in-plane to an out-of-plane magnet has been exploited, e.g., in studies of the current induced domain wall motion and spin-orbit torque discussed below in Sections~\ref{STT} and \ref{SOT} \cite{Yamanouchi:2004_a,Wang:2010_a,Curiale:2012_a,Ranieri:2012_a,Fang:2010_a}. 

The cubic magnetocrystalline anisotropy $K_{c}$ reflects the zinc-blende crystal structure of the host semiconductor. The origin of the additional uniaxial anisotropy component along the in-plane diagonal $K_{u}$ is associated with a more subtle symmetry breaking mechanism introduced during the epilayer growth \cite{Kopecky:2011_a,Mankovsky:2011_a,Birowska:2012_a}.  The sizable magnitudes of $K_c$ and $K_{u}$ and the different doping trends of these two in-plane magnetic anisotropy constants (see Fig.~\ref{stiffness_etc}(a)) are crucial for the micromagnetics of the in-plane magnetized (Ga,Mn)As materials. The cubic anisotropy $K_c$ dominates at very low dopings and the easy axis aligns with the main crystal axis [100] or [010]. At intermediate dopings, the uniaxial anisotropy $K_{u}$ is still weaker but comparable in magnitude to $K_c$. In these samples the two equilibrium easy-axes are tilted towards the [1$\bar1$0] direction and their angle is sensitive to changes of temperature (the ratio of  $K_u/K_c$ tends to increase with temperature \cite{Wang:2005_e}) or externally applied electrostatic or piezo-voltages which has been exploited in numerous studies of spintronics effects and device functionalities in (Ga,Mn)As \cite{Ohno:2000_a,Chiba:2003_a,Chiba:2008_a,Olejnik:2008_a,Owen:2008_a,Stolichnov:2008_a,Rushforth:2008_a,Overby:2008_a,Goennenwein:2008_a,Ranieri:2012_a}. The origin of the magnetocrystalline anisotropies is in the spin-orbit coupling of the valence band  holes mediating the ferromagnetic Mn-Mn coupling, as described on a qualitative or  semi-quantitative level by the model, kinetic-exchange Hamiltonian theory \cite{Dietl:2001_b,Abolfath:2001_a,Zemen:2009_a}. 

A systematic doping trend of the Gilbert damping constant is also found across the series of optimized materials (see Fig.~\ref{stiffness_etc}(b)). The magnitudes of $\alpha\sim 0.1-0.01$ and the doping dependence are consistent with Gilbert damping constants in conventional transition metal ferromagnets. In metals, $\alpha$ typically increases with increasing resistivity and is enhanced in alloys with enhanced spin-orbit coupling \cite{Ingvarsson:2002_a,Rantschler:2007_a,Gilmore:2008_a}. Similarly in (Ga,Mn)As the increase of $\alpha$ correlates with an increase of the resistivity in the lower Mn-doped samples. Moreover, the spin-orbit coupling effects tend to be stronger in the lower doped samples with lower filling of the hole bands and with the carriers closer to the metal-insulator transition.  Theory ascribing magnetization relaxation to the kinetic-exchange coupling of Mn moments with the spin-orbit coupled holes yields a comparable range of values of $\alpha$ as observed in experiment Fig.~\ref{stiffness_etc}(b) \cite{Sinova:2004_b,Nemec:2012_b}.

The direct measurement of the spin stiffness requires a rather delicate balance between thin enough epilayers whose material quality can be optimized  and thick enough films allowing to observe the higher-index  Kittel spin-wave modes \cite{Kittel:1958_a} of a uniform thin-film ferromagnet. The magneto-optical pump-and-probe technique \cite{Nemec:2012_b} has an advantage that, unlike ferromagnetic resonance (FMR), it is not limited to odd index spin wave modes \cite{Kittel:1958_a}. The ability to excite and detect the $n=0$, 1, and 2 resonances is essential for the observation of the Kittel modes in the optimized  (Ga,Mn)As epilayers whose thickness $L$ is limited to $\sim50$~nm. The modes in the optimized films show the expected quadratic scaling with $n$ and with  $1/L$, and could be fitted by one set of magnetic anisotropy constants and spin-stiffness constant $D$  \cite{Nemec:2012_b}. In the optimized series of (Ga,Mn)As epilayers a consistent, weakly increasing trend in $D$ with increasing doping is observed (see Fig.~\ref{stiffness_etc}(b)) with values of $D$ between $\sim 2$ and 3~meVnm$^2$. Similar to the Gilbert damping constant, the measured spin stiffness constant in the optimized (Ga,Mn)As epilayers is comparable to the spin stiffness in conventional transition metal ferromagnets \cite{Collins:1969_a}. The  values of the spin stiffness of the order meVnm$^2$ are consistent with calculations based on the model kinetic-exchange and tight-binding Hamiltonians, or the {\em ab initio} electronic structure of (Ga,Mn)As \cite{Konig:2001_a,Brey:2003_a,Bouzerar:2006_c,Werpachowska:2010_a}.     

To conclude Section~\ref{material}, the micromagnetic parameters of optimized  (Ga,Mn)As epilayers are characteristic of common band ferromagnets and the semiconducting and magnetic properties summarized in Figs.~\ref{doping_trends} -- \ref{stiffness_etc} are consistent with the model Hamiltonian or {\em ab initio}  theories of the electronic structure of (Ga,Mn)As. The materials research reviewed  in Section~\ref{material} establishes the overall view of  (Ga,Mn)As  as a well behaved and understood degenerate semiconductor and band ferromagnet. Combined with the tuneability of its electronic and magnetic properties, strong exchange and spin-orbit interactions in the carrier bands, special symmetries of the host zinc-blende lattice, and the compatibility with established III-V semiconductor microfabrication techniques,  this makes (Ga,Mn)As an ideal model system for spintronics research.

\section{Phenomena and device concepts for spintronics}
\label{spintronics}

\subsection{Non-relativistic versus relativistic based spintronics concepts }
\label{Mott-Dirac}
Most of the spintronic devices discussed in Section~\ref{spintronics} can be associated with one of two basic physical principles. The first one stems from Mott's two-spin-channel picture of transport in ferromagnets with exchange-split bands \cite{Mott:1936_a} and we will label it a Mott spintronics principle. Phenomena which follow from the Mott picture can be typically understood using the non-relativistic band structure with momentum-independent spin quantization axis. The second paradigm is due to the quantum-relativistic spin-orbit coupling \cite{Strange:1998_a} and we will label it a Dirac principle. Spintronics effects based on the Dirac principle stem from a relativistic  band structure comprising states with momentum dependent spin expectation values. Mott devices require that spins are transported between at least two non-collinear parts of a non-uniform magnetic structure with the magnetization in one part serving as a reference to the other one.  Dirac devices, on the other hand, can rely on a single uniform magnetic component and the reference for detecting or manipulating spins by charge carriers is provided internally by  the  spin-orbit coupling. 

\begin{figure}
\includegraphics[width=.85\columnwidth,angle=0]{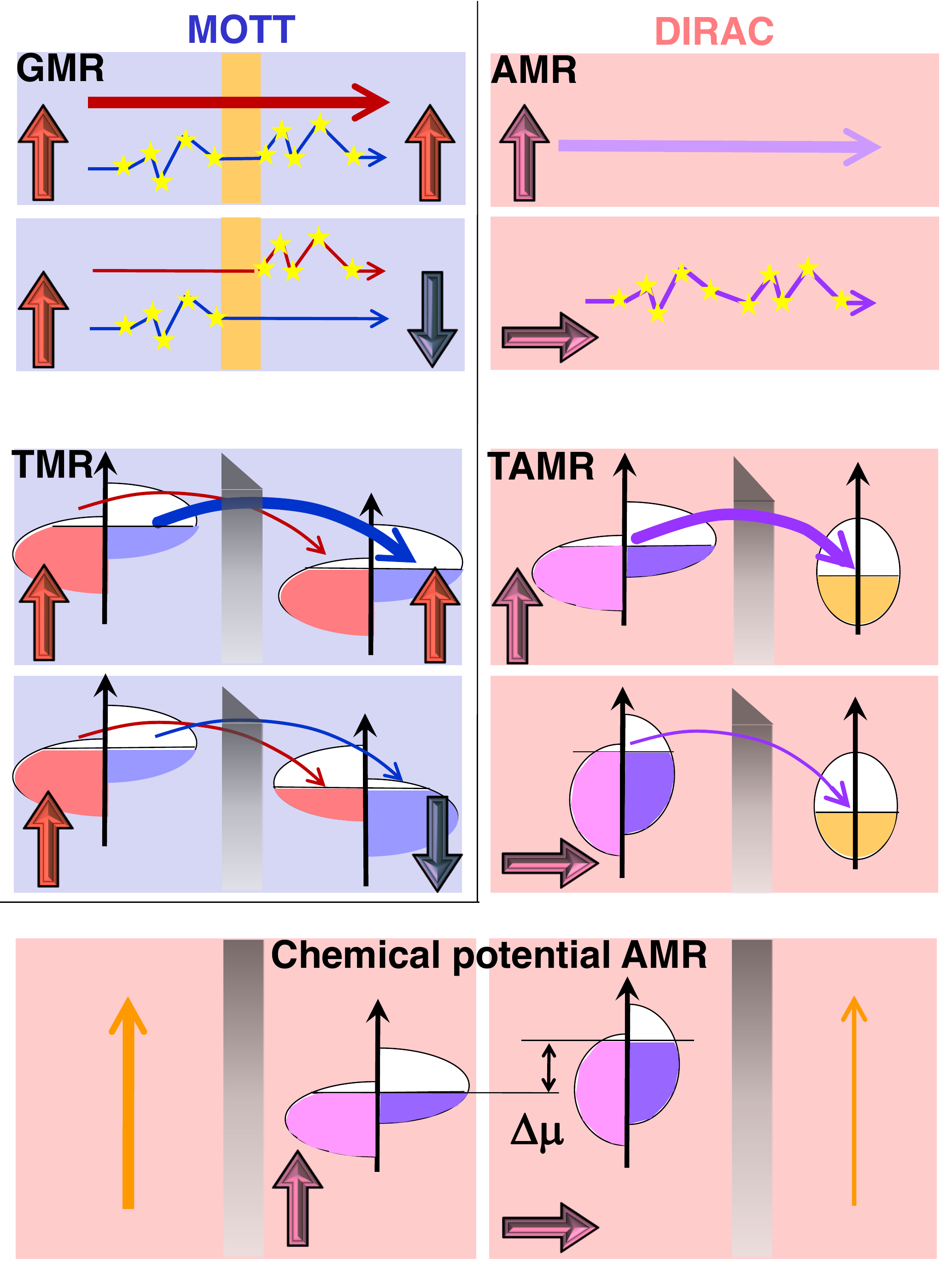}
\caption{(Color online) Schematic comparison of ohmic Mott (GMR) and Dirac (AMR) devices and tunneling Mott (TMR) and Dirac (TAMR) devices. At the bottom of the figure we show a Dirac device based on the chemical potential anisotropy (e.g. CB-AMR) which has no immediate counterpart in Mott spintronics. In GMR, the thick short arrows show the magnetization orientations in two metallic ferromagnets separated by a non-magnetic metallic (yellow) spacer. The straight long arrow illustrates a highly conductive spin channel in the parallel magnetization configuration. In the antiparallel configuration, non of the spin channels is highly conductive in both ferromagnets. Broken lines illustrate stronger scattering. In AMR, the majority and minority bands cannot be assigned to the spin-up and spin-down channels because spin-orbit coupling mixes the up and down spins. The figure illustrates that due to spin-orbit coupling the scattering strength, and therefore conduction, depends on the orientation of magnetization with respect to current direction or crystal axes.  Middle panels with insulating (grey) barriers illustrate the more direct relation between magnetoresistance and spin-up and spin-down bands in the tunneling device (TMR), as compared to the ohmic GMR. Similarly, the relation between magnetoresistance and spin-orbit coupled bands is more direct in case of the TAMR as compared to AMR. In GMR and TMR, at least two separate magnetic components have to be connected by spin current. AMR and TAMR, on the other hand, require only one magnet which does not have to be connected to another reference magnet by spin current. Chemical potential AMR illustrated in the bottom panel is a Dirac spintronic device which can operate with no spin current within the magnetic component. The electrical current depicted by  yellow arrows is moved from the magnetic component to a capacitively coupled conventional charge channel. The charge current is still sensitive to the orientation of the magnetization in the magnetic gate due to the spin-orbit coupling induced shifts of the internal chemical potential in the magnet.
}
\label{mott-dirac}
\end{figure}

The archetype ohmic Mott device, schematically illustrated in Fig.~\ref{mott-dirac}, is based on the giant-magnetoresistance (GMR)  of a ferromagnet/normal-metal/ferromagnet multilayer in which magnetizations in the ferromagnets are switched between parallel and anti-parallel configurations \cite{Baibich:1988_a,Binasch:1989_a}. The archetype ohmic Dirac device (see Fig.~\ref{mott-dirac}), which is discussed below in Section~\ref{amr},  is based on  the relativistic AMR of a uniform magnetic conductor in which magnetization is rotated with respect to the current direction or crystal axes \cite{Thomson:1857_a,McGuire:1975_a}.   In early 1990's the AMR and subsequently the GMR sensors were introduced in hard disk drive read-heads launching the field of applied spintronics \cite{Chappert:2007_a}. In these ohmic devices, the exchange-split and, in case of the AMR also  spin-orbit coupled, bands   enter the physics of spin transport in a complex way via electron scattering which is often difficult to control and accurately model. 

A more direct connection between spin dependent transport and band structure is realized in tunneling devices. Here the TMR  stack with two ferromagnetic electrodes \cite{Julliere:1975_a,Moodera:1995_a,Myiazaki:1995_a} operates on the Mott principle and the TAMR stack with one magnetic electrode \cite{Gould:2004_a,Brey:2004_a,Giraud:2005_a,Sankowski:2006_a,Ciorga:2007_a,Moser:2006_a,Gao:2007_a,Park:2008_a,Park:2010_a}, discussed below in Section~\ref{tamr}, is the corresponding Dirac spintronics device (see Fig.~\ref{mott-dirac}). The more direct connection between transport and electronic structure in tunneling devices implies that tunneling spintronics effects can be significantly larger than their ohmic counterparts. The large TMR signals are used, e.g., to represent logical 0 and 1 in MRAMs \cite{Chappert:2007_a}. 

CB-AMR devices discussed in Section~\ref{cbamr} represent an ultimate simplification in the relation between the magneto-transport  and the relativistic exchange-split band structure. Transport is governed here by a single electronic structure parameter which is the magnetization-direction dependent chemical potential, resulting in a huge magnetoresistance response of the device \cite{Wunderlich:2006_a}. A CB-AMR device with the spin-orbit coupled magnet forming a gate-electrode of the SET \cite{Ciccarelli:2012_a} illustrates that the Dirac spintronics principle not only works without a spin-current connecting two separate magnetic electrodes but also with the spin-orbit-coupled magnetic component completely removed from the transport channel (see Fig.~\ref{mott-dirac}). Such a spintronic device operating without spin-current cannot be realized within the more commonly considered Mott spintronics principle which may explains why it falls beyond the Wikipedia's  definition of spintronics as "a portmanteau meaning spin transport electronics" (http://en.wikipedia.org/wiki/Spintronics). 

The Mott GMR and TMR effects have their spin-caloritronic counterparts in the giant magneto-thermopower (GMT) \cite{Sakurai:1991_a} and TMT \cite{Walter:2011_a,Liebing:2011_a}. A similar correspondence is between the Dirac electrical transport AMR and TAMR effects and the spin-caloritronic AMT \cite{Pu:2006_a,Wisniewski:2007_a,Tang:2011_a,Mitdank:2012_a,Anwar:2012_a} and TAMT \cite{Naydenova:2011_a}, discussed in Section~\ref{TAMT}.
 
The distinction between Mott and Dirac spintronics can be analogously applied to the inverse magneto-transport effects (spin-torques), discussed below in Sections~\ref{STT} and \ref{SOT}. The STT \cite{Slonczewski:1996_a,Berger:1996_a,Zhang:2004_c,Ralph:2007_a} applied to switch the magnetization of a free layer by a vertical current driven through the TMR stack is a Mott spin-torque effect. The in-plane current induced SOT in a uniform magnet with a broken space-inversion symmetry  \cite{Bernevig:2005_c,Manchon:2008_b,Chernyshov:2009_a,Miron:2010_a} is the Dirac spin-torque counterpart. Similarly the optical STT and SOT \cite{Rossier:2003_a,Nunez:2004_b,Nemec:2012_a,Tesarova:2012_b} reviewed in Section~\ref{LIT} can be viewed as Mott and Dirac phenomena arising from the interaction of spin with light.
 
Observations of  the ohmic AMR in an antiferromagnetic metal FeRh \cite{Marti:2014_a} and antiferromagnetic semiconductor Sr$_2$IrO$_4$ \cite{Marti:2013_a}, and of the TAMR in tunnel junctions with a magnetic electrode made of a metal antiferromagnet IrMn \cite{Park:2010_a,Wang:2012_a} illustrate that the Dirac approach to spintronics can be equally applicable to spin-orbit coupled ferromagnets and antiferromagnets. The anisotropic magnetoresistance phenomena make in principle no difference between the parallel-aligned moments in ferromagnets and antiparallel-aligned moments in antiferromagnets because they are an even function of the microscopic magnetic moments. In non-magnetic conductors the SHE is an example of a spintronic phenomenon converting a normal electrical current into a spin-current or {\em vice versa} \cite{Kato:2004_d,Wunderlich:2004_a,Valenzuela:2006_a,Jungwirth:2012_a}. It has a similar microscopic physics origin to the AHE \cite{Hall:1881_a,Nagaosa:2010_a} in uniform spin-orbit coupled ferromagnets and the SHE can be therefore regarded as an example of the Dirac spintronic phenomenon in non-magnetic systems. The relevance of the research in (Ga,Mn)As to these Dirac spintronic phenomena observed in antiferromagnetic and non-magnetic conductors will be also discussed in the following sections. 

\subsection{Interaction of spin with electrical current}
\label{AHE-AMR}

\subsubsection{Anomalous and spin Hall effects}
\label{AHE-SHE}
Advanced computational techniques and experiments in new unconventional ferromagnets have recently led to a significant progress in coping with the subtle nature of the magnetoresistance  effects based on relativistic spin-orbit coupling. There are two  distinct relativistic MR coefficients in uniformly magnetized ohmic devices, the AHE \cite{Hall:1881_a} and the AMR \cite{Thomson:1857_a}.
The AHE is the antisymmetric transverse MR coefficient obeying $\rho_{xy}({\bf M})=-\rho_{xy}({\bf -M})$, where
the magnetization vector ${\bf M}$ is pointing perpendicular to the plane of the Hall bar sample.
The AMR, discussed in the following section, is the
symmetric  MR coefficient with the
longitudinal and transverse resistivities obeying, $\rho_{xx}({\bf
M})=\rho_{xx}({\bf -M})$ and $\rho_{xy}({\bf M})=\rho_{xy}({\bf -M})$, where
${\bf M}$ has an arbitrary orientation. Note, that in this review we use the term transverse AMR rather than the alternative term planar Hall effect \cite{Tang:2003_a} to clearly distinguish this symmetric off-diagonal magnetoresistance coefficient which is even in ${\bf M}$ from the above antisymmetric off-diagonal Hall coefficient which is odd in ${\bf M}$.

(Ga,Mn)As has become one of the favorable test-bed systems for the investigation of the AHE. Here the unique position of (Ga,Mn)As ferromagnets stems from their tunability and the relatively simple, yet strongly spin-orbit coupled and exchange split carrier bands. The principles of the microscopic description of  the AHE in the metallic (Ga,Mn)As materials, based on the scattering independent intrinsic mechanism \cite{Luttinger:1958_a,Onoda:2002_a,Jungwirth:2002_a}, have been successfully applied to explain the effect in other itinerant ferromagnets \cite{Yao:2004_a,Fang:2003_a,Lee:2004_a,Dugaev:2005_a,Kotzler:2005_a,Sinitsyn:2005_a,Haldane:2004_a}, including conventional transition metals such as iron and cobalt, a pattern that has since then been repeated for other relativistic magneto-transport effects. The advances in the understanding of the AHE are discussed in several reviews \cite{Chien:1980_a,Dietl:2003_c,Sinova:2004_c,Jungwirth:2006_a,Nagaosa:2010_a}. Here we recall the link between the AHE and SHE.

\begin{figure}
\includegraphics[width=1\columnwidth,angle=0]{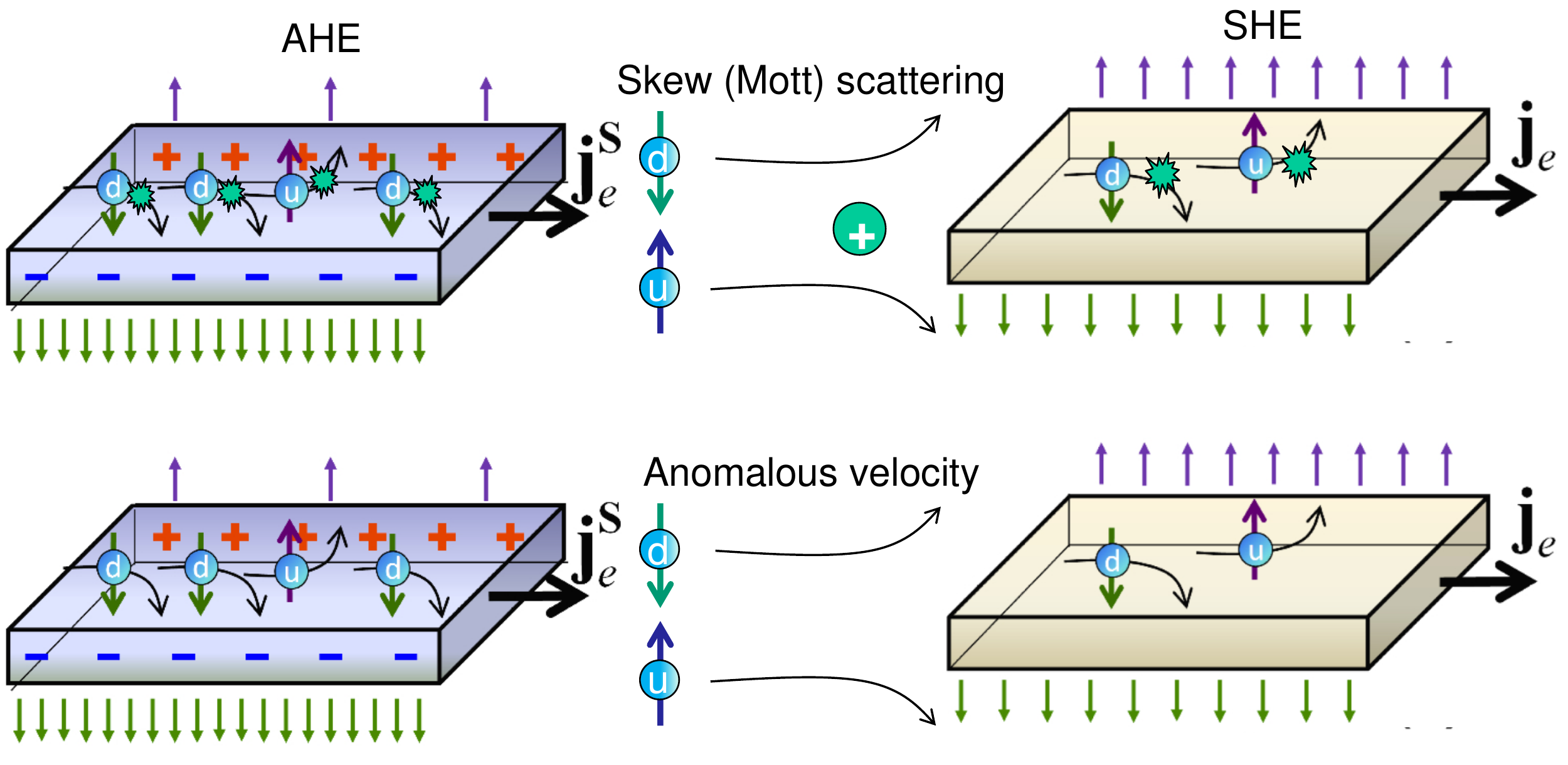}
\caption{(Color online) Schematic illustrations of the skew (Mott) scattering AHE and SHE (top panels) and the intrinsic AHE and SHE due to the anomalous transverse component of the spin-dependent velocity originating  from the spin-orbit coupled band structure in a clean crystal (bottom panels). In the AHE, an electrical current driven through a ferromagnetic conductor $j_e^s$ is spin-polarized and the spin-dependent transverse deflection of electrons produces a transverse voltage. In the SHE, an unpolarized electrical current $j_e$ is driven through a normal conductor and the spin-dependent transverse deflection of electrons produces a transverse spin-current. Opposite spins accumulate at opposite edges but unlike the AHE the transverse voltage remains zero.
}
\label{intrinsic_ahe_she}
\end{figure}

Since the 1881 discovery of the AHE by Hall in Ni and Co the phenomenon has been extensively employed in polarimetry measurements of electron spins in ferromagnets. One line of physical descriptions, illustrated in Fig.~\ref{intrinsic_ahe_she}, associates the AHE with the same physical mechanism as the electron spin-dependent scattering from heavy nuclei  which is used in polarimetry of high-energy electron beams in accelerators.  This relativistic spin-dependent skew-scattering mechanism is referred to as  Mott scattering \cite{Mott:1929_a}. (To avoid confusion we point out that Mott scattering \cite{Mott:1929_a} is unrelated to the other work of Mott on the non-relativistic two-channel description of transport in ferromagnets \cite{Mott:1936_a} mentioned earlier; the AHE and SHE physics discussed here is relativistic in nature and falls within the family of Dirac spintronics phenomena, in the terminology used in the previous section.) The applicability of the Mott skew scattering mechanism to electrons scattering from heavy nuclei in the vacuum environment of accelerators as well as  to electrons scattering off impurities in the solid-state environment of ferromagnets implies the presence of the same mechanism in non-magnetic conductors. This was recognized in 1971 by Dyakonov and Perel in their theoretical prediction of the skew-scattering SHE \cite{Dyakonov:1971_a}. 

A complementary line of research, also illustrated in Fig.~\ref{intrinsic_ahe_she} and prompted by AHE experiments in the highly-doped metallic (Ga,Mn)As epilayers \cite{Jungwirth:2002_a,Jungwirth:2003_b,Chun:2006_a,Glunk:2009_b}, ascribes the AHE to a scattering-independent based mechanism in which the anomalous transverse component of the spin-dependent velocity stems directly from the spin-orbit coupled band structure in a clean crystal. In analogy with the skew-scattering AHE and SHE, a link was proposed between the scattering-independent mechanism of the AHE and a corresponding intrinsic SHE \cite{Murakami:2003_a,Sinova:2004_a}, followed by experimental discoveries of the SHE \cite{Kato:2004_d,Wunderlich:2004_a}. We will come back to the physical description of these phenomena in Section~\ref{SOT}  where the link is extended from the AHE and SHE to the SOT. 

\subsubsection{Anisotropic magnetoresistance}
\label{amr}

Phenomenologically, the AMR has "non-crystalline" and "crystalline" components \cite{Doring:1938_a,McGuire:1975_a}. The former  corresponds to the dependence of the resistance of the ferromagnet on the angle between magnetization and the direction of the electrical current while the latter depend on the angle between magnetization and crystal axes. The non-crystalline AMR is the only component contributing to the AMR in polycrystalline samples in which the crystal axes directions average out. It is the component identified in Kelvin's seminal AMR measurements in Ni and Fe \cite{Thomson:1857_a}. The crystalline AMR components can be isolated in single-crystal materials patterned into a Corbino-disk microdevice geometry for which the averaging over the radial current lines eliminates all effects originating from a specific direction of the current. This was demonstrated in experiments in (Ga,Mn)As \cite{Rushforth:2007_a}.  The measurements took advantage of the near perfect single-crystal epilayers of (Ga,Mn)As and, simultaneously, of the low carrier density and mobility (compared with single crystal metals) resulting in large source-drain resistances compared with the contact resistances even in the short current-line Corbino geometry.  Moreover, the strong spin-orbit coupling in the (Ga,Mn)As electronic structure yields sizable and tuneable crystalline AMR components which in the lower conductive (Ga,Mn)As materials can even dominate over the non-crystalline AMR component \cite{Rushforth:2007_a}. In contrast, crystalline AMR components in common transition metal ferromagnets have been extracted indirectly from fitting the total AMR angular dependencies \cite{vanGorkom:2001_a}.

Apart from the distinct phenomenologies there is also a qualitative difference between the microscopic origins of the non-crystalline and crystalline AMR components. Since the former component depends only on the angle between magnetization and current, the effects of the rotating magnetization on the equilibrium electronic structure of the ferromagnet do not contribute to the non-crystalline AMR. Instead, in the leading order, the non-crystalline AMR reflects the difference between transport scattering matrix elements of electrons with momentum parallel to the current for the current parallel or perpendicular to {\bf M}. 

Unlike the non-crystalline AMR, the crystalline AMR originates from the changes in the equilibrium relativistic electronic structure induced by the rotating magnetization with respect to crystal axes. The picture applies not only to the ohmic crystalline AMR but also to the TAMR and CB-AMR discovered in (Ga,Mn)As \cite{Gould:2004_a,Wunderlich:2006_a}.  In the CB-AMR case, the anisotropy of the electronic structure with respect to the magnetization angle, or more specifically the anisotropy of the DOS and the corresponding position of the chemical potential, provides a direct quantitative description of the measured transport effect \cite{Wunderlich:2006_a,Ciccarelli:2012_a}. In the case of the TAMR or the crystalline ohmic AMR, the quantitative relativistic transport theory requires to combine the calculated DOS anisotropy with the tunneling or scattering matrix elements, respectively \cite{Brey:2004_b,Giddings:2004_a,Elsen:2007_a,Jungwirth:2003_b}. Due to the anisotropy of the electronic structure with respect to the magnetization angle  the matrix elements may also change when magnetization is rotated. 

A physically appealing
picture has been used to explain the positive sign of the
non-crystalline AMR (defined as the relative difference between resistances for current parallel and perpendicular to {\bf M}) observed in most transition metal
ferromagnets \cite{McGuire:1975_a,Smit:1951_a}. The interpretation is
based on the model of the spin-up and spin-down two-channel
conductance corrected for perturbative spin-orbit coupling
effects. In the model most of the current is carried by the
light-mass $s$-electrons which experience no spin-orbit coupling and a
negligible exchange splitting but can scatter to the heavy-mass
$d$-states. AMR is then explained by considering the spin-orbit potential
which mixes the exchange-split spin-up and spin-down $d$-states in a
way which leads to an anisotropic scattering rate of the current carrying
$s$-states \cite{McGuire:1975_a,Smit:1951_a}. Controversial
interpretations, however, have appeared in the literature based on
this model \cite{Smit:1951_a,Potter:1974_a} and no clear connection
has been established between the intuitive picture of the AMR the
model provides and the numerical {\em ab
initio} transport theories \cite{Banhart:1995_a,Ebert:2000_a,Khmelevskyi:2003_a}.

\begin{figure}[h]
\hspace*{0cm}\includegraphics[width=.8\columnwidth,angle=0]{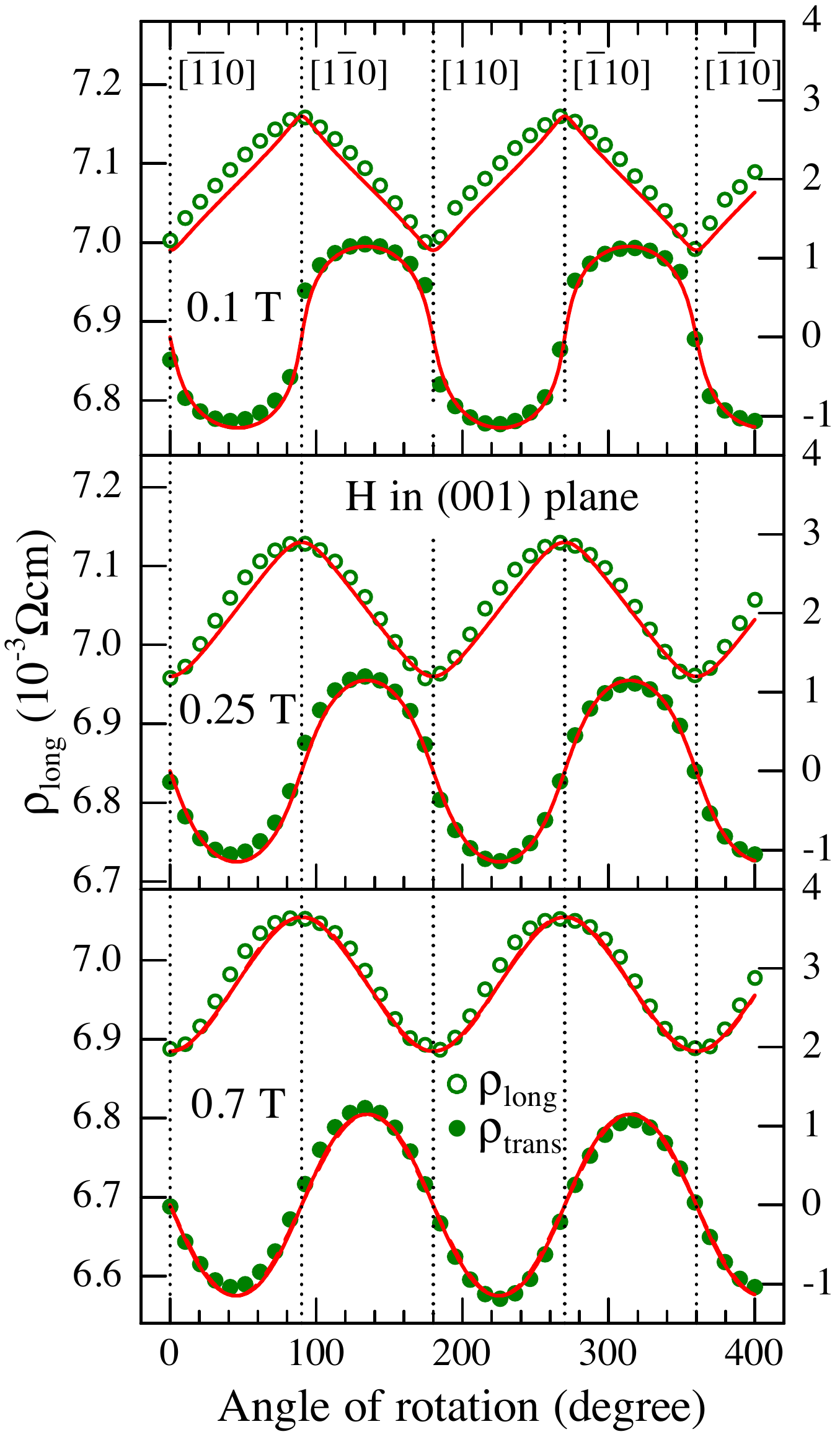}
\caption{Measured longitudinal and transverse in-plane AMR
curves at external fields smaller than the saturation field (0.1 and 0.25~T) and larger than the
saturation field (0.7~T). The solid lines represent fits to the experimental data.
From~\cite{Limmer:2006_a}.
}
\label{limmer}
\end{figure}

Among the remarkable AMR features of (Ga,Mn)As  are the
opposite sign of the non-crystalline component, as compared to most
metal ferromagnets, and the  sizable crystalline terms reflecting the
rich magnetocrystalline anisotropies of (Ga,Mn)As \cite{Baxter:2002_a,Jungwirth:2003_b,Tang:2003_a,Matsukura:2004_a,Goennenwein:2004_a,Wang:2005_c,Limmer:2006_a,Rushforth:2007_a}.
In Fig.~\ref{limmer} we show an example of AMR data from  a systematic
experimental and phenomenological study of the AMR coefficients in (Ga,Mn)As films
grown on (001)- and (113)A-oriented GaAs substrates at
non-saturating and saturating in-plane and out-of-plane magnetic fields \cite{Limmer:2006_a}. 
In the following paragraphs we describe the AMR phenomenology in (Ga,Mn)As in more
detail and explain the basic microscopic physics origin of the non-crystalline AMR in (Ga,Mn)As.
For simplicity we focus on the AMR in saturating magnetic fields, for ${\bf M}$ oriented in the plane of the device, and for (Ga,Mn)As films grown on the (001)-GaAs substrate.

The phenomenological decomposition of the AMR of (Ga,Mn)As into various terms allowed by symmetry is obtained by
extending the standard phenomenology \cite{Doring:1938_a}
to systems with the cubic  and in-plane uniaxial anisotropy. The corresponding AMR is then phenomenologically described as \cite{Rushforth:2007_a,Ranieri:2008_a},
\begin{eqnarray}
\frac{\Delta\rho_{xx}}{\rho_{av}}
&=& C_I\cos2\phi + C_U\cos2\psi + C_C\cos4\psi \nonumber \\
&+& C_{I,C}\cos(4\psi-2\phi)\;,
\label{rho_xx}
\end{eqnarray}
where $\Delta\rho_{xx}=\rho_{xx}-\rho_{av}$, $\rho_{av}$ is the  $\rho_{xx}$ averaged over 360$^{o}$ in the plane of the film, $\phi$ is the angle between the magnetization unit vector ${\bf \hat{M}}$ and the current ${\bf I}$,
and $\psi$ the angle between  ${\bf \hat{M}}$ and the [110] crystal direction.
The four contributions are the non-crystalline term, the lowest order uniaxial and cubic crystalline terms, and a crossed non-crystalline/crystalline term.
The purely crystalline terms are excluded by symmetry for the transverse AMR and one obtains \cite{Rushforth:2007_a,Ranieri:2008_a},
\begin{equation}
\frac{\Delta\rho_{xy}}{\rho_{av}}
=C_I\sin2\phi - C_{I,C}\sin(4\psi-2\phi)\; .
\label{rho_xy}
\end{equation}

\begin{figure}[h]
\vspace*{-0cm}

\hspace*{-0cm}\includegraphics[width=.9\columnwidth,angle=0]{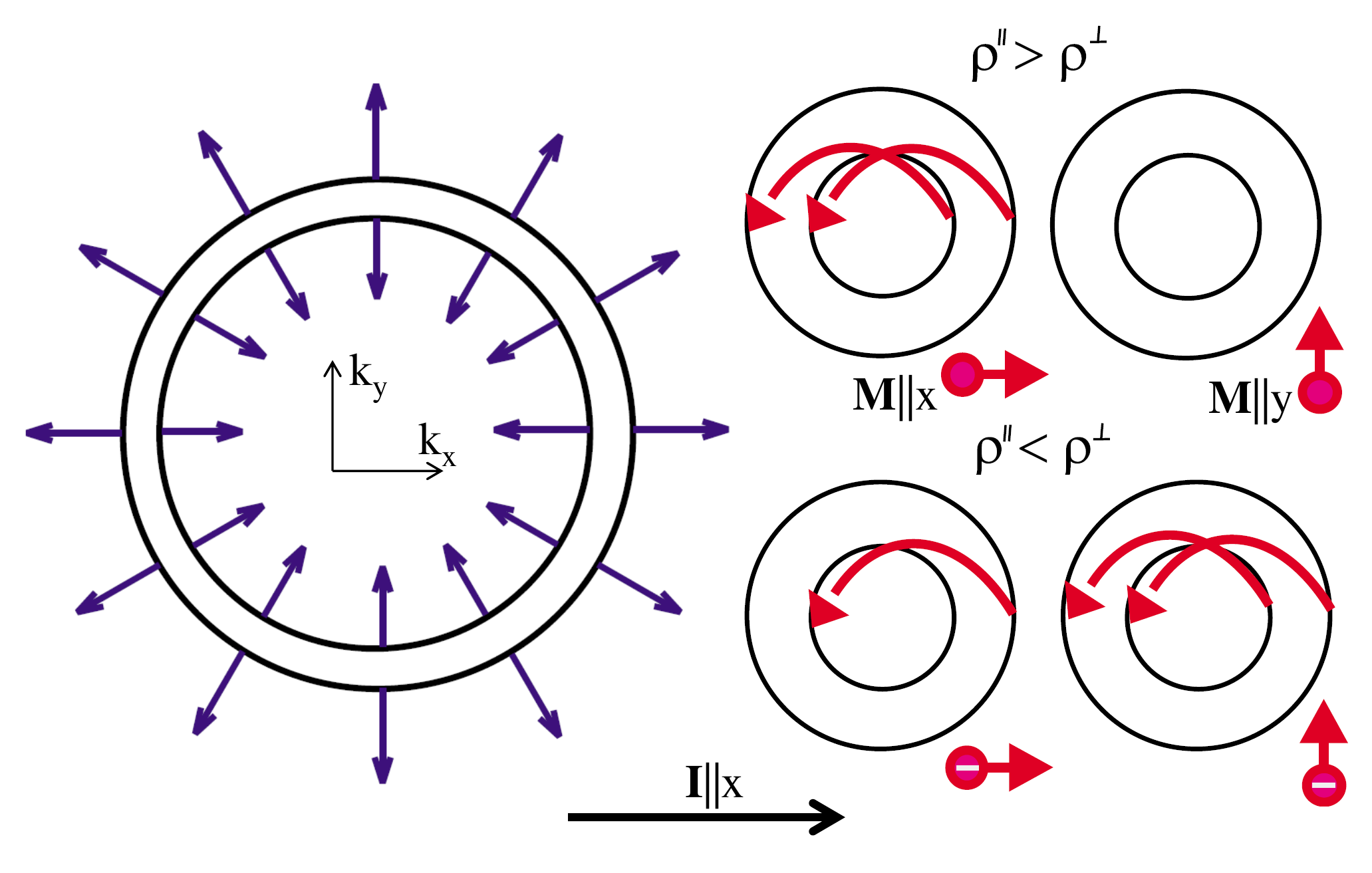}

\vspace*{-0cm}
\caption{(Color online) Left panel: Cross-section (parallel to the $k_x,k_y$ plane) of the 3D radial spin texture belonging to the two heavy-hole bands of (Ga,Mn)As in a spherical approximation. Right top panel: Non-zero back-scattering elements when neglecting the non-magnetic part of the Mn-impurity potential. The corresponding AMR has a positive sign. The purely magnetic Mn-impurity is illustrated by a red dot with an arrow. Right bottom panel: Non-zero back-scattering elements for the same strengths of the non-magnetic and magnetic parts of the Mn-impurity potential. The corresponding AMR has a negative sign. The combined ionized-acceptor and magnetic nature of the Mn-impurity is illustrated by a red dot with a negative sign and an arrow. (Electrical current ${\bf I}\parallel x$.) From~\cite{Trushin:2009_a}.}
\label{amr_theory}
\end{figure}

Microscopic numerical simulations \cite{Jungwirth:2002_c,Jungwirth:2003_b,Rushforth:2007_a,Vyborny:2009_a}
consistently describe the sign and magnitudes of the non-crystalline AMR in 
(Ga,Mn)As materials with metallic conductivities and
capture the presence of the more
subtle crystalline terms \cite{Jungwirth:2002_c,Matsukura:2004_a}.
Based on the numerical simulations the origin and sign of the non-crystalline AMR in (Ga,Mn)As was qualitatively explained using a simplified model in which carriers, represented by the heavy-hole Fermi
surface  in the spherical spin-texture approximation (see Fig.~\ref{amr_theory}), scatter off random Mn impurity potential approximated by   $\propto(r\mathds{1}+ {\bf \hat{M}}\cdot {\bf s})$. Here ${\bf  s}={\bf j}/3$ is the carrier spin-operator in the spherical approximation  with ${\bf j}$ representing the total angular momentum operator of heavy holes ($j=3/2$), and $r$ effectively models the ratio of non-magnetic (Coulomb and central cell) and magnetic ($p-d$ kinetic exchange) parts of the Mn impurity potential \cite{Rushforth:2007_a,Trushin:2009_a,Vyborny:2009_a}. 

The qualitative AMR considerations focus on scattering matrix elements of state with momentum along the current ${\bf I}$ and, in particular, on the strongest contribution to the transport life-time which comes from back-scattering (see Fig.~\ref{amr_theory}) \cite{Rushforth:2007_a,Trushin:2009_a,Vyborny:2009_a}. When neglecting the non-magnetic part of the impurity potential ($r=0$), non-zero back-scattering matrix elements 
occur only for ${\bf M}\parallel{\bf I}$ and in the notation of Fig.~\ref{amr_theory} they correspond to the elements $\langle\rightarrow|j_x|\rightarrow\rangle$ and $\langle\leftarrow|j_x|\leftarrow\rangle$. For ${\bf M}\perp{\bf I}$, all back-scattering elements $\langle\rightarrow|j_y|\rightarrow\rangle=0$, $\langle\leftarrow|j_y|\rightarrow\rangle=0$, etc., i.e., the back-scattering is completely suppressed. The picture changes when the non-magnetic part of the Mn-impurity potential is included, as illustrated in Fig.~\ref{amr_theory} for $r=1/2$. For ${\bf M}\parallel{\bf I}$, the coherent scattering of the non-magnetic and magnetic parts interferes constructively or destructively leaving only one of the back-scattering elements non-zero (see Fig.~\ref{amr_theory}). For ${\bf M}\perp{\bf I}$, the non-magnetic and magnetic parts do not interfere and now the non-magnetic part of the scattering potential results in two non-zero back-scattering elements (see Fig.~\ref{amr_theory}). As a result the resistivity $\rho_{xx}^{\parallel}$ for ${\bf M}\parallel{\bf I}$ is larger than $\rho_{xx}^{\perp}$ for ${\bf M}\perp{\bf I}$ when $r=0$ and $\rho_{xx}^{\parallel}$ is smaller than $\rho_{xx}^{\perp}$ when $r=1/2$. The presence of the non-magnetic part of the impurity potential can, therefore, flip the sign of the AMR from the positive which is seen in common transition-metal ferromagnets to the negative which is typical of (Ga,Mn)As. The negative sign is obtained in the above simplified model  for $r>1/\sqrt{20}$ which is safely satisfied in (Ga,Mn)As \cite{Rushforth:2007_a,Trushin:2009_a,Vyborny:2009_a}.

\subsubsection{Tunneling anisotropic magnetoresistance}
\label{tamr}
The electrical response to changes in the magnetic state is
strongly enhanced in layered structures consisting of alternating ferromagnetic
and non-magnetic materials. The GMR and TMR effects
which are widely exploited in  metal spintronics technologies
reflect the large difference between resistivities
in configurations with parallel and antiparallel polarizations of ferromagnetic layers in
magnetic multilayers, or trilayers like spin-valves and magnetic tunnel junctions \cite{Chappert:2007_a,Gregg:2002_a}.
The effect relies on transporting spin information between the layers.
In (Ga,Mn)As, functional magnetic tunnel junction devices can be built, as
demonstrated by the measured large TMR effects \cite{Tanaka:2001_a,Chiba:2004_a,Chiba:2004_b,Brey:2004_b,Saito:2005_a,Mattana:2005_a,Sankowski:2006_a,Saffarzadeh:2006_a,Ohya:2006_a,Elsen:2007_a}.

Here we focus on the physics of the
TAMR which was discovered 
in (Ga,Mn)As based
tunnel devices \cite{Gould:2004_a,Brey:2004_b,Ruester:2004_a,Saito:2005_a,Giraud:2005_a,Sankowski:2006_a,Elsen:2007_a,Ciorga:2007_a}. TAMR, like AMR, arises from spin-orbit coupling and reflects the dependence of the tunneling density of states of the ferromagnetic layer on the orientation of the magnetization. The effect does not rely on spin-coherence
in the tunneling process and requires only one ferromagnetic contact.
\begin{figure}[h]

\vspace{0cm}
\hspace*{0cm}\includegraphics[width=1\columnwidth,angle=0]{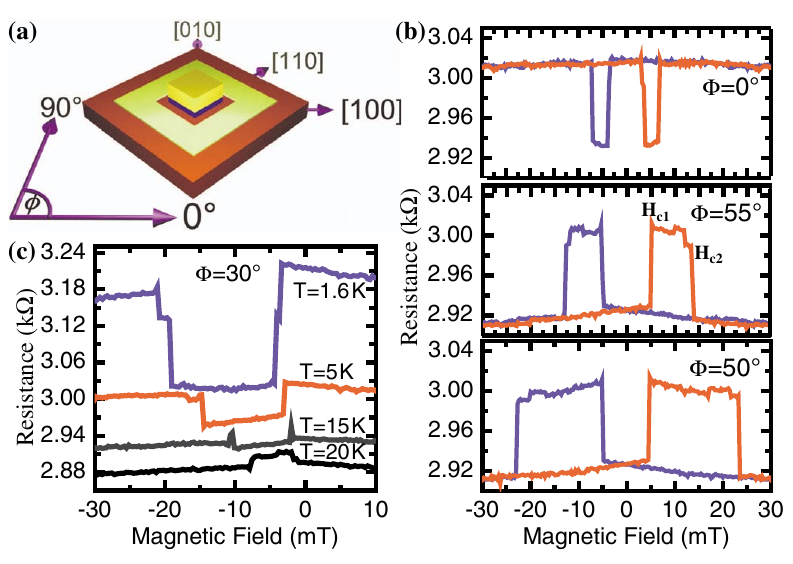}
\vspace{-0cm}
\caption{(Color online) (a) Device schematic showing the contact geometry and the crystallographic directions. (b) Hysteretic magnetoresistance curves acquired at 4.2~K with 1~mV bias by sweeping the magnetic field along the 0$^\circ$, 50$^\circ$, and 55$^\circ$ directions. Spin-valve-like features of varying widths and signs are clearly visible, delimited by two switching events labeled $H_{c1}$ and $H_{c2}$. The magnetoresistance is independent of the bias direction or amplitudes up to 1~meV. (c) TAMR along 30$^\circ$ for temperatures from 1.6 to 20~K, showing a change of sign of the signal. The curves are vertically offset for clarity.
From~\cite{Gould:2004_a}.}
\label{tamr1}
\end{figure}

In Fig.~\ref{tamr1} we show the TAMR signal which was measured in a (Ga,Mn)As/AlO$_x$/Au vertical
tunnel junction \cite{Gould:2004_a,Ruster:2005_a}.
For the in-plane magnetic field applied at an angle 50$^{\circ}$ off the [100]-axis the magnetoresistance is reminiscent
of the conventional spin-valve signal with hysteretic high resistance states at low fields and low resistance states
at saturation. Unlike the TMR or GMR, however, the sign changes when the field is applied along the [100]-axis.
Complementary SQUID magnetization measurements confirmed that for the sample measured in Fig.~\ref{tamr1}, the high resistance state corresponds to magnetization
in the (Ga,Mn)As contact aligned along the [100]-direction and the low resistance state along the [010]-direction,
and that this TAMR effect reflects the underlying magnetocrystalline anisotropy between the ${\bf M}\parallel$[100] and
${\bf M}\parallel$[010] magnetic states of the specific (Ga,Mn)As material used in the study. Since the field is rotated in the plane perpendicular to the current,
the Lorentz force effects on the tunnel transport can be ruled out. Microscopic calculations consistently showed that the spin-orbit coupling induced density-of-states anisotropies with
respect to the magnetization orientation can produce TAMR effects in (Ga,Mn)As of the order
$\sim 1$\% to $\sim 10$\% \cite{Gould:2004_a,Ruster:2005_a}.

All-semiconductor TAMR devices with a single ferromagnetic electrode
were realized in $p$-(Ga,Mn)As/$n$-GaAs Zener-Esaki diodes \cite{Giraud:2005_a,Ciorga:2007_a}.
For magnetization rotations in the (Ga,Mn)As plane \cite{Ciorga:2007_a}  comparable TAMR ratios
were detected as in the (Ga,Mn)As/AlO$_x$/Au tunnel junction. About an order of magnitude larger TAMR (40\%) was
observed when magnetization was rotated out of the (Ga,Mn)As plane towards the current direction \cite{Giraud:2005_a}.

\begin{figure}[h]

\hspace*{0cm}\includegraphics[width=.8\columnwidth,angle=-0]{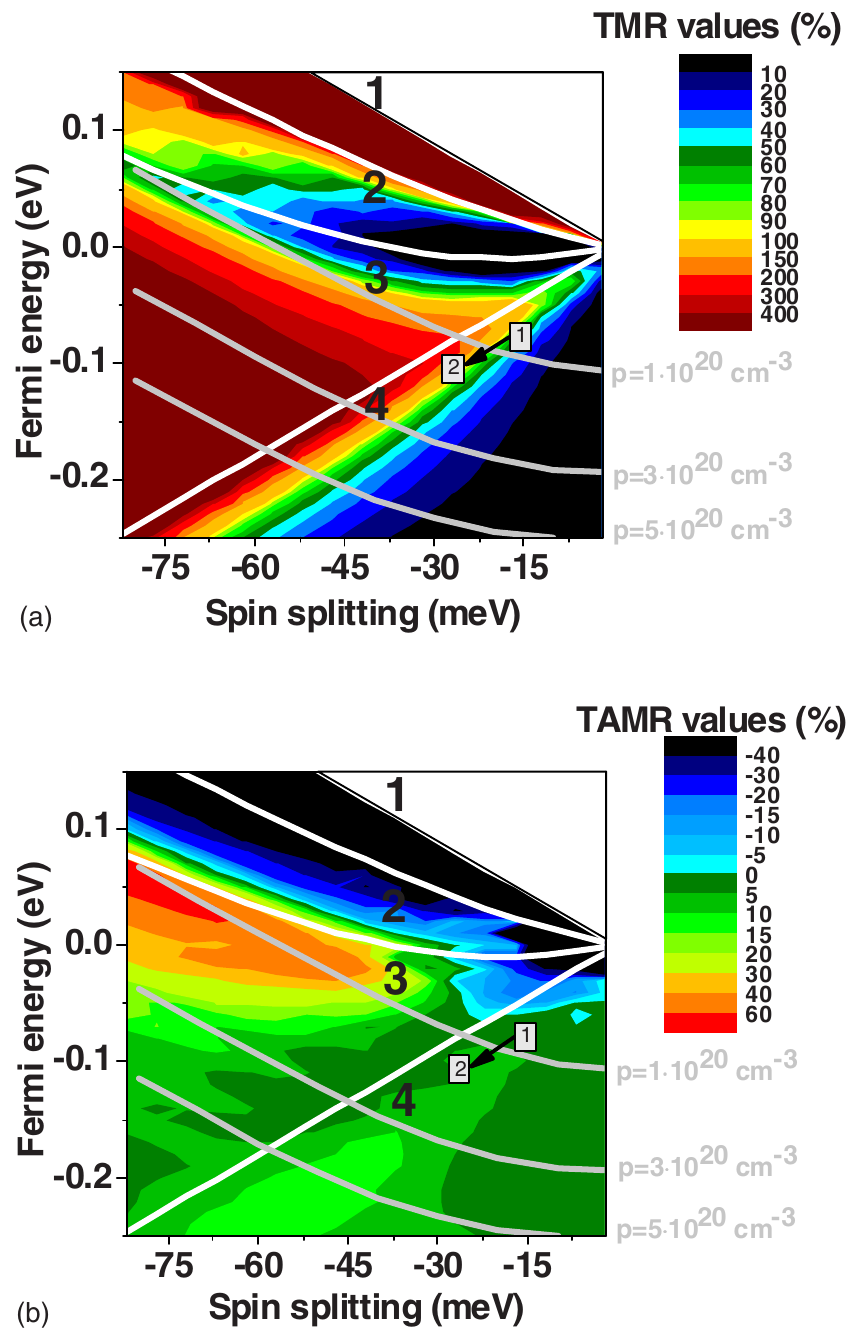}
\caption{(Color online) Calculated TMR values (a)
and TAMR values (b) represented
as a function of the Fermi and spin splitting energy
for a 6 nm (In,Ga)As barrier with a band offset of 450 meV.
White lines represent the 4 bands at the center of the
Brillouin zone. Gray lines indicate the Fermi energy for different
hole concentrations.
From~\cite{Elsen:2007_a}.}
\label{tamr2}
\end{figure}

\begin{figure}[h]
\hspace*{-0cm}\includegraphics[width=1\columnwidth,angle=0]{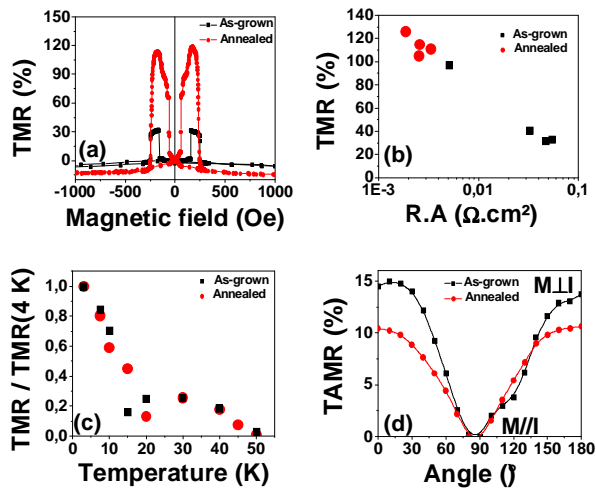}
\caption{(Color online) (a) TMR measurements as a function of the magnetic field at 1~mV and 3~K
for a 128~$\mu$m$^2$ junction. (b) TMR
measurements as a function of Resistance.Area product at 3~K for
4 (un)annealed junctions. (c) TMR at 1~mV
as a function of the temperature before and after annealing.
(d) TAMR measurements
as a function of the magnetic field at 1~mV and 3~K.
From~\cite{Elsen:2007_a}.}
\label{tamr3}
\end{figure}

Several detailed numerical studies have been performed
based on microscopic tight-binding or  kinetic-exchange models of the (Ga,Mn)As electronic structure and
the Landauer-B\"uttiker quantum transport theory \cite{Brey:2004_b,Giddings:2004_a,Sankowski:2006_a,Elsen:2007_a}.
Besides the Zener-Esaki diode geometry \cite{Sankowski:2006_a} the simulations consider magnetic tunnel junctions
with two ferromagnetic (Ga,Mn)As contacts and focus on comparison between the TMR and TAMR signals in
structures with different barrier materials and (Ga,Mn)As parameters \cite{Brey:2004_b,Sankowski:2006_a,Elsen:2007_a}.
Fig.~\ref{tamr2} shows the theoretical dependence of the TMR ratio for parallel and
antiparallel configurations of the two (Ga,Mn)As contacts and ${\bf M}$ along the [100]-direction and the TAMR ratio for
parallel magnetizations in the (Ga,Mn)As films  and ${\bf M}$ along the [100]-direction and the [001]-direction
(current direction) in a tunneling device with an {InGaAs} barrier \cite{Elsen:2007_a}. The corresponding experimental
measurements are shown in Fig.~\ref{tamr3}. There is an overall agreement between the theory and experiment,
seen also in tunnel junctions with other barrier materials, showing that the TMR  is typically 10$\times$ larger than
the TAMR. Both the theory and experiment also find that the TMR signal is always positive,
 i.e., the magnetoresistance increases as
the field is swept from saturation to the switching field. The TAMR can have both signs depending on the field angle
but also depending on the parameters of the (Ga,Mn)As film such as the hole concentration and polarization, on
the barrier characteristics, or on the temperature \cite{Gould:2004_a,Elsen:2007_a}.

At very low temperatures and bias voltages huge TAMR signals were observed \cite{Ruster:2005_a}
in a (Ga,Mn)As/GaAs/(Ga,Mn)As tunnel junction which are not described by the one-body theories of anisotropic tunneling transmission coefficients. The
observation was interpreted as a consequence of electron-electron correlation effects near the metal-insulator
transition \cite{Pappert:2006_a}. Large anisotropic magnetoresistance effects were also measured in lateral nano-constriction 
devices fabricated in ultra-thin (Ga,Mn)As materials \cite{Giddings:2004_a,Ruester:2003_a,Schlapps:2006_a}. The comparison of the anisotropic magnetoresistance signals in the unstructured part of
the device and in the nano-constriction  showed a significant
enhancement of the signal in the constriction \cite{Giddings:2004_a}.
Subsequent studies of these nano-constrictions with an additional side-gate patterned along the constriction,
discussed in detail in the following section \cite{Wunderlich:2006_a,Wunderlich:2007_a,Wunderlich:2007_b,Schlapps:2009_a}, indicated that single-electron charging effects were responsible for
the  observed large anisotropic magnetoresistance signals.

Before moving on to the (Ga,Mn)As-based field effect transistors we conclude this section with a remark on the impact
of the TAMR discovery in (Ga,Mn)As on spintronics research in other magnetic materials.
{\em Ab initio} relativistic calculations of the anisotropies in the density of states predicted sizable TAMR
effects in transition metal ferromagnets \cite{Shick:2006_a}. Landauer-B\"uttiker transport theory calculations for a Fe/vacuum/Cu structure pointed out
that apart from the density-of-states anisotropies in the ferromagnetic metal itself,
the TAMR in the tunnel devices
can arise from spin-orbit coupling  induced anisotropies of resonant surface or interface states \cite{Chantis:2006_a}. Experimentally, several reports of metal TAMR devices have already appeared in the literature including
Fe, Ni, and Co lateral break-junctions \cite{Bolotin:2006_a,Viret:2006_a} which showed comparable
($\sim 10$\%) low-temperature TMR and TAMR signals, Fe/GaAs/Au and Fe/n-GaAs vertical tunnel junctions \cite{Moser:2006_a,Uemura:2009_a}
with  a $\sim$1\% TAMR at low temperatures reflecting the spin-orbit fields and symmetries at the metal/semiconductor interface, a Co/Al$_2$O$_3$/NiFe magnetic tunnel junction with a 15\% TAMR at room temperature \cite{Grigorenko:2006_a}, reports of strongly bias dependent TAMRs in devices with CoFe \cite{Gao:2007_a} and CoPt electrodes \cite{Park:2008_a}, and larger than 100\% TAMRs in tunneling devices with an antiferromagnetic IrMn electrode \cite{Park:2010_a,Wang:2012_a}.

\subsubsection{Transistor and chemical potential anisotropy devices}
\label{cbamr}

\begin{figure}[h!]
\hspace{0cm}\includegraphics[width=.8\columnwidth,angle=0]{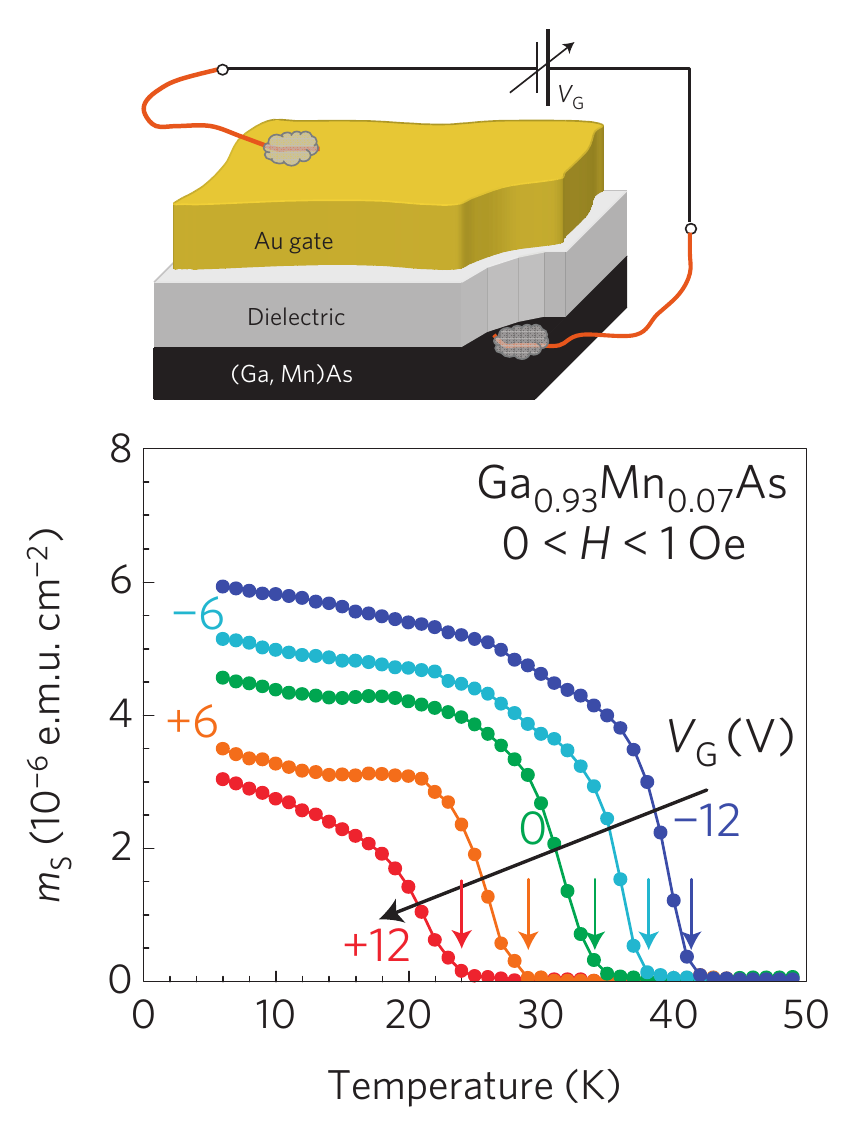}
\caption{(Color online)
Top panel: Schematics of a capacitor with an ultrathin (3.5~nm) (Ga,Mn)As layer. Bottom panel: Experimental temperature dependence of the spontaneous moment for selected values of gate voltage. Temperatures at which moment disappears define the Curie temperature $T_c$, as marked by arrows. From~\cite{Sawicki:2009_a}.
}
\label{fet}
\end{figure}

As mentioned in the Introduction, 
(In,Mn)As, (Ga,Mn)As, and (Ga,Mn)(As,P) based field effect transistors  were fabricated to demonstrate the
electric field control of ferromagnetism. It was shown
that changes in the carrier density and distribution in thin ferromagnetic semiconductor films
due to an applied gate voltage can change the Curie temperature, as illustrated in Fig.~\ref{fet}, and thus reversibly induce the
ferromagnetic/paramagnetic transition \cite{Ohno:2000_a,Chiba:2006_b,Stolichnov:2008_a,Riester:2009_a,Sawicki:2009_a}. Another remarkable effect
observed in these transistors is the
electric field control of the  magnetization orientation \cite{Chiba:2003_a,Chiba:2006_b,Wunderlich:2007_a,Chiba:2008_a,Olejnik:2008_a,Owen:2008_a,Stolichnov:2008_a,Niazi:2013_a,Chiba:2013_a}.
This 
functionality  is based on the dependence of the magnetic anisotropies
on the gate voltage, again through the modified charge density
profile in the ferromagnetic semiconductor thin film.

\begin{figure}[h]
\vspace*{0cm}
\includegraphics*[width=1\columnwidth,angle=0]{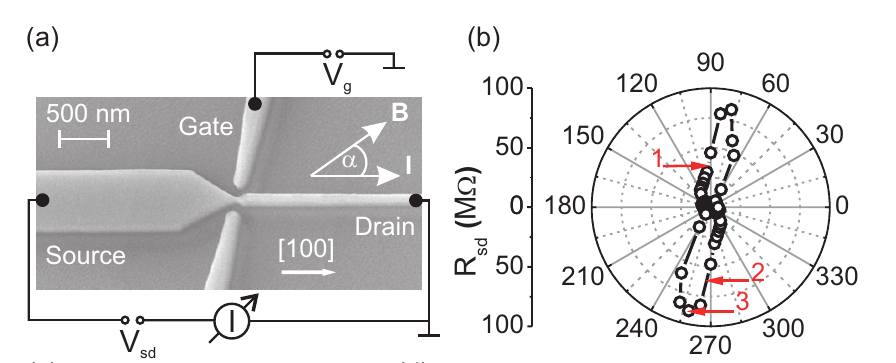}
\vspace*{0cm}
\caption{(Color online) (a) Electron micrograph of the central part of a (Ga,Mn)As SET device. (b) Polar plot of  the source-drain resistance $R_{sd}$ at 1.6 K showing the
strong anisotropy as a function of the magnetization direction. From~\cite{Schlapps:2009_a}.}
\label{GaMnAs_SET_1}
\end{figure}

For  a spintronic transistor, the magnetoresistance is another 
key characteristic which should be controllable by
the gate electric field. Large and voltage-dependent AMR effects were reported in ohmic (Ga,Mn)(As,P) channels with an integrated polymer ferroelectric
gate \cite{Mikheev:2012_a} and CB-AMR effects were demonstrated in  (Ga,Mn)As SETs \cite{Wunderlich:2006_a,Wunderlich:2007_a,Wunderlich:2007_b,Schlapps:2009_a,Ciccarelli:2012_a}, as illustrated in Fig.~\ref{GaMnAs_SET_1}.

In the conventional SET, the transfer of an electron from a source lead to a drain lead via a small, weakly-coupled island is blocked due to the charging energy of $e^2/2C_{\Sigma}$, where $C_{\Sigma}$ is the total capacitance of the island \cite{Likharev:1999_a}. Applying a voltage $V_G$ between the source lead and a gate electrode changes the electrostatic energy  function of the charge  $Q$ on the island to $Q^2/2C_{\Sigma} + QC_GV_G/C_{\Sigma}$  which has a minimum at $Q_0=-C_GV_G$. By tuning the continuous external variable $Q_0$ to $(n+1/2)e$, the energy associated with increasing the charge $Q$ on the island from $ne$ to $(n+1)e$ vanishes and electrical current can flow between the leads. Changing the gate voltage then leads to CB oscillations in the source-drain current 
where each period corresponds to increasing or decreasing the charge state of the island by one electron. The energy can be written as a sum of the internal, electrostatic charging energy term and the term associated with, in general, different chemical potentials of the lead and of the island:
\begin{equation}
U=\int_0^Q dQ^{\prime} \Delta V_D(Q^{\prime}) + Q\Delta\mu/e \; ,
\end{equation}
where $\Delta V_D(Q)=(Q+C_GV_G)/C_{\Sigma}$. The Gibbs energy $U$ is minimized at $Q_0=-C_G(V_G+V_M)$.

The ferromagnetic SETs with (Ga,Mn)As in the transport channel of the transistor \cite{Wunderlich:2006_a,Schlapps:2009_a} were fabricated by  trench-isolating a side-gated narrow (10's nm) channel in a thin-film (Ga,Mn)As epilayer. The narrow channel technique is a simple approach to realize a SET and was used previously to produce non-magnetic thin film Si and GaAs-based SETs in which disorder potential fluctuations create small islands in the channel without the need for a lithographically defined island \cite{Kastner:1992_a,Tsukagoshi:1998_a}.
The non-uniform carrier concentration produces differences between chemical potentials $\Delta\mu$ of the lead and of the island in the constriction. There are two mechanisms through which $\Delta\mu$ depends on the
magnetic field. One is caused by the direct Zeeman coupling of the external magnetic field and leads to a CB magnetoresistance
previously observed in ferromagnetic metal SETs \cite{Ono:1997_a}. 

The CB-AMR effect, discovered in the  (Ga,Mn)As SETs, is attributed to the spin-orbit coupling induced anisotropy of the carrier chemical potential, i.e., to magnetization orientation dependent differences between chemical potentials of the lead and of the island in the constriction \cite{Wunderlich:2006_a}.
For the CB-AMR effect, the magnetization orientation dependent shift of the CB oscillations is given by $V_M=C_{\Sigma}/C_G \, \Delta\mu({\bf M})/e$. Since $|C_GV_M|$ has to be of order $|e|$ to cause a marked shift
in the oscillation pattern, the corresponding $|\Delta\mu({\bf M})|$ has to be similar to $e^2/C_{\Sigma}$, {\em i.e.},
of the order of the island single-electron charging energy. The fact that CB-AMR occurs when the anisotropy in a band structure derived parameter is comparable to an independent scale (single-electron charging energy) makes the effect distinct and potentially much larger in magnitude as compared to the  AMR and TAMR. Indeed, resistance variations by more than 3 orders of magnitude were   observed in the (Ga,Mn)As SETs.

\begin{figure}[h]
\includegraphics*[width=1\columnwidth,angle=0]{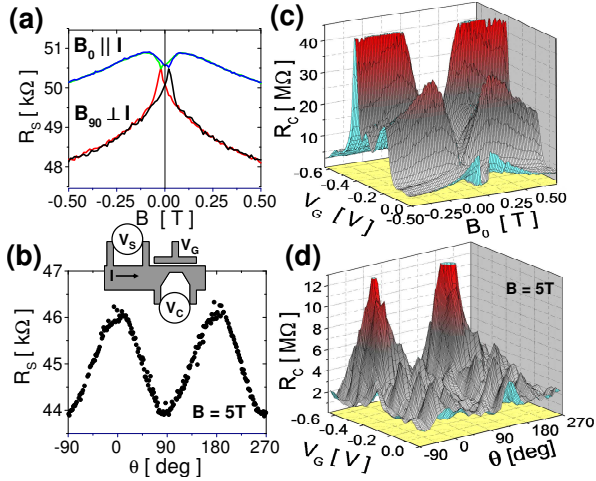}
\caption{(Color online) (a) Resistance $R_{S}=V_S/I$ of the unstructured bar (see schematic diagram) vs up and down sweeps of in-plane magnetic field parallel (blue/green) and perpendicular (red /black) to the current direction. (b) $R_S$ vs the angle between the current direction and an applied in-plane magnetic field of 5 T, at which ${\bf M} \parallel {\bf B}$. (c) Channel resistance $R_C$ vs gate voltage and down sweep of the magnetic field parallel to current.  (d) $R_C$  vs. gate voltage and the angle between the current direction and an applied in-plane magnetic field of 5~T.
From~\cite{Wunderlich:2006_a}.}
\label{CBAMR}
\end{figure}

The sensitivity of the magnetoresistance to the orientation of the applied magnetic field is
 an indication of the anisotropic magnetoresistance origin of the effect. This is confirmed by the observation of comparably large and gate-controlled magnetoresistance in a field-sweep experiment and when the saturation magnetization is rotated with respect to the crystallographic axes. The
 field-sweep and rotation measurements are shown in Figs.~\ref{CBAMR}(c) and (d) and compared with
 analogous measurements of the ohmic AMR in the unstructured part of the (Ga,Mn)As bar, plotted in Figs.~\ref{CBAMR}(a) and (b) \cite{Wunderlich:2006_a}.
In the unstructured bar, higher or lower resistance states correspond to magnetization along or perpendicular to the current direction. Similar behavior is seen in the SET part of the device at, for example, $V_G=-0.4V$, but the anisotropic magnetoresistance is now hugely increased and depends strongly on the gate voltage.

\begin{figure}[h]
\includegraphics*[width=1\columnwidth,angle=0]{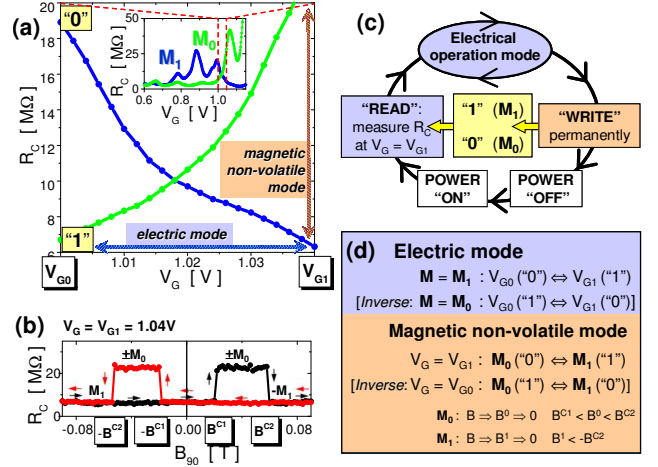}
\caption{(Color online) (a) Two opposite transistor characteristics (blue and green) in a gate-voltage range ~V ($V_{G0}$) to 1.04~V ($V_{G1}$) for two different magnetization orientations ${\bf M_0}$ and ${\bf M_1}$; corresponding Coulomb blockade oscillations in a larger range of $V_G = 0.6$ to 1.15~V are shown in the inset. Switching between low-resistance ("1") and high-resistance ("0") states can be performed electrically or magnetically. (b) Hysteretic magnetoresistance at constant gate voltage $V_{G1}$ illustrating the non-volatile memory effect in the magnetic mode. (c) Illustration of integrated transistor (electric mode) and permanent storage (magnetic mode) functions in a single nanoscale element. (d) The transistor characteristic for ${\bf M}={\bf M_1}$ is reminiscent of an n-type field effect transistor and is inverted (reminiscent of a p-type field effect transistor) for ${\bf M}={\bf M_0}$; the inversion can also be realized in the non-volatile magnetic mode.
From \cite{Wunderlich:2007_a}.}
\label{cbamr_trans}
\end{figure}

The huge magnetoresistance signals can be also hysteretic which shows that CB-AMR SETs can act as a non-volatile memory/transistor element. In non-magnetic SETs, the CB "on" (low-resistance) and "off" (high-resistance) states can represent logical "1" and "0" and the switching between the two states can be realized by applying a gate voltage, in analogy with a standard field-effect transistor. The CB-AMR SET can be addressed also magnetically with comparable "on" to "off" resistance ratios in the electric and magnetic modes. The functionality is illustrated in Fig.~\ref{cbamr_trans} \cite{Wunderlich:2007_a}. The inset of Fig.~\ref{cbamr_trans}(a) shows two CB oscillation curves corresponding to two different magnetization states ${\bf M_0}$ and ${\bf M_1}$. As illustrated in Fig.~\ref{cbamr_trans}(b),  ${\bf M_0}$  can be achieved by performing a small loop in the magnetic field, $B\rightarrow B_0\rightarrow0$ where $B_0$ is larger than the first switching field $B_{c1}$ and smaller than the second switching field $B_{c2}$, and ${\bf M_1}$ is achieved by performing the large field-loop,  $B\rightarrow B_1\rightarrow0$ where $B_1<-B_{c2}$. The main plot of Fig.~\ref{cbamr_trans}(a) shows that the high resistance 0 state can be set by either the combinations $({\bf M_1},V_{G0})$ or $({\bf M_0},V_{G1})$ and the low resistance 1 state by  $({\bf M_1},V_{G1})$ or $({\bf M_0},V_{G0})$. One can therefore switch between states 0 and 1 either by changing $V_G$ in a given magnetic state (the electric mode) or by changing the magnetic state at fixed $V_G$ (the magnetic mode). Due to the hysteresis, the magnetic mode represents a non-volatile memory effect.
The diagram in Fig.~\ref{cbamr_trans}(c) illustrates one of the new functionality concepts the device suggests in which low-power electrical manipulation and permanent storage of information are realized in one physical nanoscale element.  Fig.~\ref{cbamr_trans}(d)  highlights the possibility to invert the transistor characteristic; for example, the system is in the low-resistance "1" state at $V_{G1}$ and in the high-resistance "0" state at $V_{G0}$ (reminiscent of an n-type field effect transistor) for the magnetization  ${\bf M_1}$ while the characteristic is inverted (reminiscent of a p-type field effect transistor) by changing magnetization to ${\bf M_0}$.

Chemical potential shifts in the relativistic band structure
of solids have rarely been discussed in the scientific
literature. This reflects the conceptual difficulty
in describing the chemical potential shifts by quantitative
theories, the lack of direct measurements of the effect,
and the lack of proposals in which the phenomenon
could open unconventional paths in microelectronic device
designs. Refs.~\cite{Wunderlich:2006_a, Shick:2010_a,Ciccarelli:2012_a} are among the few attempts
to quantify chemical potential anisotropies with respect
to the spin orientation in semiconductor and metal magnets
using relativistic model Hamiltonian or full-potential
density-functional band structure calculations. The theories
could account for chemical potential shifts due to
the distortion in the dispersion of the spin-orbit coupled
bands but for principle reasons omit possible shifts of the
vacuum level with respect to band edges, in other words,
possible shifts in band line-ups in realistic heterostructure
systems. 

In experiments described above and in other related measurements, the magnetic
materials have been integrated in a conventional
design of a magneto-electronic device, i.e. embedded in
the transport channel, and the chemical potential shifts
could have been inferred only indirectly from the measured
data \cite{Ono:1997_a,Deshmukh:2002_a,vanderMolen:2006_a,Wunderlich:2006_a,Tran:2009_a,Schlapps:2009_a,Bernand-Mantel:2009_a}. One exception is the work discussed in more detail below, which has demonstrated  direct measurements of chemical potential
shifts in a spin-orbit coupled ferromagnet \cite{Ciccarelli:2012_a}. The corresponding 
spintronic device operates without spin
currents, i.e, it demonstrates a  functionality which goes beyond the common
concepts of spintronics.  The device represents an unconventional spin transistor where the charge state of the transport channel is sensitive to the spin state of its magnetic gate.

\begin{figure}[h]
\includegraphics*[width=1\columnwidth,angle=0]{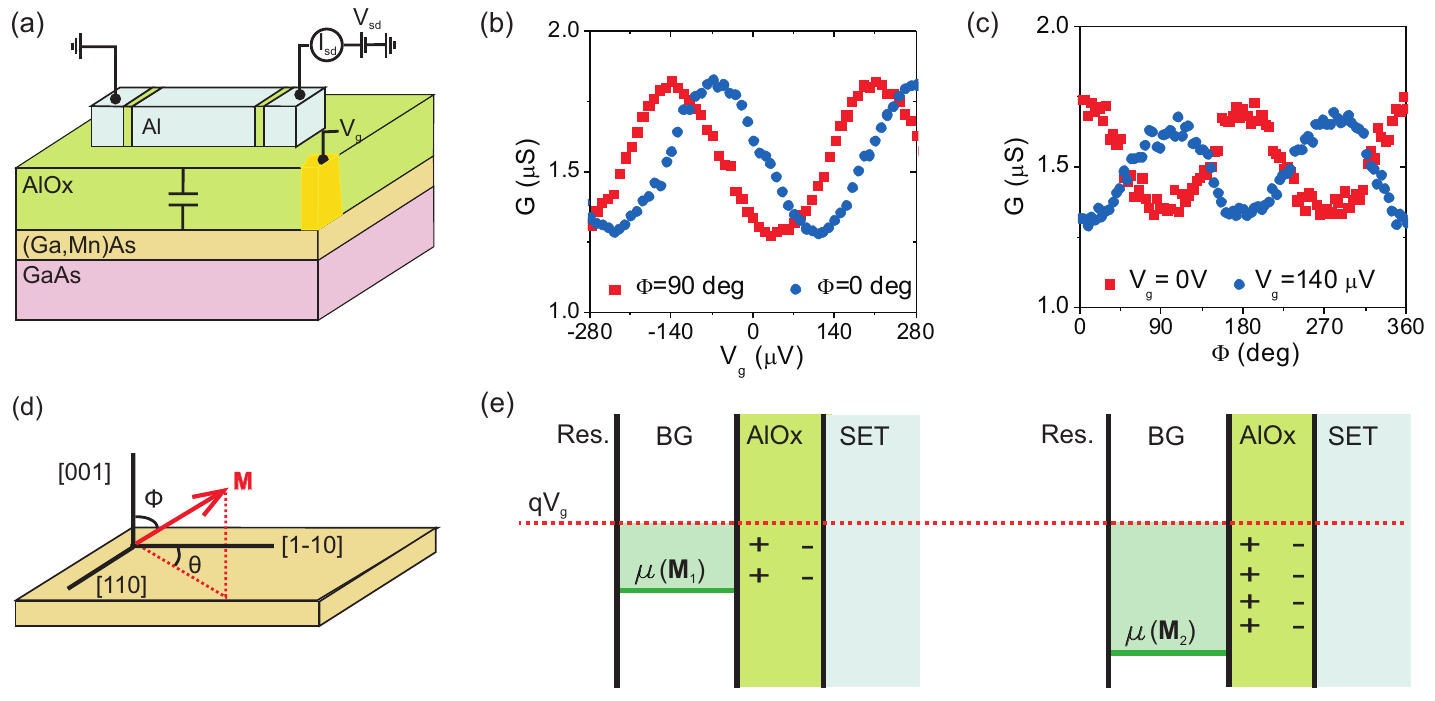}
\caption{(Color online) (a) Schematic showing the SET channel separated by AlO$_x$ dielectric from the ferromagnetic (Ga,Mn)As back-gate. The SET comprises Al leads and island, and AlO$_x$ tunnel barriers. (b) Coulomb oscillations for the SET on Ga$_{0.97}$Mn$_{0.03}$As for two different polar angles $\Phi$ of the magnetization. (c) Magneto-Coulomb oscillations shown by the same SET by varying the angle of magnetization for two different gate voltages.   (d) Magnetization vector with respect to (Ga,Mn)As crystal axes. (e) Schematic explaining the spin gating phenomenon: reorientation of the magnetization from \textbf{M$_1$} to \textbf{M$_2$} causes a change in the chemical potential of the (Ga,Mn)As back-gate (BG). This causes charge to flow onto the back-gate  from the reservoir (Res.). The net effect is to alter the charge on the back-gate and therefore the SET conductance. The externally applied electrochemical potential on the gate $\mu_{ec}=qV_g$ is held constant.
From~\cite{Ciccarelli:2012_a}.}
\label{spin_gating}
\end{figure}

The  SET from Ref.~\cite{Ciccarelli:2012_a} has a micron-scale Al island separated by aluminum oxide tunnel junctions from Al source and drain leads (Fig.~\ref{spin_gating}(a)). It is fabricated on top of an epitaxially grown (Ga,Mn)As layer, which is electrically insulated from the SET by an alumina dielectric, and act as a spin-back-gate to the SET.  
By sweeping the externally applied potential to the SET gate ($V_g$) one  obtains the conductance oscillations that characterize the CB, as shown in Fig.~\ref{spin_gating}(b). Due to the magnetic gate a shift is observed in these oscillations by an applied saturating magnetic field which rotates the magnetization in the (Ga,Mn)As gate. Fig.~\ref{spin_gating}(b) shows measurements for the in-plane ($\Phi=90^\circ$) and for the perpendicular-to-plane ($\Phi=0^\circ$) directions of magnetization. Alternatively, Fig.~\ref{spin_gating}(c) shows the channel conductance as a function of the magnetization angle $\Phi$ for a fixed external potential $V_g$ applied to the gate. The oscillations in $\Phi$ seen in Fig.~\ref{spin_gating}(c) are of comparable amplitude as the oscillations in $V_g$ in Fig.~\ref{spin_gating}(b).

Since the (Ga,Mn)As back-gate is attached to a charge reservoir, any change in the internal chemical potential of the gate induced by the rotating magnetization vector causes an inward, or outward, flow of charge in the gate, as illustrated in Fig.~\ref{spin_gating}(e). This change in back-gate charge offsets the Coulomb oscillations (Fig.~\ref{spin_gating}(b)) and changes the conductance of the transistor channel for a fixed external potential applied to the gate (Fig.~\ref{spin_gating}(c)). 

In the case of the SET with
the magnetic gate no capacitance scaling factors are
required and the chemical potential shift may be directly
read off as a shift in gate voltage. This removes a source
of systematic error, present in  experiments
on the magneto-Coulomb effect \cite{Ono:1997_a,Deshmukh:2002_a,vanderMolen:2006_a} or chemical potential
anisotropy in SETs with the ferromagnet forming part of the transport channel (lead or island) \cite{Wunderlich:2006_a,Schlapps:2009_a,Bernand-Mantel:2009_a,Tran:2009_a}, where the gate voltage shift must
be scaled due to the presence of a capacitive divider.

In agreement with experiment, the theoretical chemical potential anisotropies in the studied (Ga,Mn)As epilayers with Mn doping of several per cent are of the order of 10-100~$\mu$eV \cite{Ciccarelli:2012_a}.  So far, the spin-gating technique was employed to accurately measure the anisotropic (and also isotropic Zeeman \cite{Ciccarelli:2012_a}) chemical potential shifts in (Ga,Mn)As. However, the technique can be applied to catalogue these effects in other magnetic materials by the simple step of exchanging the magnetic gate electrode.

\subsubsection{Spin torques and spin pumping}
\label{torques}

When spin polarized carriers are injected into a magnetic region whose moments are misaligned with the injected spin polarization of the carriers,   STTs can act on the magnetic moments \cite{Ohno:2007_a,Ralph:2007_a}. The phenomena belong to an important area of spintronics research focusing on the means for manipulating magnetization by electrical currents and are the basis of the emerging technologies for scalable MRAMs \cite{Chappert:2007_a}. Apart from STTs in non-uniform magnetic structures, whose research in (Ga,Mn)As is reviewed later in Section~\ref{STT}, experiments in (Ga,Mn)As devices established the presence of current-induced spin torques in uniform magnetic structures originating from the internal spin-orbit coupling. These current-induced SOT phenomena are reviewed in Section~\ref{SOT}, and in Sections~\ref{OSTT} and \ref{OSOT} we discuss the optical counterparts of the STT and SOT which were also discovered in (Ga,Mn)As. A theory framework outlined in this section can be used to highlight the key common and distinct characteristics of all these spin torque phenomena \cite{Ralph:2007_a,Zhang:2004_c,Vanhaverbeke:2007_a,Rossier:2003_a,Nemec:2012_a,Tesarova:2012_b,Ranieri:2012_a}. At the end of this section we also introduce the Onsager related reciprocal effects to the STT (spin-pumping) and to the SOT \cite{Tserkovnyak:2005_a,Hals:2010_a}.

The framework for describing spin torque phenomena treats  the non-equilibrium spin density of carriers ${\bf s}$ and magnetization  of the ferromagnet as separate degrees of freedom and explores their coupled dynamics. The dilute moment ferromagnetic semiconductor  (Ga,Mn)As is a model system in which the separation is well justified microscopically; magnetization is primarily due to  Mn $d$-orbital local moments while the carrier states near the top of the valence band (or bottom of the conduction band) are dominated by As $p$-orbitals (or Ga $s$-orbitals). 

The carrier Hamiltonian can be written as 
\begin{equation}
H=H_0+H_{ex}+H_{so}\,,
\label{H} 
\end{equation}
where $H_0$ is the spin-independent part of the Hamiltonian, the kinetic-exchange term 
\begin{equation}
H_{ex}=J{\bf M}\cdot{\bm\sigma}
\label{H_ex}
\end{equation}
where $J$ is the exchange coupling constant (in units of energy$\cdot$volume), ${\bf M}=cS\hat{\bf M}$ ($S=5/2$) is the spin density of Mn local-moments, $\hat{\bf M}$ is the magnetization unit vector, and ${\bm\sigma}$ is the carrier spin operator, and $H_{so}$ is the spin-orbit coupling Hamiltonian. The current-induced and optical STT phenomena are  determined by the following dynamics equations for the non-equilibrium carrier spin density {\bf s} and for the magnetic moment density ${\bf M}$,
\begin{eqnarray}
\label{carrier_STT}
\frac{d{\bf s}}{dt}&=&\frac{J}{\hbar}{\bf s}\times{\bf M}+P{\bf n}-\frac{\bf s}{\tau_s}\\
\frac{d{\bf M}}{dt}&=&\frac{J}{\hbar}{\bf M}\times{\bf s}\,.
\label{M_STT}
\end{eqnarray} 
The first term on the right-hand side of Eq.~(\ref{carrier_STT}) is obtained from the Hamiltonian dynamics,
\begin{equation}
\frac{d\langle{\bm\sigma}\rangle}{dt}=\frac{1}{i\hbar}\langle[{\bm\sigma},H]\rangle\,,
\label{Ham_dyn}
\end{equation}
where $\langle\cdot\cdot\cdot\rangle$ represents quantum-mechanical averaging over the non-equilibrium carrier states, $\langle{\bm\sigma}\rangle={\bf s}$, and $H_{so}$ was neglected in $H$ for the STT effects which are basically non-relativistic. The second term in Eq.~(\ref{carrier_STT}) is the rate $P$ of carriers with spin polarization along a unit vector ${\bf n}$ injected from an external polarizer. In the current induced STT, the external polarizer may be, e.g.,  an adjacent magnetic layer in a multilayer structure. In the optical STT, $P$ and ${\bf n}$ of non-equilibrium photo-carrier spins are governed again by the properties of an external polarizer which are the intensity, propagation axis and helicity of the circularly polarized pump laser pulse. The last term in Eq.~(\ref{carrier_STT}) reflects a finite spin-lifetime of the non-equilibrium carriers in the ferromagnet. 

Two components of the STT can be distinguished when considering two limiting cases of Eq.~(\ref{carrier_STT}) \cite{Ralph:2007_a,Zhang:2004_c,Vanhaverbeke:2007_a,Rossier:2003_a,Nemec:2012_a}. One limit is when  the carrier spin lifetime $\tau_{\rm s}\gg\tau_{\rm ex}$ where the carrier precession time $\tau_{\rm ex}=\hbar/JcS$.  In this limit the last term on the right-hand side of Eq.~(\ref{carrier_STT}) can be neglected and introducing the steady-state solution of Eq.~(\ref{carrier_STT}) ($d{\bf s}/dt=0$),
\begin{equation}
{\bf s}=P\tau_{ex}({\bf n}\times\hat{\bf M})\,,
\label{adiabatic_s}
\end{equation}
 into Eq.~(\ref{M_STT}) yields the anti-damping adiabatic STT \cite{Slonczewski:1996_a,Berger:1996_a},
\begin{equation}
\frac{d{\bf M}}{dt}=P\hat{\bf M}\times({\bf n}\times\hat{\bf M})\,.
\label{adiabatic_STT}
\end{equation}
(Recall that the form of this torque is the same as the damping term in the Landau-Lifshitz-Gilbert equation.) In this adiabatic STT the entire spin angular momentum of the injected carriers is transferred to the magnetization, independent of $\tau_{\rm s}$, $\tau_{\rm ex}$, and other parameters of the system. The adiabatic STT has been considered since the seminal theory works \cite{Slonczewski:1996_a,Berger:1996_a} on carrier induced magnetization dynamics which opened a large field ranging from metal magnetic tunnel junctions switched by current to tuneable oscillators \cite{Ralph:2007_a} and ultrafast photo-magnetic laser excitations of ferromagnetic semiconductors \cite{Rossier:2003_a,Nemec:2012_a}

In the opposite limit of $\tau_{\rm s}\ll\tau_{\rm ex}$, the first term on the right-hand side of Eq.~(\ref{carrier_STT}) can be neglected resulting in the field-like non-adiabatic STT \cite{Zhang:2004_c}, 
\begin{equation}
\frac{d{\bf M}}{dt}=\frac{\tau_s}{\tau_{ex}}P(\hat{\bf M}\times{\bf n})\,,
\label{non-adiabatic_STT}
\end{equation}
The non-adiabatic STT is perpendicular to the adiabatic STT and only a fraction $\tau_{\rm s}/\tau_{\rm ex}$ of the injected spin angular momentum is transferred to the magnetization.  For intermediate ratios $\tau_{\rm ex}/\tau_{\rm s}$, both the non-adiabatic and adiabatic torques are present and the ratio of their magnitudes (non-adiabatic to adiabatic) is given by $\beta=\tau_{\rm ex}/\tau_{\rm s}$ \cite{Zhang:2004_c,Vanhaverbeke:2007_a,Rossier:2003_a}.  The non-adiabatic STT plays a crucial role in current induced domain wall (DW) motion \cite{Zhang:2004_c,Vanhaverbeke:2007_a,Metaxas:2007_a,Mougin:2007_a} and, as we discuss below, (Ga,Mn)As is a favorable material for exploring the effects of the non-adiabatic and adiabatic STTs. 

The SOT is distinct from the STT as it is a relativistic phenomenon in which magnetization dynamics is induced in a uniform spin-orbit coupled ferromagnet in the absence of the external polarizer \cite{Bernevig:2005_c,Manchon:2008_b,Manchon:2009_a,Chernyshov:2009_a,Garate:2009_a,Miron:2010_a,Endo:2010_a,Fang:2010_a,Gambardella:2011_a,Kurebayashi:2013_a,Tesarova:2012_b}. The Hamiltonian spin-dynamics described by Eq.~(\ref{Ham_dyn}) with the $H_{so}$ term included in the carrier Hamiltonian implies that Eq.~(\ref{carrier_STT}) is replaced with, 
\begin{equation}
\frac{d{\bf s}}{dt}=\frac{J}{\hbar}{\bf s}\times{\bf M}+\frac{1}{i\hbar}\langle[{\bm\sigma},H_{so}]\rangle\,.
\label{carrier_SOT}
\end{equation} 
The SOT is obtained by introducing the steady-state solution of Eq.~(\ref{carrier_SOT}) into Eq.~(\ref{M_STT}),
\begin{equation}
\frac{d{\bf M}}{dt}=\frac{J}{\hbar}{\bf M}\times{\bf s}=\frac{1}{i\hbar}\langle[{\bm\sigma},H_{so}]\rangle\,.
\label{SOT_eq}
\end{equation} 

In the current-induced SOT  the absence of an external polarizer implies that the effect can be observed when electrical current is driven through a uniform magnetic structure \cite{Bernevig:2005_c,Manchon:2008_b,Manchon:2009_a,Chernyshov:2009_a,Garate:2009_a,Miron:2010_a,Endo:2010_a,Fang:2010_a,Gambardella:2011_a,Kurebayashi:2013_a}. The optical SOT analogy of the absence of an external polarizer is in that the non-equilibrium photo-carriers are excited by helicity independent pump laser pulses which do not impart angular momentum \cite{Tesarova:2012_b}. 

The electrical and optical SOTs may differ in the specific contributions to $H_{so}$ which dominate the effect. This can be illustrated considering the Boltzmann linear-response transport theory of the current induced SOT. Here $\langle\cdot\cdot\cdot\rangle$ represents quantum-mechanical averaging constructed from the equilibrium eigenstates of $H$ and with the non-equilibrium steady state entering through an asymmetric redistribution of the occupation numbers of these eigenstates on the Fermi surface due to the applied electrical drift and relaxation. Because of this specific form of the asymmetric non-equilibrium charge redistribution with a conserved total number of carriers, the current induced SOT requires broken inversion symmetry terms in $H_{so}$ \cite{Manchon:2008_b,Manchon:2009_a,Garate:2009_a,Chernyshov:2009_a,Miron:2010_a,Fang:2010_a}. The  optical SOT is caused by optical generation and relaxation of photo-carriers without an applied drift (without a defined direction of the carrier flow) and without conserving the equilibrium number of carriers in dark. Therefore, the broken inversion symmetry in the crystal is not required, and inversion symmetric $H_{so}$ plus the time-reversal symmetry breaking exchange-coupling term in the carrier Hamiltonian are sufficient for observing the optical SOT.  

In the STT, spin-angular momentum is transferred from the carriers to the magnet, applying a torque to the magnetization. Via the STT, the injected spin current is able to excite magnetization dynamics. A reciprocal effect to the STT is the spin-pumping phenomenon in which pure spin-current is generated from magnetization precession \cite{Mizukami:2001_a,Tserkovnyak:2005_a}.  The spin-pumping has been measured, e.g., in ferromagnet/normal-metal/ferromagnet GMR structures \cite{,Heinrich:2003_a,Woltersdorf:2007_a} or in ferromagnet/normal-metal bilayers \cite{Saitoh:2006_a,Czeschka:2011_a}. In the latter structure, the  inverse SHE in the spin-orbit coupled paramagnet adjacent to the ferromagnet serves as a spin-charge converter and provides direct means for detecting the spin pumping phenomenon electrically. Spin pumping can, therefore, be used  not only for  probing magnetization dynamics in ferromagnets but also  spin physics in paramagnets, e.g., for measuring the SHE angles.  Magnetization dynamics of ferromagnetic resonance also produces electrical signals in the ferromagnetic layer through galvanomagnetic effects. Experiments in a (Ga,Mn)As/p-GaAs model system, where sizable galvanomagnetic effects are present, have demonstrated that neglecting the galvanomagnetic effects in the ferromagnet can lead to a large overestimate of the SHE angle in the paramagnet. The study has also shown a method to separate voltages of these different origins in the spin-pumping experiments in the ferromagnet/paramagnet bilayers \cite{Chen:2013_a}.  

The Onsager reciprocity relations imply that, as for the STT/spin-pumping, there exists a reciprocal phenomenon of the SOT in which electrical signal is  generated from magnetization precession in a uniform, spin-orbit coupled magnetic system with broken spatial inversion symmetry \cite{Hals:2010_a,Tatara:2013_a}. In this reciprocal SOT effect no secondary spin-charge conversion element is required and, as for the SOT, (Ga,Mn)As with broken inversion symmetry in its bulk crystal structure and strongly spin-orbit coupled holes represents a favorable model system to explore this phenomenon.

\subsubsection{Current induced spin-transfer torque}
\label{STT}
In this section we focus on the current-induced STT studies in (Ga,Mn)As. The dilute-moment ferromagnet (Ga,Mn)As has a low saturation magnetization, as compared to conventional dense-moment metal ferromagnets. Together with the high degree of spin polarization of carriers it implies that electrical currents required to excite magnetization by STT in (Ga,Mn)As are also comparatively low. In magnetic tunnel junctions with (Ga,Mn)As electrodes, STT induced switching was observed at current densities of the order $10^{4}-10^{5}$~Acm$^{-2}$ \cite{Chiba:2004_b}, consistent with theory expectations \cite{Sinova:2004_b}. These are 1-2 orders of magnitude lower current densities than in the STT experiments in common dense-moment metal ferromagnets.

\begin{figure}[h]
\includegraphics*[width=1\columnwidth,angle=0]{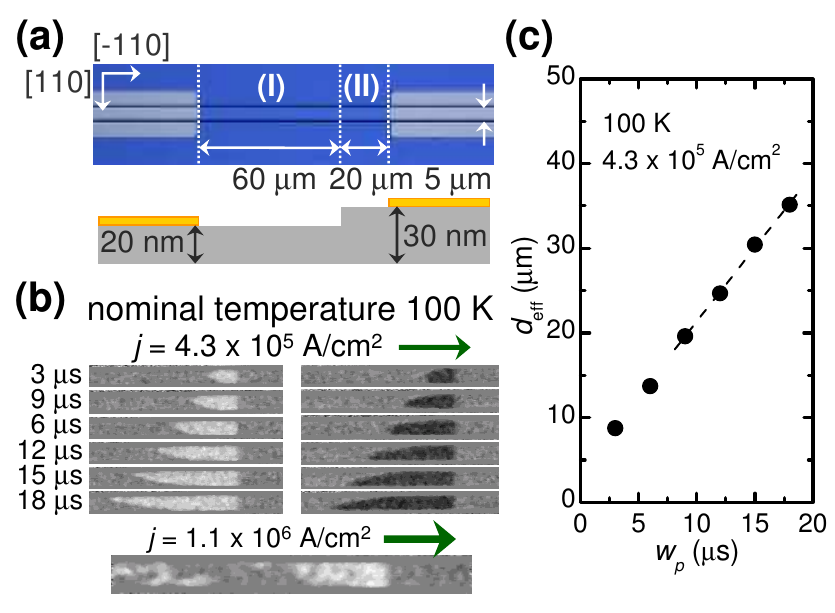}
\caption{(Color online) (a) Layout of the device showing the
5~$\mu$m mesa and step for DW pinning in perpendicular magnetic
anisotropy (Ga,Mn)As film. (b) 7~$\mu$m wide magneto-optical
images with a 5~$\mu$m mesa in the center show that DW moves
in the opposite direction to current independent of the initial
magnetization orientation, and that DW displacement is proportional
to pulse duration (c). The lowest panel in (b) shows
destruction of ferromagnetic phase by Joule heating.
From ~\cite{Yamanouchi:2006_a}.}
\label{ohno_dw}
\end{figure}

Current induced  DW motion in the creep regime at $\sim 10^{5}$~Acm$^{-2}$ current densities was reported and thoroughly explored 
in perpendicularly
magnetized (Ga,Mn)As thin film devices, shown in Fig.~\ref{ohno_dw} \cite{Yamanouchi:2004_a,Chiba:2006_a,Yamanouchi:2006_a,Yamanouchi:2007_a}.
The perpendicular magnetization geometry was achieved by growing the films under a tensile strain on a (In,Ga)As substrate and allowed for a direct
magneto-optical Kerr-effect imaging of the magnetic domains, as illustrated in Fig.~\ref{ohno_dw}.  

\begin{figure}[h]
\includegraphics*[width=1\columnwidth,angle=0]{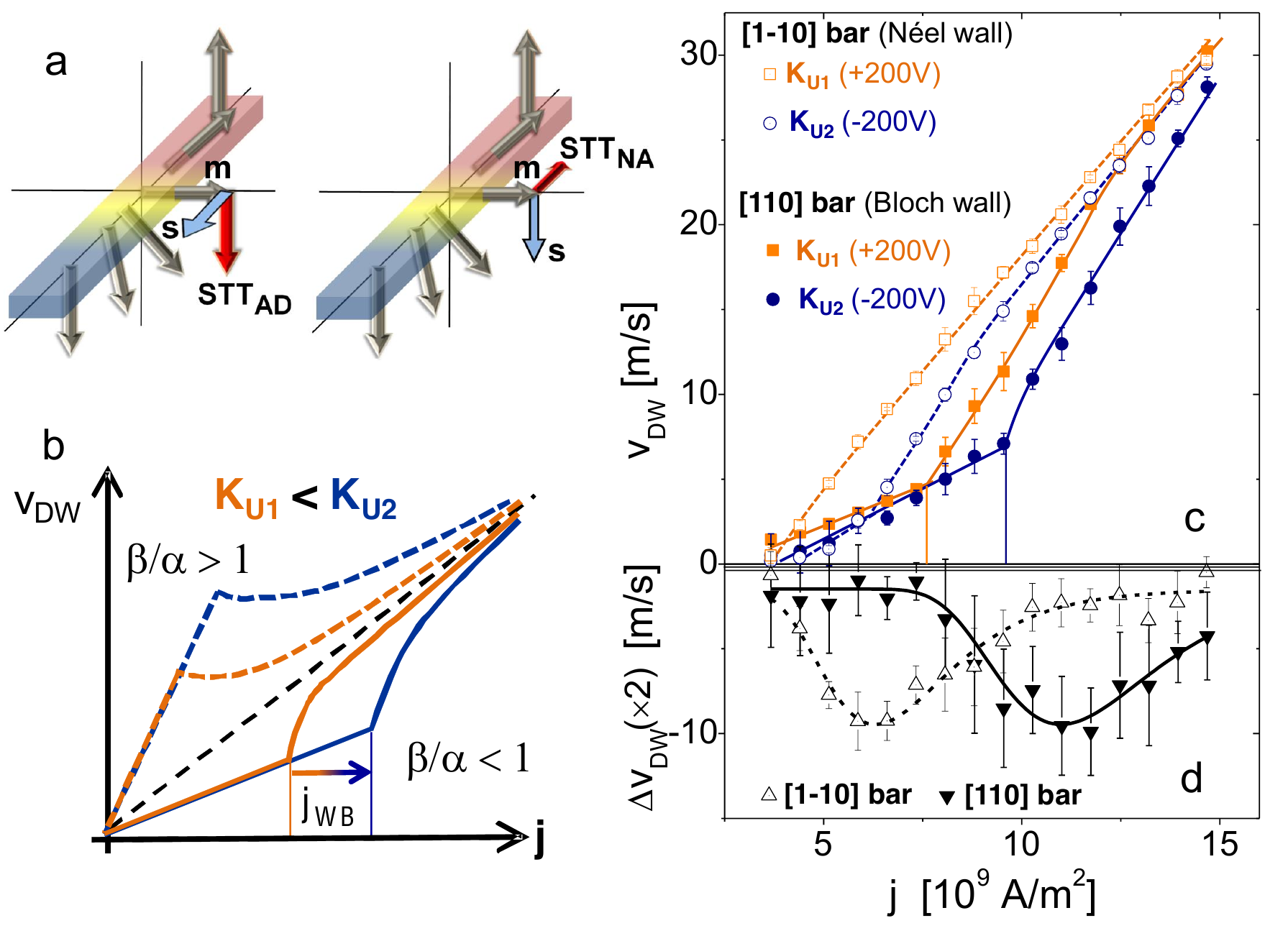}
\caption{(Color online) (a) Illustration of the steady-state non-equilibrium carrier spin polarization {\bf s} and corresponding adiabatic STT (STT$_{\rm AD}$) acting on magnetization {\bf m} in the $\tau_s\gg\tau_{ex}$ limit (left) and non-adiabatic STT (STT$_{\rm NA}$) in the $\tau_s\ll\tau_{ex}$ limit (right). (b) Schematic diagram of the predicted DW velocity as a function of the driving current in the presence of adiabatic and non-adiabatic STTs and $\beta/\alpha<1$ or $\beta/\alpha>1$, and of the predicted shift of the WB threshold current $j_{WB}$ for two values of the in-plane magnetocrystalline constant ${K_{u,1}<K_{u,2} }$, controlled {\em in situ} by a piezo-stressor. (c) Measured DW velocity vs. driving current density  at piezo-voltages -200~V or +200~V, strengthening or weakening the [1$\bar{1}$0] in-plane easy axis, respectively. Open symbols correspond to the  [1$\bar{1}$0]-oriented microbar with less internally stable N\'eel DW and filled symbols to the [110]-oriented microbar with more internally stable Bloch DW.  The character of the measured data, including the shift of the WB threshold current, imply STTs with $\beta/\alpha<1$.  (d) $\Delta v_{DW} = v_{DW}(+200V)-v_{DW}(-200V)$ vs. current density illustrates the piezo-electric control of the DW mobility achieved starting from lower currents in  the  [1$\bar{1}$0]-oriented microbar with less internally stable DW.   
From~\cite{Ranieri:2012_a}.}
\label{DW_2}
\end{figure}

Alternatively, tensile-strained perpendicularly magnetized films for DW studies were grown on a GaAs substrate with P added into the magnetic film \cite{Wang:2010_a,Curiale:2012_a,Ranieri:2012_a}. In high crystal quality (Ga,Mn)(As,P)/GaAs epilayers the viscous flow regime was achieved over a wide current  range  allowing to observe \cite{Ranieri:2012_a} the lower-current  steady DW motion regime separated from a  higher-current precessional regime by the Walker breakdown (WB) \cite{Thiaville:2005_a, Metaxas:2007_a,Mougin:2007_a}. This in turn enabled to assess the ratio of adiabatic and non-adiabatic  STTs in the current driven DW motion. When the non-adiabatic STT is strong enough that $\beta/\alpha>1$, where $\alpha$ is the DW Gilbert damping parameter, the mobility of a DW (velocity  divided by the DW driving current) is larger below the WB. For  $\beta/\alpha<1$, on the other hand, the DW mobility is larger above the WB critical current. From the experiments in (Ga,Mn)(As,P) samples, shown in Fig.~\ref{DW_2}, it was concluded that $1>\beta/\alpha\gtrsim 0.5$ \cite{Ranieri:2012_a}, i.e., that the non-adiabatic STT plays a significantly more important role  than in conventional transition metals where typically $\beta/\alpha\ll 1$ \cite{Zhang:2004_c}. Relatively large values of $\beta=\tau_{ex}/\tau_s$, compared to common dense-moment ferromagnets, are both due to larger $\tau_{ex}$ in the dilute-moment ferromagnetic semiconductors and due to smaller $\tau_s$ of the strongly spin-orbit coupled holes in the ferromagnetic semiconductor valence band \cite{Garate:2008_d,Hals:2008_a,Adam:2009_a,Curiale:2012_a,Ranieri:2012_a}. 

The combination of low saturation moment and strong spin-orbit coupling has yet another key advantage which is the dominant role of magnetocrystalline anisotropy fields over the shape anisotropy fields. It allows to control the internal DW structure and stability {\em ex situ} by strain relaxation in (Ga,Mn)As microstructures  \cite{Wunderlich:2007_c} or {\em in situ} by a piezo-electric stressor attached to the ferromagnetic semiconductor epilayer \cite{Ranieri:2012_a}. As a result, the WB critical current can be tuned \cite{Roy:2011_a} resulting in the observed 500\% variations of the DW mobility induced by the applied piezo-voltage  \cite{Ranieri:2012_a}.

\subsubsection{Current induced spin-orbit torque}
\label{SOT}
\begin{figure}[h]
\includegraphics*[width=1\columnwidth,angle=0]{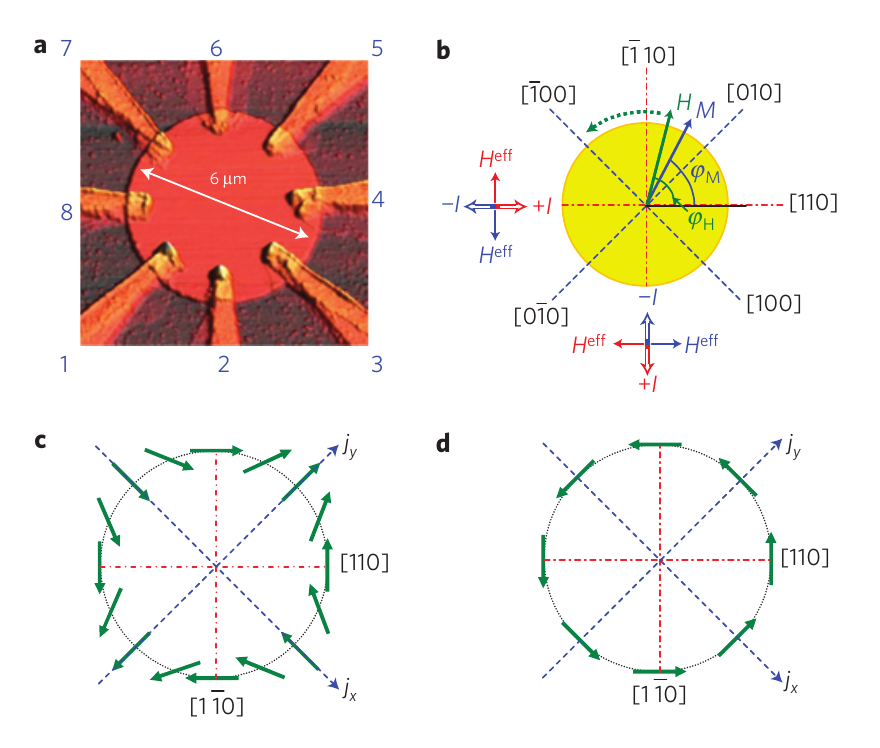}
\caption{(Color online) (a) Atomic force micrograph of the studied sample  with eight non-magnetic metal
contacts. (b) Diagram of device orientation with respect to crystallographic
axes, with easy and hard magnetization axes marked with blue dashed and
red dot-dash lines, respectively. Measured directions of ${\bf H}_{eff}$ field are
shown for different current directions. (c),(d) Orientation of effective SOT
field with respect to current direction for Dresselhaus (c) and Rashba (d)
spin-orbit interactions.   
From \cite{Chernyshov:2009_a}.}
\label{SOT_Rokhinson_1}
\end{figure}

\begin{figure}[h]
\includegraphics*[width=.9\columnwidth,angle=0]{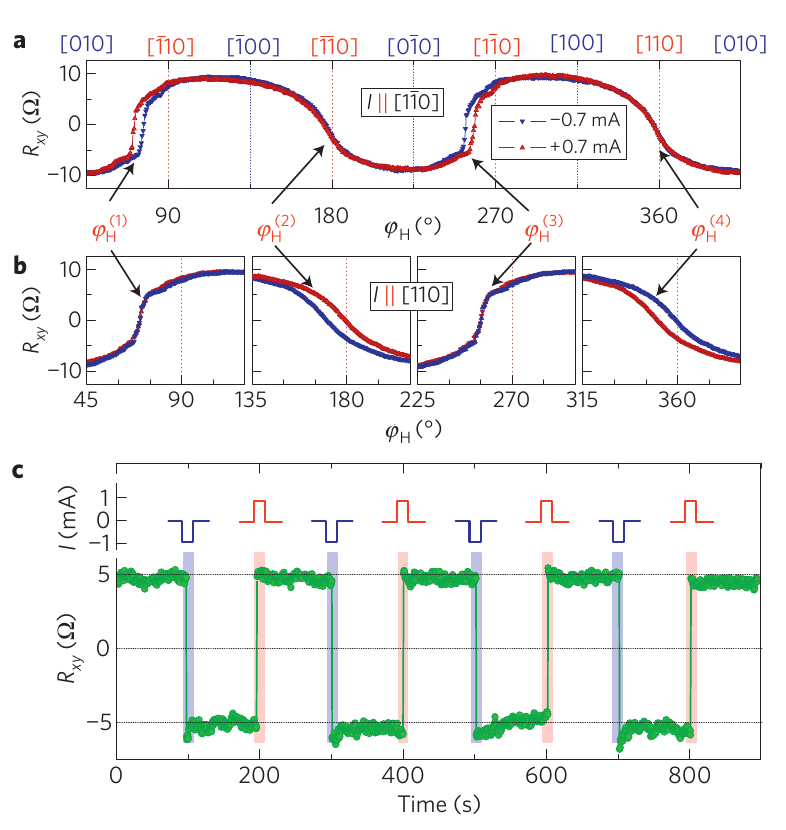}
\caption{(Color online) (a),(b) Transverse anisotropic
magnetoresistance $R_{xy}$ as a function of external field direction $\varphi_H$ for
$H=10$~mT and current $I=\pm0.7$~mA. The angles $\varphi_H^{(i)}$ mark
magnetization switchings. (c) Magnetization
switches between the [010] and [$\bar{1}$00] directions when alternating $I=\pm1$~mA
current pulses are applied with the current ${\bf I}\parallel [1\bar{1}0]$. The pulses have 100~ms duration and are shown
schematically above the data curve. $R_{xy}$ is measured with $I=10$~$\mu$A. Adapted from \cite{Chernyshov:2009_a}.}
\label{SOT_Rokhinson_2}
\end{figure}

Following the theoretical prediction for III-V zinc-blende crystals with broken inversion symmetry \cite{Bernevig:2005_c}, the experimental discovery of the SOT was reported in a (Ga,Mn)As device whose image is shown Fig.~\ref{SOT_Rokhinson_1}(a) \cite{Chernyshov:2009_a}. The sample
was patterned into a circular device with eight non-magnetic
ohmic contacts (Fig.~\ref{SOT_Rokhinson_1}a). In the presence of a saturating external magnetic field $H$,
the magnetization of the (Ga,Mn)As sample is aligned with the
field. For weak fields, however, the direction of magnetization
is primarily determined by magnetic anisotropy. As a small field
($5 < H < 20$~mT) is rotated in the plane of the sample, the
magnetization is re-aligned along the easy axis closest to the field
direction. Such rotation of magnetization by an external field is
demonstrated in Fig.~\ref{SOT_Rokhinson_2}a,b. For the current ${\bf I}\parallel [1\bar{1}0]$, the measured transverse AMR ($R_{xy}$) is
positive for  ${\bf M}\parallel [100]$ and negative for ${\bf M}\parallel [010]$. The switching angles 
where $R_{xy}$ changes sign are denoted as $\varphi_H^{(i)}$ on the plot. The data can be qualitatively understood if one considers an
extra current-induced effective magnetic field $H_{eff}$, as shown
schematically in Fig.~\ref{SOT_Rokhinson_1}b. The symmetry of the measured $H_{eff}$ with respect
to the direction of current is sketched in Fig.~\ref{SOT_Rokhinson_1}c and this current-induced SOT field has been shown to allow for reversibly switching magnetization between the [010] and [$\bar{1}$00] directions at a fixed magnetic field when applying positive and negative current pulses with the current ${\bf I}\parallel [1\bar{1}0]$, as shown in Fig.~\ref{SOT_Rokhinson_2}c. It was also demonstrated that the SOT in (Ga,Mn)As can generate a 180$^\circ$ magnetization reversal in the absence of an external magnetic field \cite{Endo:2010_a}. Apart from the current-induced magnetization switching of a uniform ferromagnet, the SOT was shown to provide means for developing an all-electrical broadband FMR technique applicable to individual nanomagnets \cite{Fang:2010_a}. The SOT-FMR was used for determining micromagnetic parameters of (Ga,Mn)As nano-bars which were not accessible by conventional FMR techniques and simultaneously allowed to perform 3D vector magnetometry on the driving SOT fields  \cite{Fang:2010_a,Kurebayashi:2013_a}.
  
The SOT fields of the Dresselhaus and Rashba symmetries shown in Figs.~\ref{SOT_Rokhinson_1}c,d, respectively, can arise in (Ga,Mn)As due to the following broken inversion symmetry terms in the spin-orbit-coupling Hamiltonian,
\begin{eqnarray}
  \label{eq:c4strain_main}
  H_{so}^{D,R}
&=& -    3C_4\left[
    \sigma_x k_x\left(\epsilon_{yy}-\epsilon_{zz}\right)-\sigma_y k_y\left(\epsilon_{xx}-\epsilon_{zz}\right)
  \right] \nonumber \\
&& -  3C_5
  \left[
    (\sigma_x k_y-\sigma_y k_x)\epsilon_{xy}
  \right] \,.
  \label{H_so^D,R}
\end{eqnarray}
The first, Dresselhaus term is due to the broken inversion symmetry of the host zinc-blende lattice combined with the growth-induced strain in the (Ga,Mn)As epilayer ($\epsilon_{xx}=\epsilon_{yy}\neq\epsilon_{zz}$) while the second, Rashba term combines the zinc-blende inversion asymmetry with a shear strain in the epilayer ($\epsilon_{xy}\neq0$) \cite{Silver:1992_a,Stefanowicz:2010_a,Chernyshov:2009_a,Fang:2010_a,Kurebayashi:2013_a}. In Ref.~\cite{Chernyshov:2009_a}, a Dresselhaus SOT field was identified corresponding to a compressively strained (Ga,Mn)As epilayer grown on a GaAs substrate. In Ref.~\cite{Fang:2010_a}, a sign change of the Dresselhaus SOT field was observed between (Ga,Mn)As/GaAs and (Ga,Mn)(As,P)/GaAs samples consistent with the change in the growth-induced strain in the epilayer from compressive  in the former sample to tensile in the latter sample. A weaker Rashba SOT field was also observed in  these experiments \cite{Fang:2010_a}. The shear-strain
component which yields the Rashba SOT field is not
physically present in the crystal structure of (Ga,Mn)As epilayers. It
has been introduced, however, in magnetization and SOT studies to effectively model the
in-plane uniaxial anisotropy present in (Ga,Mn)As epilayers \cite{Fang:2010_a,Sawicki:2004_a,Zemen:2009_a}.

The correspondence between the in-plane Dresselhaus and Rashba spin-orbit Hamiltonian terms in Eq.~(\ref{H_so^D,R}) and the in-plane SOT fields shown in Figs.~\ref{SOT_Rokhinson_1}c,d can be understood from Eq.~(\ref{SOT_eq}) within the Boltzmann transport theory description of  the non-equilibrium state $\langle\cdot\cdot\cdot\rangle$. In this semiclassical transport theory, the linear response of the carrier system to the applied electric field is described by the non-equilibrium  distribution function of carrier eigenstates which are considered to be unperturbed by the electric field. The form of the non-equilibrium distribution function is obtained by accounting for the combined effects of the carrier acceleration in the field and of scattering. In particular, the non-equilibrium distribution function is used here to evaluate the current induced SOT. 

\begin{figure}[h]
\includegraphics*[width=.8\columnwidth,angle=0]{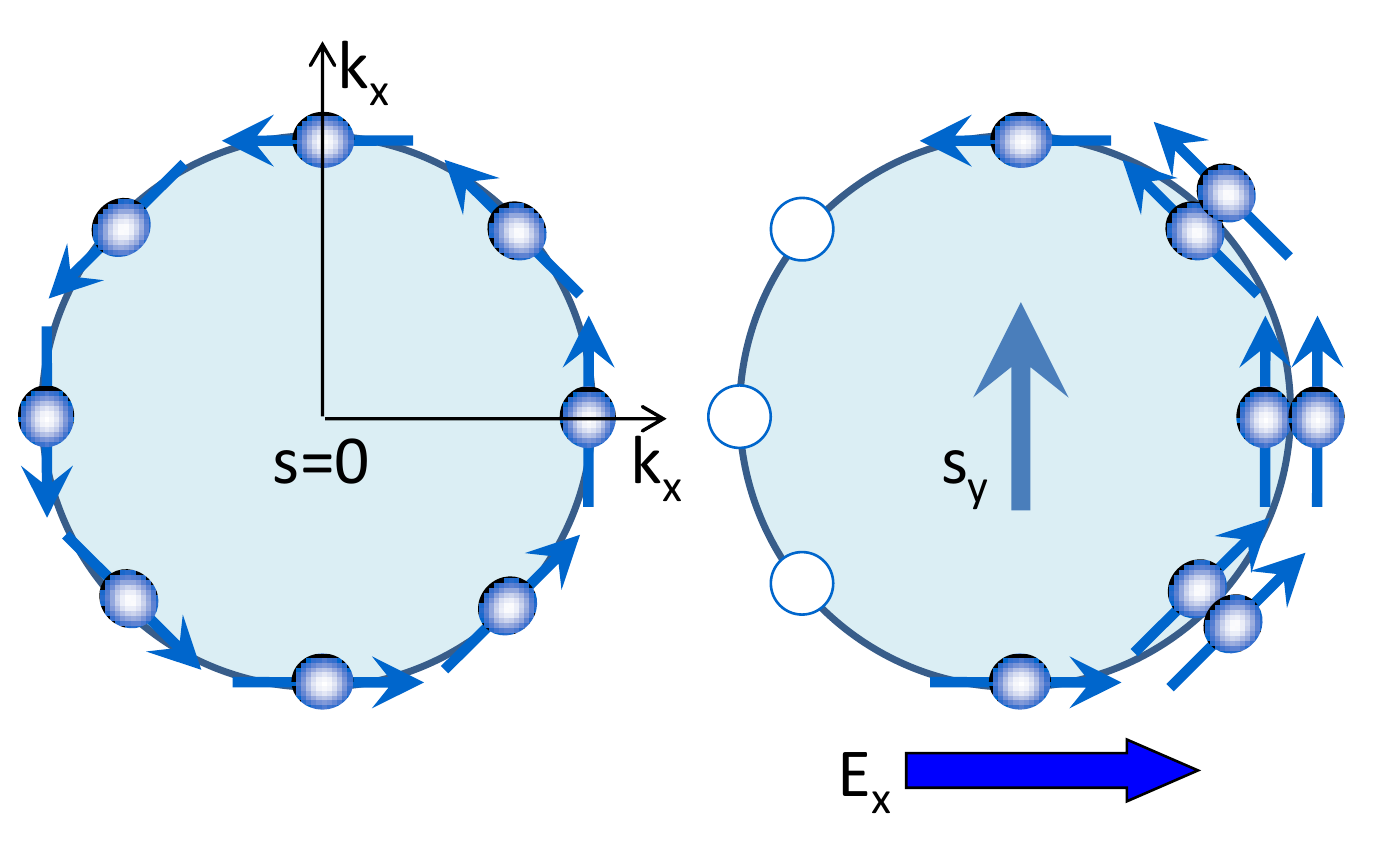}
\caption{(Color online) Left panel: Rashba spin-texture in equilibrium with zero net spin-density. Right panel: Non-equilibrium redistribution of eigenstates in applied electric field resulting in a non-zero spin-density due to broken inversion symmetry of the spin-texture.}
\label{field-like_SOT}
\end{figure}

Eq.~(\ref{SOT_eq}) explicitly shows that the SOT is non-zero only when both the exchange and spin-orbit fields act on the carrier states. However, when evaluating the SOT from $\frac{J}{\hbar}{\bf M}\times\langle{\bm\sigma}\rangle$
where part of the effect of the exchange field is explicitly factored out in the expression, an approximate form of the SOT can be obtained by considering in $\langle{\bm\sigma}\rangle$ eigenstates of  the Hamiltonian $H$ with $H_{ex}$ neglected. Since the resulting 
\begin{equation}
{\bf s}=\langle{\bm\sigma}\rangle=\frac{1}{V}\sum_{n,{\bf k}}{\bm\sigma}_{n,{\bf k}}g_{n,{\bf k}}
\label{field-like-SOT}
\end{equation}
is independent of {\bf M} this approximate form describes a pure field-like SOT whose origin is illustrated in Fig.~\ref{field-like_SOT} for  the Rashba spin-orbit coupling (analogous cartoons apply for the Dresselhaus or another broken inversion symmetry $H_{so}$). The non-equilibrium spin-density in the $H_{ex}=0$ approximation is a direct consequence of an electric-field and scattering induced redistribution of carriers  $g_{n,{\bf k}}$ on the Fermi surface whose texture of spin expectation values ${\bm\sigma}_{n,{\bf k}}$ has a broken inversion symmetry. For the Rashba spin-orbit coupling, the in-plane non-equilibrium spin polarization is perpendicular to the applied electric field for all crystal directions of the electric field. For the Dresselhaus spin-orbit coupling the relative angle between the in-plane non-equilibrium spin polarization and the applied electric field depends on the crystal direction of the electric field (see Fig.~\ref{SOT_Rokhinson_1}c). This current induced spin-polarization phenomenon was discussed in non-magnetic  semiconductors \cite{Aronov:1989_a,Edelstein:1990_a,Ganichev:2002_b} prior to the SOT experiments in (Ga,Mn)As. Analogous field-like SOT mechanism was subsequently considered in non-magnetic/ferromagnetic transition metal bilayers with broken structural inversion symmetry at the interface \cite{Manchon:2008_a,Manchon:2009_a,Miron:2010_a}.

\begin{figure}[h!]
\vspace*{0cm}
\includegraphics[width=1\columnwidth,angle=0]{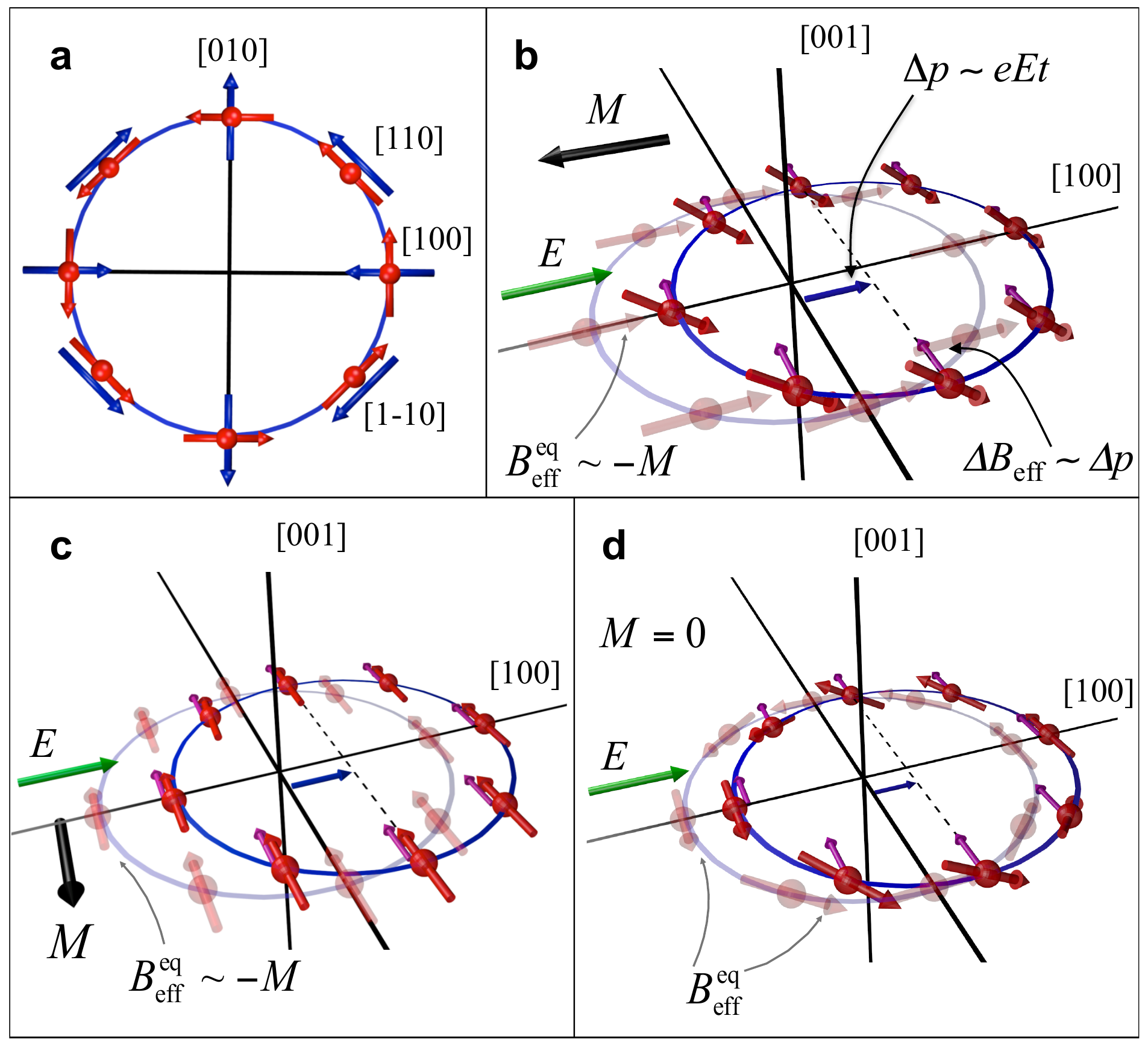}
\vspace*{-0cm}
\caption{(Color online) (a) Rashba (red) and Dresselhaus (blue) spin textures.
(b) For the case of a Rashba-like symmetry, the out-of-plane non-equilibrium carrier spin-density that generates the intrinsic anti-damping SOT has a maximum for $\mathbf{E}$ (anti)parallel to $\mathbf{M}$. In this configuration the equilibrium effective field $B^{eq}_{eff}$ and the additional field $\Delta B_{eff}\perp \mathbf{M}$ due to the acceleration are perpendicular to each other causing all spins to tilt in the same out-of-plane direction.
(c) For the case of a Rashba-like symmetry, the out-of-plane non-equilibrium carrier spin-density is zero for $\mathbf{E}\perp \mathbf{M}$ since $B^{eq}_{eff}$ and $\Delta B_{eff}$ are parallel to each other.
(d)  The analogous physical phenomena for zero magnetization induces a tilt of the spin out of the plane that has opposite sign for momenta pointing to the
left or the right of the electric field, inducing in this way the intrinsic  SHE. From \cite{Sinova:2004_a} and \cite{Kurebayashi:2013_a}.}
\label{fig_berry_1}
\end{figure}

Studies of the SOT in (Ga,Mn)As have identified an additional, anti-damping SOT contribution which has a common microscopic origin with the intrinsic SHE \cite{Kurebayashi:2013_a}. Unlike the above scattering-related field-like SOT, described within the semi-classical Boltzmann theory, the presence of an anti-damping SOT with a scattering-independent origin is captured by the time-dependent quantum-mechanical perturbation theory. Here the linear response theory considers the equilibrium distribution function and  the applied electric field perturbs the carrier wavefunctions. This can be visualized by solving the Bloch equations of the carrier spin dynamics during the acceleration of the carriers in the applied electric field, i.e., between the scattering events, as shown in Fig.~\ref{fig_berry_1} \cite{Kurebayashi:2013_a}. In the limit of large $H_{ex}$ compared to $H_{so}$ the spins are approximately aligned with the exchange field in equilibrium. During the acceleration, the field acting on the carriers acquires a time-dependent component due to $H_{so}$, as illustrated in Fig.~\ref{fig_berry_1}b for the Rashba spin-orbit coupling. This yields a non-equilibrium spin reorientation. In the linear response, i.e. for small tilts of the spins from equilibrium,  the carriers acquire a time and momentum independent out-of-plane component, resulting in a net out-of-plane spin density proportional to the strength of the spin-orbit field and inverse proportional to the strength of the exchange field \cite{Kurebayashi:2013_a}. 

As illustrated in Figs.~\ref{fig_berry_1}b,c, the non-equilibrium out-of-plane spin density $s_z$ depends on the direction of the magnetization $\mathbf{M}$ with respect to the applied electric field. For the Rashba spin-orbit coupling it has a maximum for $\mathbf{M}$ (anti)parallel to $\mathbf{E}$ and vanishes for $\mathbf{M}$ perpendicular to $\mathbf{E}$.  For a general angle $\theta_{\mathbf{M-E}}$ between $\mathbf{M}$ and $\mathbf{E}$,  $s_z\sim\cos\theta_{\mathbf{M-E}}$. The non-equilibrium spin polarization produces an out-of-plane field which exerts a torque on the in-plane magnetization given by Eq.~(\ref{SOT_eq}). This intrinsic SOT is anti-damping-like,
\begin{equation}
\frac{d\mathbf{M}}{dt}=\frac{J}{\hbar}(\mathbf{M}\times s_z\hat{z})\sim \mathbf{M}\times([\mathbf{E}\times\hat{z}]\times\mathbf{M})\,.
\label{SOT_R}
\end{equation}
For the Rashba spin-orbit coupling, Eq.~(\ref{SOT_R}) applies to all directions of the applied electric field with respect to crystal axes.  In the case of the Dresselhaus spin-orbit coupling, the symmetry of the anti-damping SOT depends on the direction of $\mathbf{E}$ with respect to crystal axes, as seen from Fig.~\ref{fig_berry_1}a. 

To highlight the analogy between the intrinsic anti-damping SOT and the intrinsic SHE \cite{Murakami:2003_a,Sinova:2004_a} the solution of the Bloch equations in the absence of the exchange Hamiltonian term is illustrated in Fig.~\ref{fig_berry_1}d \cite{Sinova:2004_a}. In the SHE case, the sense of the out-of-plane spin rotation depends on the carrier momentum resulting in a non-zero transverse spin-current but no net non-equilibrium spin density.

\begin{figure}[h!]
\vspace*{0cm}
\includegraphics[width=1\columnwidth,angle=0]{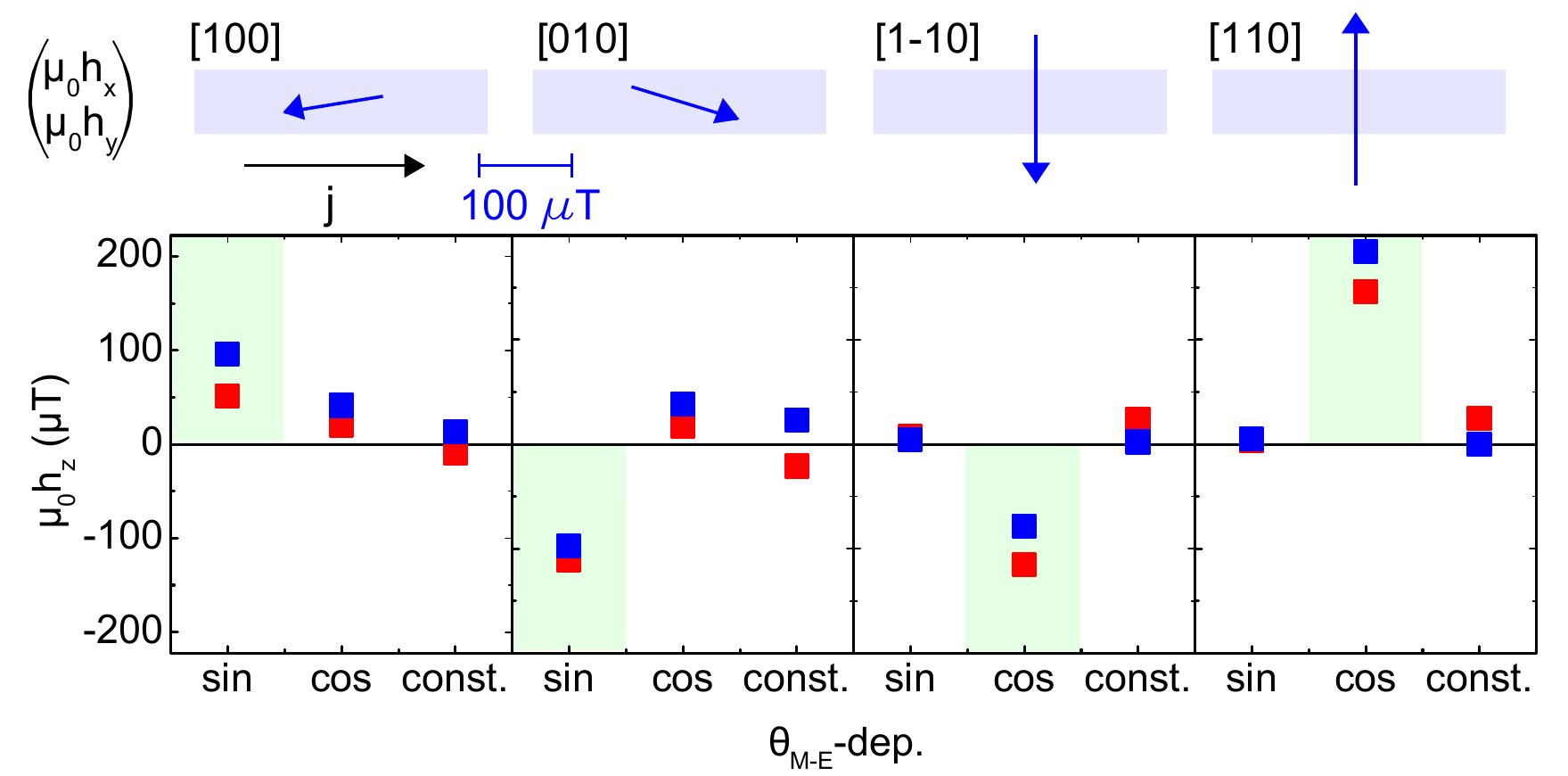}
\vspace*{-0cm}
\caption{(Color online) Measured in-plane and out-of-plane SOT fields in (Ga.Mn)As. In-plane spin-orbit field and coefficients of the $\cos{\theta_{\mathbf{M-E}}}$ and $\sin{\theta_{\mathbf{M-E}}}$ fits to the angle-dependence of out-of-plane SOT field for our sample set. For the in-plane fields, a single sample in each micro-bar direction is shown (corresponding to the same samples that yield the blue out-of-plane data points). In the out-of-plane data, 2 samples are shown in each micro-bar direction. The symmetries expected for the anti-damping SOT, on the basis of the theoretical model for the Dresselhaus term in the spin-orbit interaction, are shown by light green shading. All data are normalised to a current density of $10^5$~Acm$^{-2}$. From \cite{Kurebayashi:2013_a}.}
\label{fig_berry_2}
\end{figure}

The anti-damping like SOT with the theoretically predicted symmetries was identified in measurements in (Ga,Mn)As, as shown in Fig.~\ref{fig_berry_2} \cite{Kurebayashi:2013_a}. The all-electrical broadband SOT-FMR technique \cite{Fang:2010_a} was applied which allowed to perform 3D vector magnetometry on the driving SOT fields. Since the magnitude of the measured out-of-plane and in-plane SOT fields are comparable, the anti-damping SOT plays an important role in driving the magnetization dynamics in (Ga,Mn)As. 

The observation of the intrinsic anti-damping like SOT in (Ga,Mn)As has direct consequences also for the physics of in-plane current induced torques in the transition metal bilayers \cite{Miron:2011_b,Liu:2012_a}. Here the anti-damping like SOT  considered at the broken inversion symmetry interface can compete with another, conceptually distinct mechanism in which the intrinsic SHE in the paramagnet generates a spin-current which upon entering the ferromagnet exerts an anti-damping STT on the magnetization \cite{Liu:2012_a}. It has been mentioned above that the non-equilibrium spin-density in the intrinsic anti-damping SOT scales with the strength of the spin-orbit field and with the inverse of the strength of the exchange field. Similarly, the SHE spin-current, which takes the role of the spin-injection rate $P$ in Eq.~(\ref{adiabatic_s}) for the non-equilibrium spin density ${\bf s}$ in the adiabatic STT, scales with the strength of the spin-orbit coupling in the paramagnetic metal \cite{Tanaka:2007_b} and ${\bf s}$ in the adiabatic STT is inverse proportional to the exchange field (Eq.~(\ref{adiabatic_s})).

\subsection{Interaction of spin with light}
\label{LIT}

\subsubsection{Magneto-optical effects}
\label{MO}

\begin{figure}[h!]
\vspace*{0cm}
\includegraphics[width=1\columnwidth,angle=0]{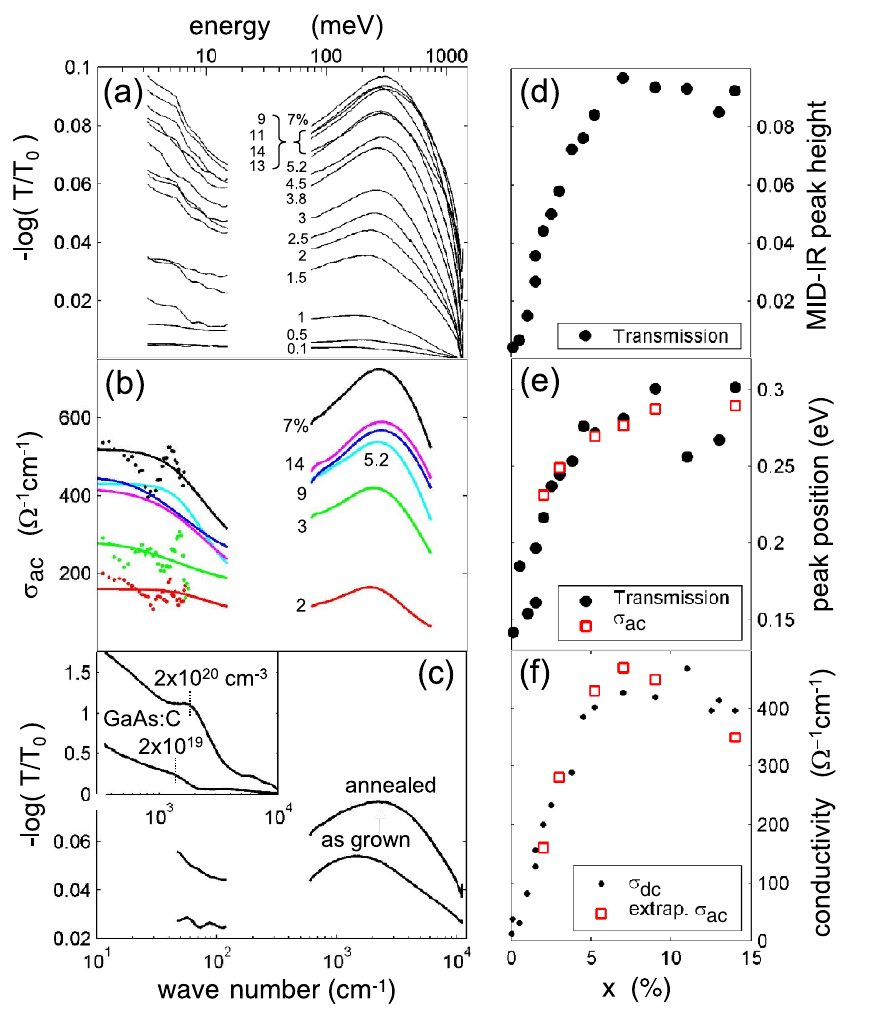}
\vspace*{-0cm}
\caption{(Color online) (a) Infrared absorption of a series of optimized (Ga,Mn)As/GaAs epilayers with nominal Mn doping $x=0.1-14$\% plotted from the measured optical transmissions of the samples ($T$) and of the reference bare GaAs substrate ($T_0$). (b) Real part of the ac conductivity (lines) obtained from the measured complex conductivity in the terahertz range (points) and from fitting the complex conductivity in the infrared range to the measured transmissions. (c) Comparison of the infrared absorption in as-grown and annealed 4.5\% doped sample. Inset: Comparison to GaAs:C samples with carbon doping densities $2\times 10^{19}$~cm$^{-3}$ and $2\times 10^{20}$~cm$^{-3}$. (d) Height of the (Ga,Mn)As mid-infrared absorption peak as a function of Mn doping. (e) Position of the peak inferred from the transmission measurements and from the fitted ac conductivities. (f) Zero frequency conductivities obtained from dc transport measurements and from extrapolated optical ac conductivities measured in the terahertz range.From \cite{Jungwirth:2010_b}.}
\label{fig_IR_absorption}
\end{figure}

Similar to the dc conductivity, the unpolarized finite-frequency absorption spectra \cite{Burch:2006_a,Chapler:2011_a,Jungwirth:2007_a,Jungwirth:2010_b} show signatures of the vicinity of the metal-insulator transition and of strong disorder effects even in the most metallic (Ga,Mn)As materials, as illustrated in Fig.~\ref{fig_IR_absorption}.  Compared to a shallow-acceptor counterpart such as, e.g., C-doped GaAs (see inset of Fig.~\ref{fig_IR_absorption}(c)), the spectral weight in (Ga,Mn)As is shifted from the low-frequency Drude peak to higher frequencies. The ac conductivity scales with the dc conductivity over a broad range of Mn dopings and does not reflect strongly the spin-dependent interactions in the system.

\begin{figure}[h!]
\vspace*{0cm}
\includegraphics[width=1\columnwidth,angle=0]{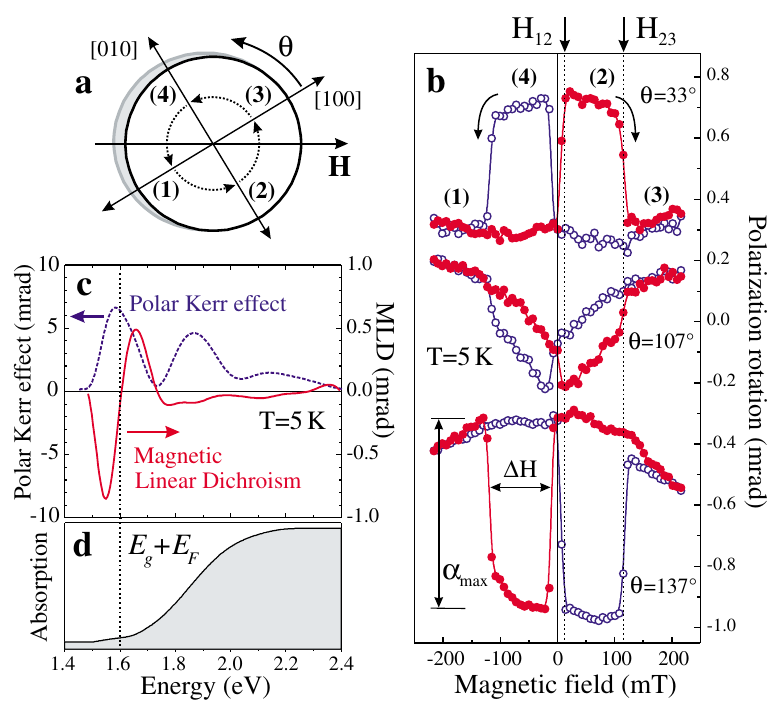}
\vspace*{-0cm}
\caption{(Color online) (a) Ga$_{0.98}$Mn$_{0.02}$As sample orientation
with respect to the applied magnetic field and the four-step
magnetization reversal process as consecutive 90$^\circ$ jumps (shown
by dotted arrows) between the four easy directions (1)Ð(4).
(b) Field dependencies of the magnetic linear dichroism for
different angles $\theta$ between the incident polarization and the
[100] crystallographic direction, measured at a wavelength of $\lambda=815$~nm. Numbers (1) to (4) correspond to the magnetization
directions indicated in (a). $H_{12}$ and $H_{23}$ are the magnetic
field values required for making jumps (1)$\rightarrow$(2) and (2)$\rightarrow$(3),
respectively. (c) Spectra of polar magneto-optical Kerr effect and
magnetic linear dichroism (at $\theta=135^\circ$); (d) absorption spectrum at 5 K. In panels (c) and (d) Fabry-Perot oscillations in the signal due to the finite buffer thickness have been removed numerically using a bandpass filter. From \cite{Kimel:2005_a}.}
\label{fig_MLD}
\end{figure}

Magneto-optical spectroscopies, on the  other hand, provide a detailed probe into the exchange-split and spin-orbit coupled electronic structure of (Ga,Mn)As \cite{Ando:1998_a,Kuroiwa:1998_a,Beschoten:1999_a,Szczytko:1999_a,Komori:2003_a,Lang:2005_a,Chakarvorty:2007_a,Ando:2008_a,Acbas:2009_a,Kimel:2005_a,Moore:2003_a,Tesarova:2012_c,Tesarova:2012_a,Tesarova:2013_a}. It implies that they can be used as sensitive optical spin-detection tools, as illustrated in Fig.~\ref{fig_MLD} \cite{Kimel:2005_a}. 

For the light propagating in the perpendicular direction to the sample surface the magneto-optical effects can be classified in the following way \cite{Tesarova:2013_a}: The magnetic circular birefringence (MCB) is given by the real  part of the difference between refractive indices of two circularly polarized modes with opposite helicities and the magnetic circular dichroism (MCD) is given by its imaginary part.  These magneto-optical coefficients are sensitive to the out-of-plane component of the magnetization, are an odd function of ${\bf M}$, and represent the finite frequency counterparts of the AHE. The magnetic linear birefringence (MLB) is given by the real  part of the difference between refractive indices of two modes linearly polarized perpendicular and parallel to the magnetization and the magnetic linear dichroism (MLD) is given by its imaginary part. These magneto-optical coefficients are sensitive to the in-plane components of the magnetization,  are an even function of ${\bf M}$, and represent the finite frequency counterparts of the AMR. 

Both the circular and linear magneto-optical effects can cause a rotation (and ellipticity) of the polarization of a transmitted or reflected linearly polarized light. For the rotation originating form the MCB/MCD the effects are referred to as the Faraday effect in transmission and Kerr effect in reflection.  For the rotation originating from the MLB/MLD the terminology is not unified across the literature \cite{Tesarova:2013_a}, however, it is clearly distinguishable from the Kerr (Faraday) rotation.  While the Kerr (Faraday) rotation is independent of the polarization angle of the incident light, the rotation originating from the MLB/MLD depends on the angle between the light polarization and the in-plane magnetization. There is a direct analogy between this magneto-optical effect and the transverse voltage in the non-crystalline off-diagonal AMR described by Eq.~(\ref{rho_xy}). The transverse voltage in the latter case and the polarization rotation in the former case  have both the $\sim\sin\phi$ form where $\phi$ is the angle between the in-plane magnetization and the applied voltage in the transverse AMR case, and between the in-plane magnetization and the incident light polarization in the case of the MLB/MLD induced rotation. 

Measurements in Fig.~\ref{fig_MLD}b used the dependence on the polarization angle to optically detect magnetization switchings between [100] and [010] crystal axes in a 2\% Mn-doped (Ga,Mn)As sample with a dominant in-plane cubic anisotropy \cite{Kimel:2005_a}. Consistent with the phenomenology of the MLB/MLD induced rotation, the largest signal is observed when the incident-light polarization is aligned with the in-plane diagonal crystal axis. Fig.~\ref{fig_MLD}c,d highlight that both the Kerr effect and the MLB/MLD induced rotation can be strong in (Ga,Mn)As for a suitably chosen frequency of the probe laser light. This allows for a sensitive optical detection of the in-plane and out-of-plane components of the magnetization.

\begin{figure}[h!]
\vspace*{0cm}
\includegraphics[width=1\columnwidth,angle=0]{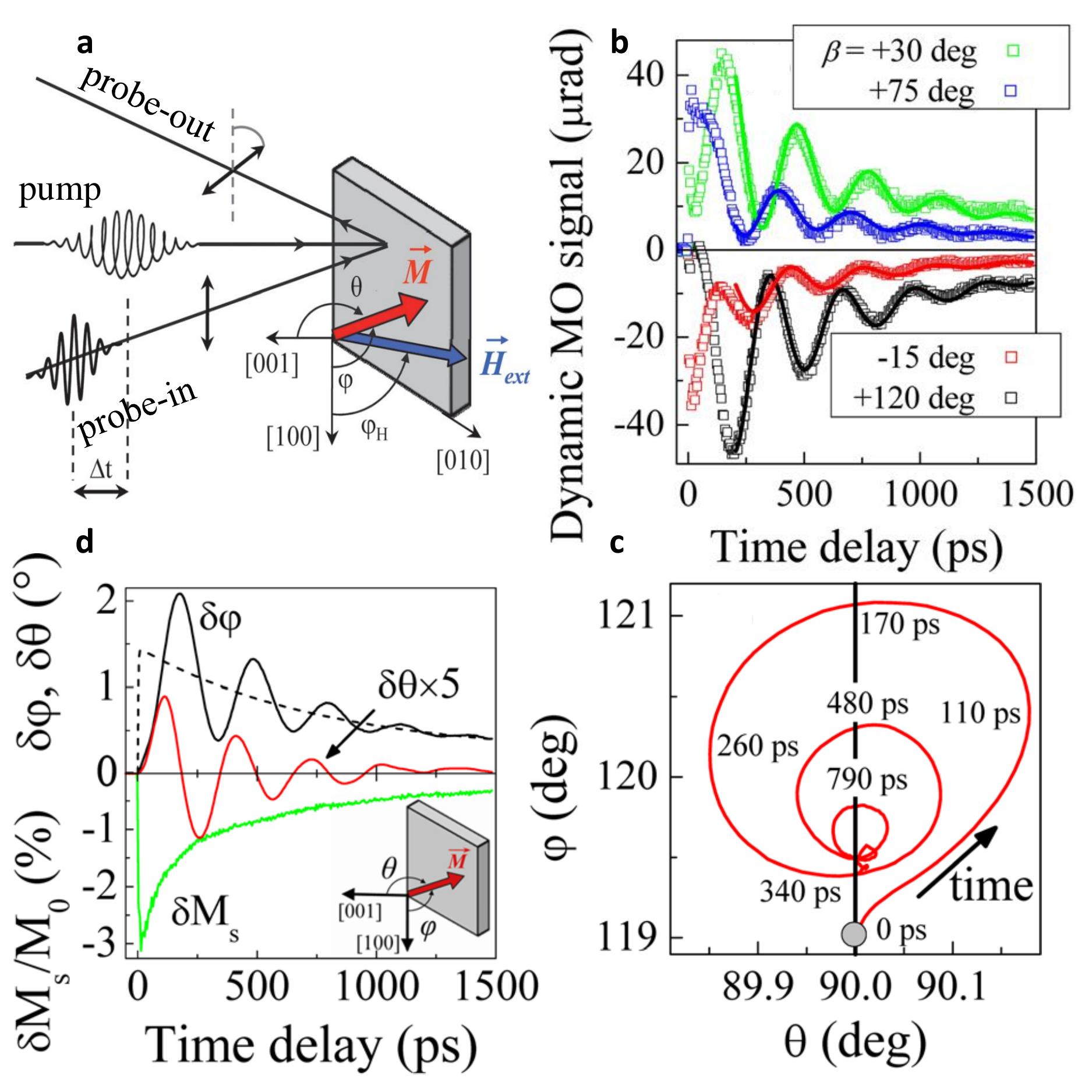}
\vspace*{-0cm}
\caption{(Color online) (a) Schematic diagram of the experimental set-up for a detection of the magnetization precession induced in (Ga,Mn)As by an impact of the femtosecond laser pump pulse. Rotation of the polarization plane of reflected linearly polarized probe pulses is measured as a function of the time delay $\Delta t$ between pump and probe pulses. The orientation of magnetization in the sample is described by the in-plane angle $\varphi$ and  the out-of-plane angle $\theta$. The external magnetic field $H_{ext}$ is applied in the sample plane at an angle $\varphi_H$. (b) Dynamics of the magneto-optical signal induced by an
impact of pump pulse on the sample that was measured by probe pulses with
different polarization orientations $\beta$. (c) Time evolution of the in-plane magnetization angle $\delta\varphi (t)$, the out-of-plane angle $\delta\theta (t)$, and the magnitude $\delta M_s(t)/M_0$; the dotted line depicts the in-plane evolution of the easy axis position around which the magnetization precesses. (d) Orientation of magnetization at different times after the impact of the pump pulse; the sample plane is represented by the vertical line and the equilibrium position of the easy axis is
depicted by the grey spot. From \cite{Tesarova:2012_a}.}
\label{optical_3D_trajectory}
\end{figure}

The decomposition of the magneto-optical signal into the MCB/MCD induced rotation due to the out-of-plane magnetization and the MLB/MLD induced rotation due to in-plane magnetization was also employed to quantitatively determine the three-dimensional magnetization vector trajectory in the time-resolved pump-and-probe magneto-optical measurements in (Ga,Mn)As, as shown in Fig.~\ref{optical_3D_trajectory} \cite{Tesarova:2012_a}. The technique helped to experimentally identify different mechanisms by which  photo-carriers can induce magnetization dynamics in the pump-and-probe experiments  in (Ga,Mn)As. The recombining photo-carriers can heat the lattice and the transient increase of temperature can trigger magnetization dynamics or, on much shorter time-scales, the photo-carriers can directly induce spin torques acting on the magnetization \cite{Oiwa:2005_a,Wang:2006_b,Takechi:2007_a,Qi:2007_a,Qi:2009_a,Rozkotova:2008_a,Rozkotova:2008_b,Hashimoto:2008_a,Hashimoto:2008_b,Kobayashi:2010,Nemec:2012_a,Tesarova:2012_a,Tesarova:2012_b}. These effects are reviewed in more detail in the following sections. We note that earlier magneto-optical pump-and-probe studies of photo-carriers exchange coupled to local magnetic moments have been performed in non-ferromagnetic (II,Mn)VI diluted magnetic semiconductors \cite{Baumberg:1994_a,Crooker:1996_a,Camilleri:2001_a}.

\subsubsection{Optical spin-transfer torque}
\label{OSTT}
A direct observation of a non-thermal photo-carrier induced spin torque was reported in a pump-and-probe optical experiment in which a coherent spin precession in a (Ga,Mn)As ferromagnetic semiconductor was excited by circularly polarized laser pulses at normal incidence \cite{Nemec:2012_a}. During the pump pulse, the spin angular momentum of photo-carriers generated by the absorbed circularly-polarized light is transferred to the collective magnetization of the ferromagnet, as described by Eqs.~(\ref{H})-(\ref{non-adiabatic_STT}) and prediced in Refs.~\cite{Rossier:2003_a,Nunez:2004_b}. 
\begin{figure}[h!]
\hspace*{-.2cm}\includegraphics*[width=0.9\columnwidth,angle=0]{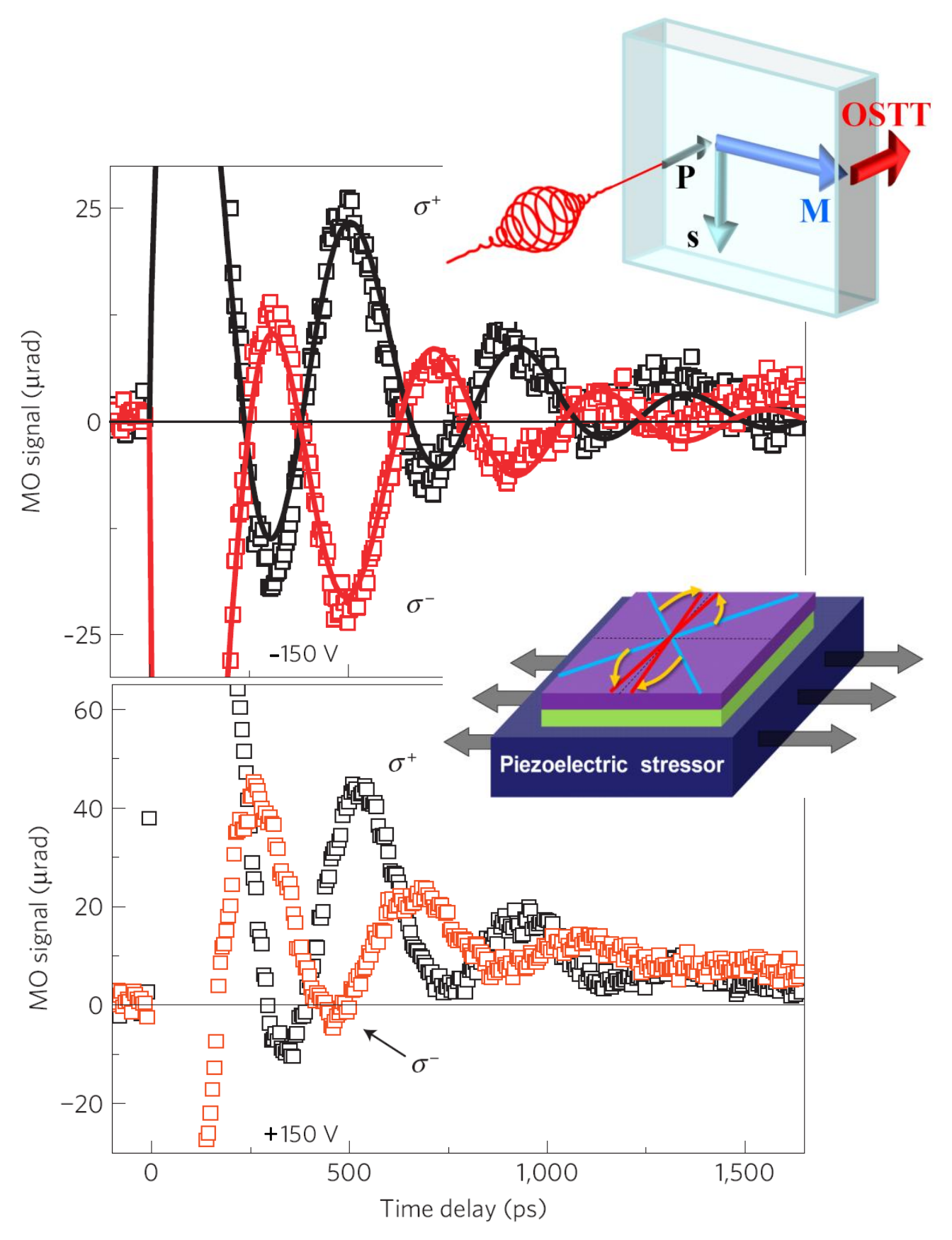}
\caption{(Color online) Schematic illustration (top inset) of the optical spin transfer torque induced by the rate $P$ of the photo-carrier spin injection along light propagation axis $\hat{\bf n}$ (normal to the sample plane). The steady state component of the non-equilibrium spin density ${\bf s}$ is oriented in the plane of the sample and perpendicular to the in-plane equilibrium magnetization vector. The (Ga,Mn)As sample is placed on a piezoelectric stressor (lower inset) which allows to control the magnetic anisotropy {\em in situ}. Top panel: Precession of the magnetization induced in (Ga,Mn)As by $\sigma^+$ and $\sigma^-$ circularly polarized pump pulses. Points are the measured rotations of the polarization plane of the reflected linearly polarized probe pulse as a function of the time delay between pump and probe pulses. The experiment was performed on the (Ga,Mn)As sample attached to a piezo-stressor at applied bias $U=-150$~V for which the $\sigma^+$ and $\sigma^-$ circularly polarized pump pulses produces signals with opposite sign corresponding to the opposite sign of the optical STT and no polarization-independent ($\sigma^+ +\sigma^-$) signal for this piezo-voltage.  Bottom panel: Same as in the top panel for a piezo-voltage $U=+150$~V. Here magnetization dynamics is excited by both the optical STT and a polarization-independent mechanism.
Adapted from~\cite{Nemec:2012_a}.}
\label{OSTT_1}
\end{figure}

The timescale  of photo-electron precession due to the exchange field produced by the ferromagnetic Mn moments is $\tau_{ex}\sim$100~fs in (Ga,Mn)As \cite{Rossier:2003_a,Nemec:2012_a}.  The major source of spin decoherence of the photo-electrons in (Ga,Mn)As is the exchange interaction with fluctuating Mn moments. Microscopic calculations of the corresponding relaxation time give a typical scale of 10's ps \cite{Rossier:2003_a}. The other factor that limits $\tau_s$ introduced in  Eq.~(\ref{M_STT}) is the photo-electron decay time which is also $\sim$10's~ps, as inferred from reflectivity measurements of the (Ga,Mn)As samples \cite{Nemec:2012_a}. Within the spin life-time, the photo-electron spins therefore precess many times around the exchange field of ferromagnetic moments. In the corresponding regime of $\tau_s\gg\tau_{ex}$, the steady-state photo-electron spin-polarization is given by Eq.~(\ref{adiabatic_s}), i.e. is perpendicular to both the polarization unit vector of the optically injected carrier spins and magnetization, and the  optical STT has the form of the adiabatic STT given by Eq.~(\ref{adiabatic_STT}), as  illustrated in the top inset of Fig.~\ref{OSTT_1}a. The precession time of holes in (Ga,Mn)As is $\sim$10's fs and the spin life-time of holes, dominated by the strong spin-orbit coupling, is estimated to $\sim$1-10~fs \cite{Rossier:2003_a}. Since $\tau_s\lesssim\tau_{ex}$ for holes, their contribution in the experiment with circularly-polarized pump-pulse is better approximated by the weaker torque which has the form of the non-adiabatic STT given by Eq.~(\ref{non-adiabatic_STT}) and can be neglected.

The experimental observation of the magnetization precession in (Ga,Mn)As excited by the optical STT, with the characteristic opposite phases of the oscillations  excited by pump pulses of opposite helicities, is shown in the top panel of Fig.~\ref{OSTT_1} \cite{Nemec:2012_a}. Since the period of the magnetization precession (0.4~ns)  is much larger than the pump-pulse duration, the action of the optical STT is reflected only in the initial phase and amplitude of the free precession of the magnetization. The decomposition of the magneto-optical signal in Fig.~\ref{OSTT_1} into MCB/MCD induced rotation due to the out-of-plane magnetization and the MLB/MLD induced rotation due to in-plane magnetization shows \cite{Nemec:2012_a} that the initial tilt of the magnetization is in the out-of-plane direction, as expected from Eq.~(\ref{adiabatic_STT}) for the adiabatic STT. The precisely opposite phase of the measured magneto-optical signals triggered by pump pulses with opposite helicities, shown in the top panel of Fig.~\ref{OSTT_1}, implies that the optical STT is  not accompanied by any polarization-independent excitation mechanism. These were intentionally suppressed in the experiment shown in the top panel of Fig.~\ref{OSTT_1} by  negatively biasing an attached piezo-stressor to the (Ga,Mn)As sample which modified the magnetic anisotropy of the ferromagnetic film.  At positive piezo-voltage, on the other hand, the polarization-independent mechanisms \cite{Oiwa:2005_a,Wang:2006_b,Takechi:2007_a,Qi:2007_a,Qi:2009_a,Rozkotova:2008_a,Rozkotova:2008_b,Hashimoto:2008_a,Hashimoto:2008_b,Kobayashi:2010} start to act along with the optical STT, as illustrated in the bottom panel of Fig.~\ref{OSTT_1} \cite{Nemec:2012_a}. The polarization-independent optical excitation mechanisms are discussed in the following section.

\subsubsection{Optical spin-orbit torque}
\label{OSOT}

In the optical STT reviewed above, the external source for injecting spin polarized photo-carriers is provided by the circularly polarized light at normal incidence which yields high degree of out-of-plane spin-polarization of injected photo-carriers due to the optical selection rules in GaAs. 
Since large optical STT requires large spin lifetime of injected carriers, i.e. spin-orbit coupling is detrimental for optical STT, the weakly spin-orbit coupled photo-electrons play the key role in this case. The optical SOT, on the other hand, originates from spin-orbit coupling of non-equilibrium photo-carriers excited by polarization-independent pump laser pulses which do not impart angular momentum. Since the effect relies on the strong spin-orbit coupling, the non-equilibrium photo-holes generated in (Ga,Mn)As valence band are essential for the optical SOT. The physical picture of the optical SOT in (Ga,Mn)As is based on the SOT formalism of Eqs.~(\ref{carrier_SOT}) and (\ref{SOT_eq}), and on the following representation of the non-equilibrium steady state spin-polarization of the photo-holes \cite{Tesarova:2012_b}: The optically injected photo-holes relax towards the hole Fermi energy of the p-type (Ga,Mn)As on a short ($\sim100$~fs) timescale \cite{Yildirim:2012_a} and the excitation/relaxation processes create a non-equilibrium excess hole density in the spin-orbit coupled, exchange-split valence band. The increased number of non-equilibrium occupied hole states, as compared to the equilibrium state in dark, can generate a non-equilibrium  spin-polarization of holes which is misaligned with the equilibrium orientation of Mn moments. This non-equilibrium photo-hole polarization  persists over the timescale of the hole recombination ($\sim$ps) during which it exerts a torque on the Mn local moments. Approximately, the non-equilibrium photo-holes  can be represented by a steady state which differs from the equilibrium state in the dark in that the distribution function has a shifted Fermi level corresponding to the extra density of the photo-holes. In this approximation, the non-equilibrium  spin-polarization of holes which is misaligned with the equilibrium orientation of Mn moments, and the corresponding optical SOT, is determined by the hole density dependent magnetocrystalline anisotropy field \cite{Tesarova:2012_b}.

\begin{figure}[h!]
\hspace*{-0.4cm}\includegraphics*[width=1.1\columnwidth,angle=0]{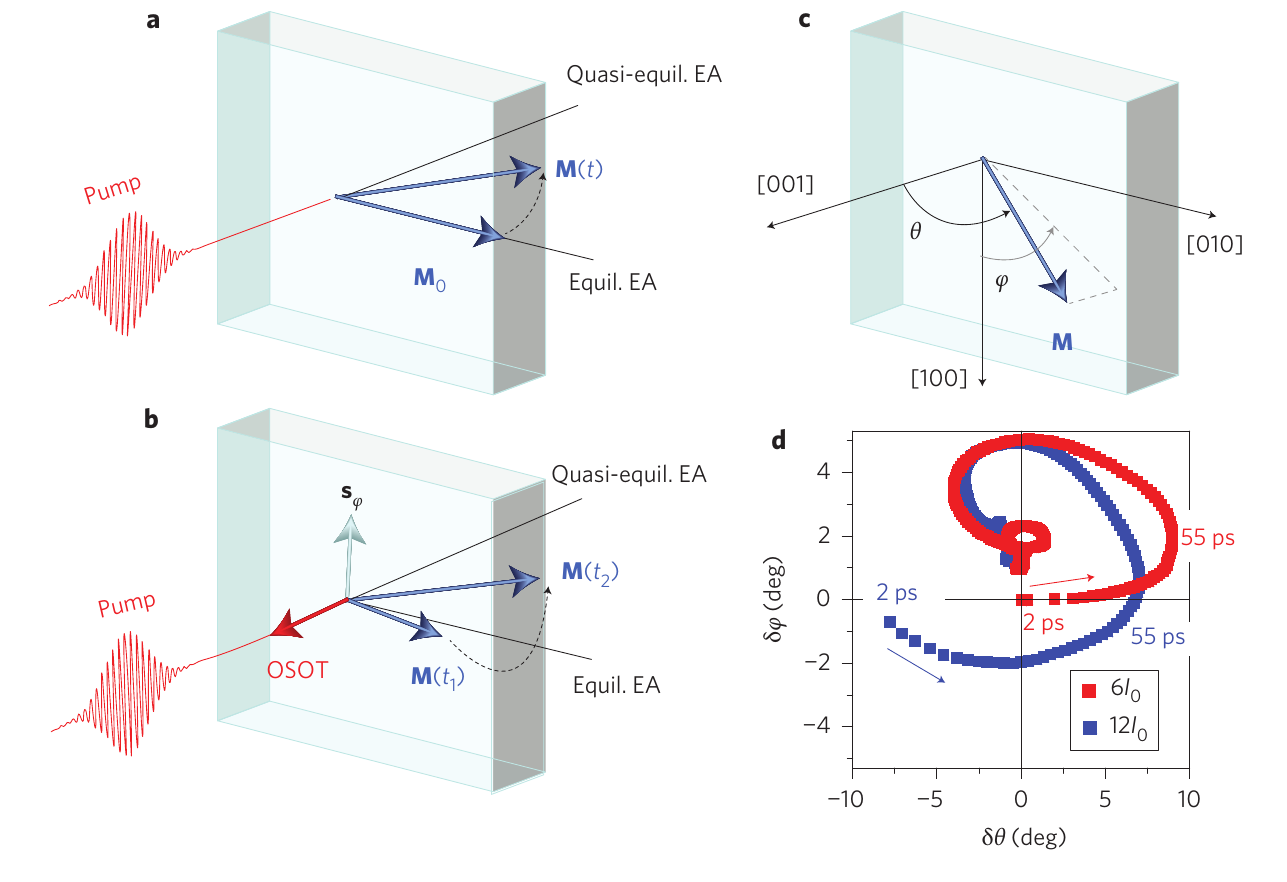}
\caption{(Color online) (a) Schematic illustration of the thermally excited precession of magnetization {\bf M}(t) around the transient quasi-equilibrium easy axis (EA). {\bf M}$_0$ is the magnetization vector aligned with  in-plane equilibrium EA before the pump pulse.  (b) Schematic illustration of optical SOT induced by the in-plane transverse component {\bf s}$_{\varphi}$ of the non-equilibrium hole spin polarization. On the time-scale of magnetization precession, optical SOT causes an instantaneous tilt of the magnetization {\bf M}(t$_1$) which allows  to clearly distinguish optical SOT from the considerably slower thermal excitation mechanism. The initial optical SOT induced tilt of magnetization can yield precession angles that are inaccessible in the thermally induced magnetization dynamics. (c) Definition of the coordinate system. (d) Time evolution of the magnetization vector measured  in a (Ga,Mn)As material with nominal Mn-doping $x=3$\%. The direction of the time increase is depicted by arrows. Magnetization tilt angles $\delta\varphi$ and $\delta\theta$ are measured with respect to  equilibrium EA. From~\cite{Tesarova:2012_b}.}
\label{OSOT_1}
\end{figure}

The experimental identification of the optical SOT \cite{Tesarova:2012_b} required to separate this non-thermal photo-magnetic effect from the competing thermal excitation mechanism of magnetization dynamics \cite{Wang:2006_d,Kirilyuk:2010_a}. The absorption of the pump laser pulse leads to photo-injection of electron-hole pairs. The non-radiative recombination of photo-electrons produces a  transient increase of the lattice temperature which builds up on the time scale of $\sim 10$~ps and persists over $\sim1000$~ps. This results in a quasi-equilibrium easy-axis (EA) orientation which is tilted from the equilibrium EA.  Consequently, Mn moments in (Ga,Mn)As will precess around the quasi-equilibrium EA, as schematically illustrated in Fig.~\ref{OSOT_1}a, with a typical precession time of $\sim 100$~ps given by the magnetic anisotropy fields in (Ga,Mn)As. The EA stays in-plane and the sense of rotation within the plane of the (Ga,Mn)As film with increasing temperature is uniquely defined by the different temperature dependences of the in-plane cubic and uniaxial anisotropy fields \cite{Zemen:2009_a,Tesarova:2012_b}. In the notation shown in Fig.~\ref{OSOT_1}c, the change of the in-plane angle $\delta\varphi$ of the magnetization during the thermally excited precession  can be only positive. 

The optical SOT, illustrated schematically in Fig.~\ref{OSOT_1}b, acts during the laser pulse (with a duration of 200~fs) and fades away within the hole recombination time ($\sim$~ps), followed by free magnetization precession. It causes an impulse tilt of the magnetization which is a signature that allowed to clearly distinguish the optical SOT from the considerably slower thermal excitation mechanism. Moreover, the initial optical SOT induced tilt of magnetization can yield precession angles that are opposite to the initial tilt of the magnetization dynamics induced by the slower thermal mechanism. 

Examples of the direct observation of the thermally governed excitation of magnetization at a lower pump pulse intensity, $6I_0$ where $I_0=7$~$\mu$J~cm$^{-2}$,  and of  the excitation at a higher intensity,  $12I_0$, with a strong contribution from the optical SOT are shown in Fig.~\ref{OSOT_1}d for a 3\% doped (Ga,Mn)As sample \cite{Tesarova:2012_b}. The distinct features of the optical SOT observed at pump intensity $12I_0$, namely the impulse tilt and precession angles inaccessible by thermal excitations seen at the lower intensity $6I_0$, are clearly visible when comparing the two measured magnetization trajectories in Fig.~\ref{OSOT_1}d. We recall that both dynamical magneto-optical signals shown in Fig.~\ref{OSOT_1}d are independent of the polarization of pump pulses which distinguishes both the slower thermal mechanism and the fast optical SOT mechanism from the optical STT. 
A complete suppression of the thermal mechanism and magnetization precession induced solely by the optical SOT was achieved by tuning the micromagnetics of the (Ga,Mn)As film {\em ex situ} by doping or {\em in situ} by applied magnetic fields \cite{Tesarova:2012_b}.

Magneto-optical pump-and-probe studies in (Ga,Mn)As demonstrated the possibility to study STT and SOT on the short time-scales achievable by the optical techniques. The relativistic  optical SOT should be observable in other systems including, e.g., antiferromagnetic semiconductors which unlike their ferromagnetic counterparts can have magnetic transition temperatures well above room temperature \cite{Jungwirth:2010_a}. It is well established that magnetocrystalline anisotropies are equally present in spin-orbit coupled antiferromagnets as in ferromagnets and in Sections~\ref{tamr} we pointed out that  the spin-orbit coupling induced anisotropic magnetotransport effects can be also strong in antiferromagnets. The optical SOT belongs to this family of relativistic effects and its exploration in antiferromagnets may open a new direction of optical spin torque studies beyond the ferromagnetic semiconductor (Ga,Mn)As.

\subsection{Interaction of spin with heat}
\label{heat}

In Section~\ref{Mott-Dirac} we have outlined the distinction between the basically non-relativistic Mott spintronic phenomena, such as the GMR or TMR, which depend on relative magnetization orientations  in  non-uniform magnetic structures, and the relativistic Dirac effects, such as the AHE, AMR, or TAMR, in uniform spin-orbit coupled magnets.  In this section we recall that the research of the relativistic spintronics effects in (Ga,Mn)As has led  to seminal results not only in magneto-transport and magneto-optical studies but also  in the research of magneto-thermopower phenomena. 

\subsubsection{Anomalous Nernst effect}
\label{ANE}
In analogy to the AHE, we consider an experimental geometry for detecting the ANE  in which the thermal gradient $\nabla T\parallel\hat{x}$, magnetization ${\bf M}\parallel\hat{z}$, and the Nernst signal is the M-antisymmetric electric field ${\bf E}\parallel\hat{y}$.
In non-magnetic systems in zero magnetic field, the charge current density is given by,
\begin{equation}
j_x=\sigma_{xx}E_x-\alpha_{xx}\partial_x T
\end{equation}
which for the open circuit geometry ($j_x=0$) yields,
\begin{equation}
E_x=\frac{\alpha_{xx}}{\sigma_{xx}}\partial_x T=S_{xx}\partial_x T\,,
\end{equation}
where $\alpha_{xx}$ is the diagonal Peltier coefficient and $S_{xx}$ is the diagonal Seebeck (thermopower) coefficient. In the presence of the $\hat{z}$-axis magnetization, an off-diagonal Peltier current is generated resulting in the ANE, 
\begin{equation}
j_y=-\alpha_{yx}\partial_x T+\sigma_{yx}E_x+\sigma_{xx}E_y\,,
\end{equation}  
and for $j_y=0$,
\begin{equation}
E_y=\frac{1}{\sigma_{xx}}(\alpha_{yx}-\sigma_{yx}S_{xx})\partial_x T=S_{yx}\partial_x T\,,
\label{S_xy}
\end{equation}
where  $\alpha_{xy}$ and $S_{xy}$ are the antisymmetric off-diagonal Peltier and Seebeck coefficients, respectively.

Thermoelectric measurements on Hall bars fabricated in (Ga,Mn)As/(Ga,In)As epilayers with perpendicular-to-plane easy-axis were performed \cite{Pu:2008_a} in order to test in a ferromagnet the validity of the Mott relation for the off-diagonal transport coefficients \cite{Wang:2001_b},
\begin{equation}
\alpha_{yx}=\frac{\pi^2k_B^2T}{3e}\left(\frac{\partial\sigma_{yx}}{\partial E}\right)_{\mu}\,,
\label{Mott_relation}
\end{equation}
and to experimentally assess the microscopic mechanism of the AHE and ANE in (Ga,Mn)As. In the same devices, the four thermoelectric coefficients, $\rho_{xx}$, $\rho_{xy}$, $S_{xx}$, and $S_{xy}$ were measured which allowed to directly fit the experimental data by the formula,
\begin{equation}
S_{yx}=\frac{\rho_{xy}}{\rho_{xx}}\left(T\frac{\pi^2k_B^2}{3e}\frac{\lambda^\prime}{\lambda}+(1-n)S_{xx}\right)\,.
\label{Mott_test}
\end{equation}
Eq.~(\ref{Mott_test}) is obtained by introducing the Mott relation (\ref{Mott_relation}) into the expression for $S_{yx}$ from Eq.~(\ref{S_xy}) and by considering a general power-law dependence of the AHE resistivity on the diagonal resistivity, 
\begin{equation}
\rho_{xy}=\sigma_{yx}/(\sigma_{xx}^2+\sigma_{xy}^2)\approx\sigma_{yx}/\sigma_{xx}^2=\lambda M_z\rho_{xx}^n\,.
\end{equation}
Here the proportionality of the AHE to $M_z$ is factored out explicitly in the power-low dependence, $\lambda$ is the remaining scaling factor ($\lambda^\prime=(\partial\lambda/\partial E)_\mu$), and
\begin{equation}
\rho_{xx}=\sigma_{xx}/(\sigma_{xx}^2+\sigma_{xy}^2)\approx 1/\sigma_{xx}\,.
\end{equation}

\begin{figure}[h!]
\hspace*{-0cm}\includegraphics*[width=.8\columnwidth,angle=0]{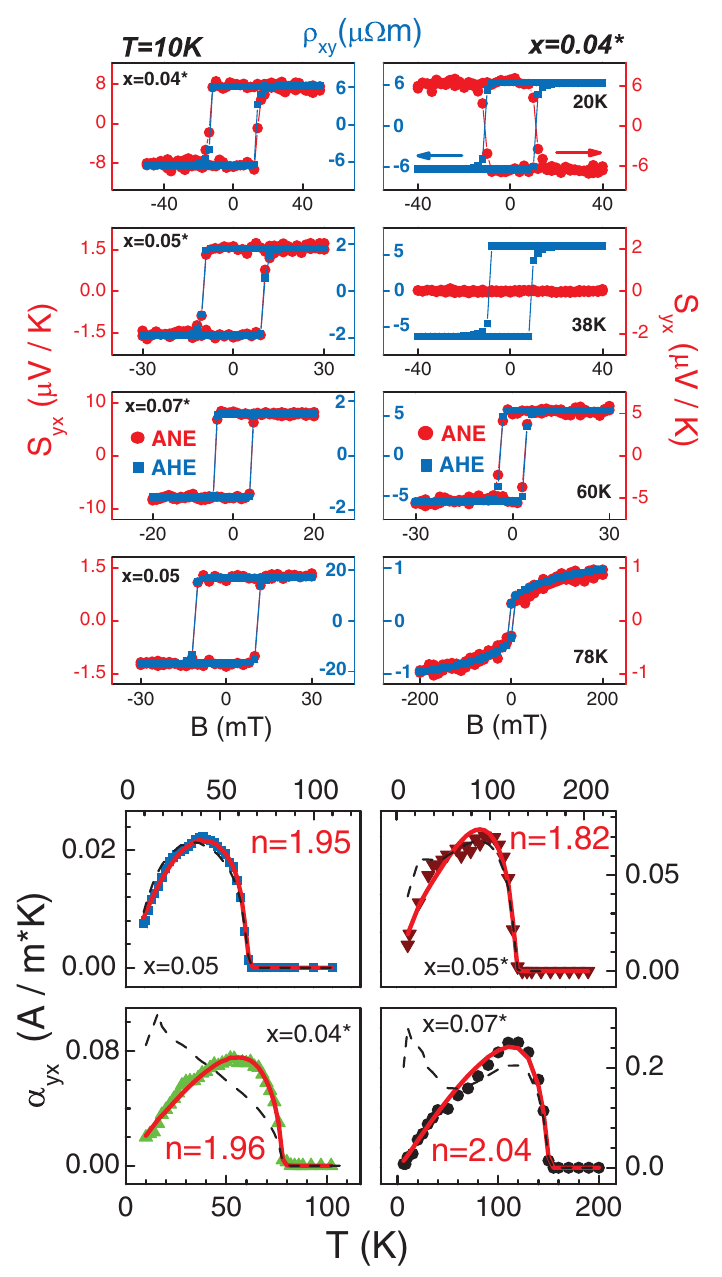}
\caption{(Color online) Top eight panels: AHE and ANE loops at $T=10$~K for different
samples (left column) and at different temperatures for the 4\%
annealed sample (right column). In the left column, ANE data of
0.04*, 0.05*, and 0.07* samples were multiplied by -1 ($\ast$ means that the sample was annealed). Bottom four panels: zero-field ANE coefficient. The solid lines are the best fits using Eq.~(\ref{Mott_test}) (or equivalently Eq.~(\ref{alpha_xy}), and the
dashed curves are the best fits with $n=1$. Adapted from~\cite{Pu:2008_a}.}
\label{nernst}
\end{figure}

The intrinsic AHE is characterized by the off-diagonal conductivity $\sigma_{yx}$ which is independent of the scattering life-time $\tau$, i.e., independent of $\sigma_{xx}$. This corresponds to the above power-law scaling with $n=2$. On the other hand, for e.g. the extrinsic skew-scattering AHE, $\sigma_{yx}\sim\tau\sim\sigma_{xx}$, which corresponds to $n=1$. The detection of both the AHE and ANE signals in (Ga,Mn)As Hall-bar samples is illustrated in the top panels of Fig.~\ref{nernst}.  The measured  $\rho_{xx}$, $\rho_{xy}$, $S_{xx}$, and $S_{xy}$ could be accurately fitted to Eq.~(\ref{Mott_test}) which confirmed the Mott relation between the AHE and ANE in a ferromagnet. Moreover, the inferred values of $n$ from the fitting were close to 2 in all measured samples (see bottom panels of Fig.~\ref{nernst}). This  confirmed the intrinsic origin of the AHE and ANE in (Ga,Mn)As. Using Eq.~(\ref{S_xy}) we can rewrite Eq.~(\ref{Mott_test}) as,
\begin{equation}
\alpha_{yx}=\sigma_{yx}\left(T\frac{\pi^2k_B^2}{3e}\frac{\lambda^\prime}{\lambda}+(2-n)S_{xx}\right)\,,
\label{alpha_xy}
\end{equation}
from which we directly obtain that for $n=2$ the intrinsic,  scattering independent AHE coefficient is accompanied by a scattering-independent ANE coefficient,   
\begin{eqnarray}
\sigma_{yx}&=&\lambda M_z \nonumber \\
\alpha_{yx}&=&\lambda^\prime M_zT\frac{\pi^2k_B^2}{3e}\,.
\end{eqnarray}

\subsubsection{Anisotropic magneto-thermopower}
\label{AMT}
\begin{figure}[h!]
\hspace*{-0cm}\includegraphics*[width=.9\columnwidth,angle=0]{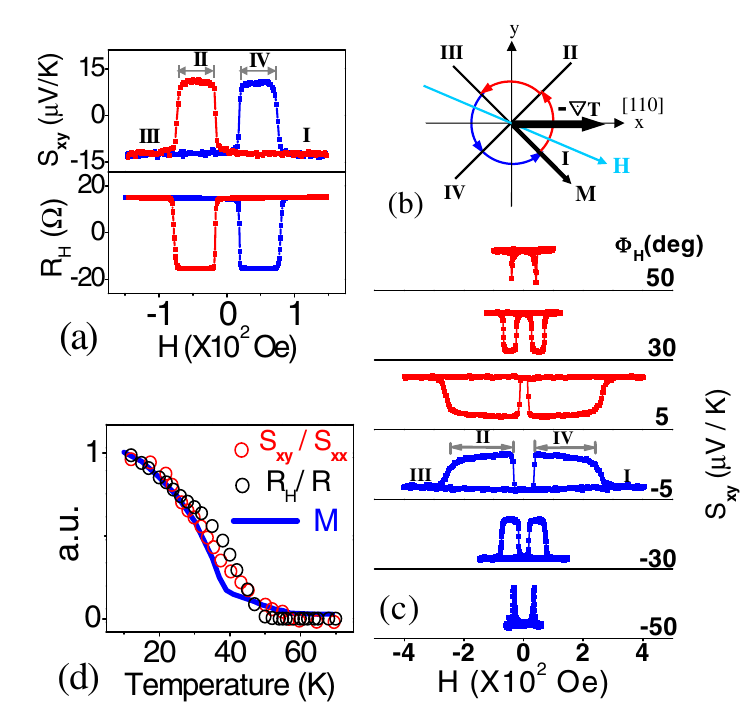}
\caption{(Color online) (a) Transverse AMT, $S_{x,y}$, and transverse AMR, $R_H$, in a 3.9\% Mn-doped (Ga,Mn)As. (b) Sketch of the relative orientation of $-\nabla T$, {\bf M} and magnetic field {\bf H}. Four directions marked as I, II, III, and IV are easy directions of {\bf M}. (c) Angular dependence of the transverse AMT. (d) Comparison of $S_{xy}/S_{xx}$ and $R_H/R$, and sample magnetization  M measured by SQUID.
From~\cite{Pu:2006_a}. Note that we use the terms transverse AMT and transverse AMR  instead of the alternative planar Nernst effect and planar Hall effect \cite{Pu:2006_a} to clearly distinguish that the effects shown here are the symmetric off-diagonal coefficients even in {\bf M}.}
\label{fig_AMT}
\end{figure}

Besides ANE, the thermoelectric measurements in (Ga,Mn)As also revealed strong AMT signals, in particular the spin-caloritronic analogue of the non-crystalline AMR \cite{Pu:2006_a}. A non-crystalline AMT as high as 6\% was measured in the longitudinal direction obeying the $\cos2\phi$ dependence as for the non-crystalline longitudinal AMR, where $\phi$ is the angle between magnetization and the applied electrical (thermal) voltage.  Simultaneously, the transverse AMT was also observed, as illustrated in Fig.~\ref{fig_AMT}, following the $\sin2\phi$ dependence of the corresponding transverse AMR coefficient. Experiments in (Ga,Mn)As marked a renewed interest in the AMT phenomenon  \cite{Ky:1966_a} which  was subsequently identified in a broad class of magnetic materials, ranging from the strongly spin-orbit coupled uranium pnictides \cite{Wisniewski:2007_a} to transition-metal based oxides \cite{Tang:2011_a,Anwar:2012_a}, and nano-wires and thin films of elemental transition metal ferromagnets \cite{Mitdank:2012_a,Anwar:2012_a}.

\subsubsection{Tunneling anisotropic magneto-thermopower}
\label{TAMT}
Similar to uniform magnetic films, in the ohmic GMR multilayers electrical and heat transport measurements can be performed in macroscopic samples in the current-parallel-to-plane geometry. This allowed to observe the GMT effect \cite{Sakurai:1991_a} shortly after the discovery of the GMR \cite{Baibich:1988_a,Binasch:1989_a} in the same type of transition-metal-multilayer samples and to show that switching from parallel to anti-parallel magnetization configurations can lead to comparatively large changes in the thermopower \cite{Sakurai:1991_a}. 

Magneto-thermopower measurements are significantly more challenging in the perpendicular-to-plane geometry of the  magnetic tunnel  junctions and the TMT effect was observed in transition metal tunnel devices \cite{Walter:2011_a,Liebing:2011_a} more than 15 years after the discovery of the TMR \cite{Moodera:1995_a,Myiazaki:1995_a}. Similar to the electrical-transport, the magneto-thermopower in the tunneling regime is much more closely related to  the exchange-split electronic structure of the ferromagnets than in the ohmic regime of the GMR multilayers and correspondingly can be in principle much stronger in the tunneling devices 
\cite{Czerner:2011_a,Liebing:2011_a}. 

\begin{figure}[h!]
\hspace*{-0cm}\includegraphics*[width=.9\columnwidth,angle=0]{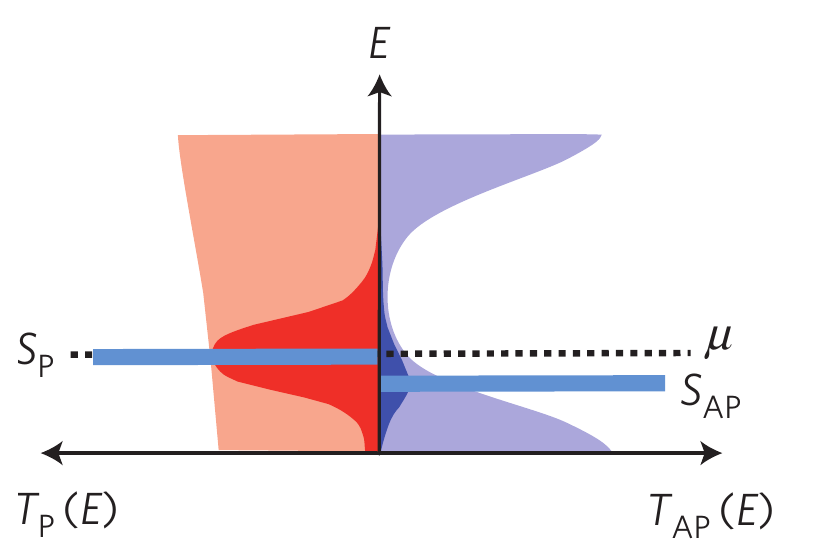}
\caption{(Color online) In magnetic tunnel junctions, thermal differences in the
electron distributions and strong asymmetry in the spin-dependent tunneling channels are depicted. $T(E)$ is the transmission of the full tunnel junction, for
which either the ferromagnetic electrodes can be a highly spin-polarized half-metal or the combination of the barrier and the ferromagnet exhibits
half-metallic characteristics. The function $T(E)(-\partial_Ef(E,\mu,T))$ is given in darker color. The thick line marks the resulting value of the geometric centre determining the Seebeck coefficient in the parallel magnetization, $S_P$, 
and antiparallel magnetization, $S_{AP}$, of the electrodes. Adapted from~\cite{Walter:2011_a}. Note that we use the term TMT instead of the alternative magneto-Seebeck effect to distinguish it clearly from the spin-Seebeck effect discussed in the following section.}
\label{fig_TMT} 
\end{figure}

The origin of the TMT effect is schematically illustrated in Fig.~\ref{fig_TMT} \cite{Walter:2011_a}. Unlike electrical conductance of the tunneling device, 
\begin{equation}
G=\frac{e^2}{h}\int T(E)(-\partial_E f(E,\mu,T))dE\,,
\end{equation}
which in the linear response is governed by the transmission function $T(E)$ multiplied by the derivative of the
electron occupation function $\partial_E f(E,\mu,T)$ at temperature $T$ and
electrochemical potential $\mu$, the Seebeck coefficient,
\begin{equation}
S=-\frac{\int T(E)(E-\mu)(-\partial_E f(E,\mu,T))dE}{eT\int T(E)(-\partial_E f(E,\mu,T))dE}\,,
\end{equation}
reflects the asymmetry in the energy dependence of the transmission around the chemical potential. As shown in Fig.~\ref{fig_TMT}, the Seebeck coefficient is the geometric centre of $T(E)(-\partial_E f(E,\mu,T))$. When this changes from the parallel to the antiparallel magnetization configurations the corresponding Seebeck coefficients are different in the two configurations resulting in the TMT. 

\begin{figure}[h!]
\hspace*{-0cm}\includegraphics*[width=.9\columnwidth,angle=0]{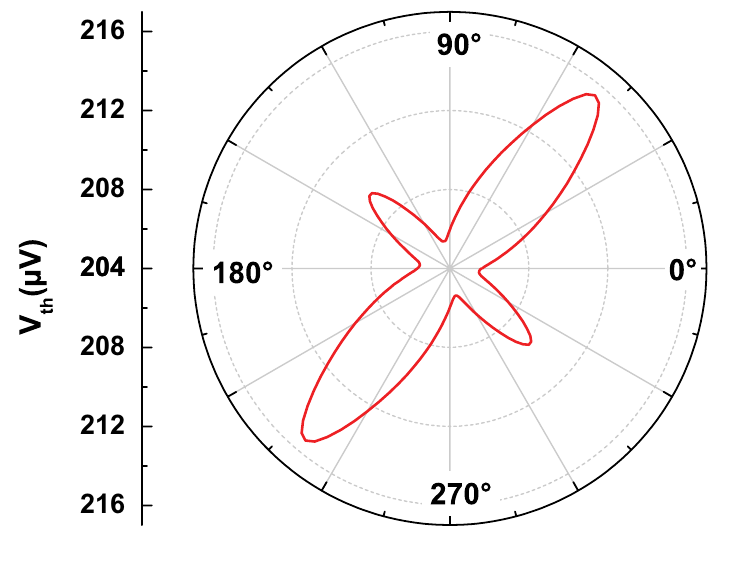}
\caption{(Color online) Thermovoltage in a (Ga,Mn)As/i-GaAs/GaAs:Si tunnel junction as a function of the magnetization angle. 0 is along [010] crystal axis. From~\cite{Naydenova:2011_a}.}
\label{fig_TAMT}
\end{figure}

The relativistic counterpart of the TMT in a tunnel junction with only one magnetic electrode is the TAMT. Observations of the TMT \cite{Walter:2011_a,Liebing:2011_a} and TAMT \cite{Naydenova:2011_a} effects were reported independently and simultaneously and, reminiscent of  the discovery of the TAMR \cite{Gould:2004_a},   the TAMT was first identified in a (Ga,Mn)As based tunnel junction \cite{Naydenova:2011_a}. The experiment was performed while rotating the magnetization in the plane of the (Ga,Mn)As layer, i.e., always perpendicular to the applied temperature gradient across the tunnel junction. As shown in Fig.~\ref{fig_TAMT}, four equivalent minima
close to the [100] and [010] crystal axes and two sets of
local maxima were observed. The symmetry of the observed TAMT reflects the competition of in-plane cubic and uniaxial magnetocrystalline anisotropies in the (Ga,Mn)As epilayer. The TAMT phenomenon originates from the changes in the energy dependence of the tunneling density of states when changing the angle of the magnetization with respect to crystal axes, i.e., has the same spin-orbit-coupled band structure origin as magnetocrystalline anisotropies and the TAMR.

\subsubsection{Spin Seebeck effect}
\label{spin-Seebeck}
Among the most intriguing spin-caloritronics effects  is the spin-Seebeck effect \cite{Uchida:2008_a,Uchida:2010_a,Jaworski:2010_a,Sinova:2010_b,Bauer:2012_a}. Instead of directly generating electrical voltages from thermal gradients, as was the case of the above discussed magneto-thermopower effects, in the spin-Seebeck effect it is primarily the difference between spin-up and spin-down chemical potentials, $\mu_\uparrow-\mu_\downarrow$, which is induced by the applied thermal voltage in a ferromagnet. An appealing picture was proposed following the first experimental observation of the spin-Seebeck effect in NiFe in which the ferromagnet functions like a thermocouple, but in the spin sector \cite{Uchida:2008_a}. In this picture, instead of two different charge Seebeck coefficients in two  metals forming the thermocouple, it is the different carrier scattering and density and the corresponding  Seebeck coefficient in the two spin channels which produce the non-zero difference $\mu_\uparrow-\mu_\downarrow$.

In this seminal work and in the subsequent experiments, the SHE in attached non-magnetic electrodes was employed to convert the difference in spin-dependent chemical potentials into electrical voltages \cite{Uchida:2008_a,Uchida:2010_a,Jaworski:2010_a}. Specifically, $|\mu_\uparrow-\mu_\downarrow|$ decreases in the non-magnetic electrode from the interface with the ferromagnet along the vertical direction. This results in a vertical spin-current in the non-magnetic electrode which is converted into an in-plane electrical voltage via the SHE. 

\begin{figure}[h!]
\hspace*{0.5cm}\includegraphics*[width=.9\columnwidth,angle=0]{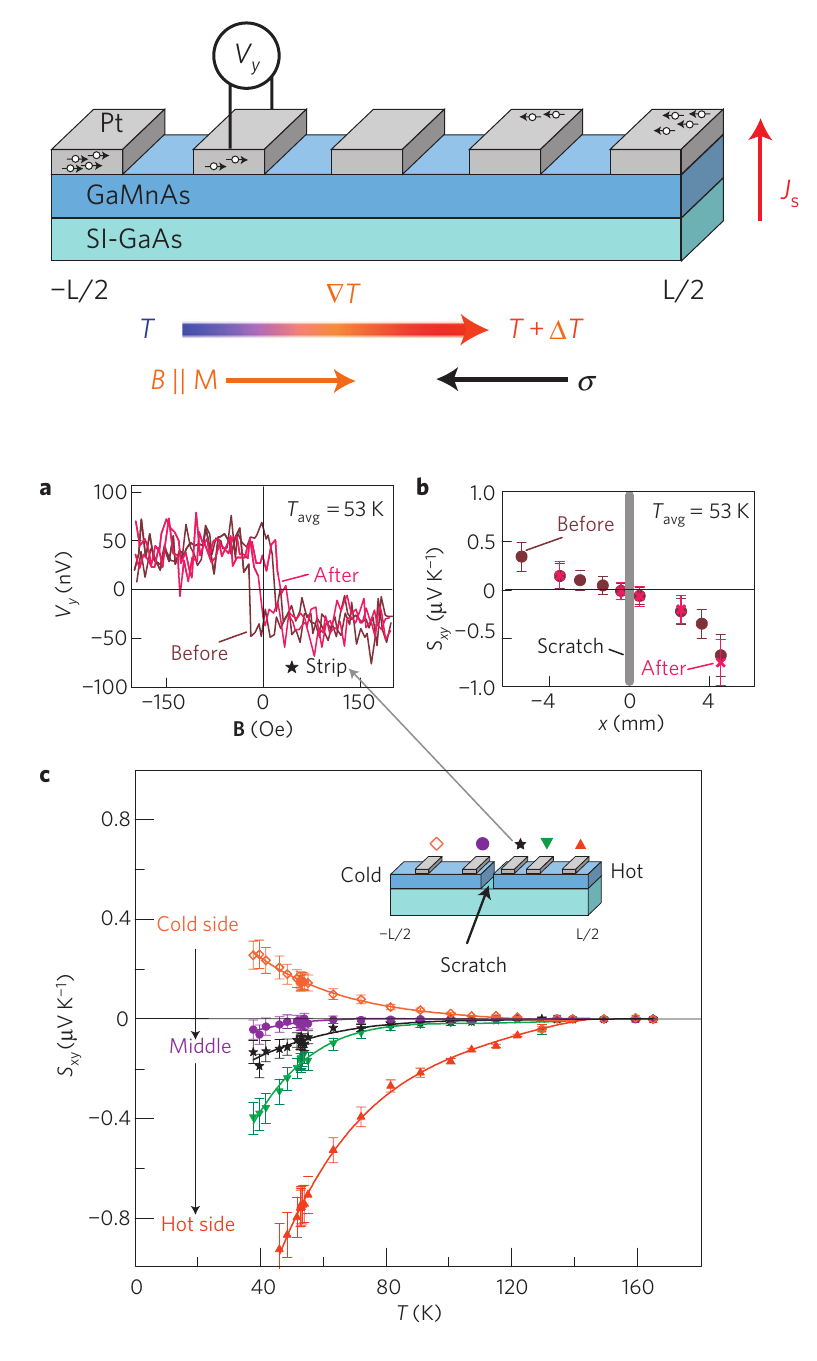}
\caption{(Color online) Top panel: Measurement geometry of the spin-Seebeck effect.  a, Transverse voltage, $V_y$, as a function of
applied field, $B$, from the strip contact 0.3~mm above the scratch (star) with
an applied $\Delta T_x$ of 0.63~K. b, Spatial dependence of the spin-Seebeck
coefficient, $S_{xy}$, before and after the scratch. The scratched region is
indicated by the shaded region. c, Temperature dependence of $S_{xy}$ after the
scratch at various positions along the sample. Adapted from~\cite{Jaworski:2010_a}.}
\label{fig_spin-Seebeck}
\end{figure}

Experiments in which the transition metal ferromagnet was replaced with the layer of a metallic (Ga,Mn)As \cite{Jaworski:2010_a} ruled out the original picture of longitudinal diffusion of electrons in the two spin channels over macroscopic distances in the ferromagnet. As shown in Fig.~\ref{fig_spin-Seebeck}, same electrical signals were detected on the SHE electrodes after scratching out the conductive (Ga,Mn)As film in the middle of the sample. The non-local character of the observed spin-Seebeck effect, i.e. the dependence of the measured SHE voltage on the position of the electrode along the sample, has been intensively discussed since the experiments in (Ga,Mn)As and the parallel observation of the spin-Seebeck effect in a ferromagnetic insulator \cite{Uchida:2010_a}. It has been argued that phonons or magnons in the ferromagnet/substrate structure may be responsible for the non-locality of the spin-Seebeck effect \cite{Bauer:2012_a,Tikhonov:2013_a}.

\section{Summary}
\label{sum}
We have reviewed  several areas of the rich physics of spintronics phenomena and device concepts explored in the ferromagnetic semiconductor (Ga,Mn)As. 
The most extensively studied transport characteristics of (Ga,Mn)As are the spin-orbit coupling related magnetoresistance effects. Experiments and calculations in (Ga,Mn)As have provided an unprecedented physical insight into the anomalous Hall effect which prompted a renewed interest and experimental discovery of the spin Hall effect. Anisotropic magnetoresistance phenomena have been identified in (Ga,Mn)As based tunneling devices and in devices sensing the anisotropy of the chemical potential.   Apart from these direct magnetoresistance phenomena, (Ga,Mn)As has become a fruitful model system for exploring the inverse magnetotransport phenomena, i.e., the current induced spin torques. The studies have provided new insight into spin-transfer torques in domain walls and led to the discovery of the current induced spin-orbit torques in uniform magnets. Moreover, optical counterparts of both the non-relativistic spin-transfer and the relativistic spin-orbit torques have been identified in (Ga,Mn)As, allowing to study these phenomena on timescales attainable in the optical pump-and-probe experiments. (Ga,Mn)As based research has also made seminal contributions to the field of spin-caloritronics by discovering the ohmic and tunneling anisotropic thermopower effects and helping to elucidate the origin of the spin-Seebeck effect.  

It is likely that (Ga,Mn)As and related ferromagnetic semiconductors will continue to inspire new avenues of magnetic materials and spintronics research in the future. Many studies, in particular of the relativistic phenomena in (Ga,Mn)As may become directly relevant to room-temperature magnetic systems with strong spin-orbit coupling
and may therefore lead to new technological applications, independent of the existing limits of the  Curie temperature in the ferromagnetic semiconductors. This knowledge transfer applies to room-temperature magnetic systems which include not only the conventional transition metal ferromagnets but also, e.g.,  a class of metal and semiconductor antiferromagnets with high N\'eel temperatures. 

\section*{List of acronyms}
\noindent {\bf ABE}: Aharonov-Bohm effect

\noindent {\bf AHE}: Anomalous Hall effect

\noindent {\bf AMR}: Anisotropic magnetoresistance

\noindent {\bf AMT}: Anisotropic magneto-thermopower

\noindent {\bf ANE}: Anomalous Nernst effect

\noindent {\bf CB}: Coulomb blockade

\noindent {\bf DOS}: Density of states

\noindent {\bf DW}: Domain wall

\noindent {\bf FMR}: Ferromagnetic resonance

\noindent {\bf GMR}: Giant magnetoresistance

\noindent {\bf GMT}: Giant magneto-thermopower 

\noindent {\bf GGA}: Generalized gradient approximations

\noindent {\bf LT-MBE}: Low temperature molecular beam epitaxy

\noindent {\bf MCB}: magnetic circular birefringence

\noindent {\bf MCD}: magnetic circular dichroism

\noindent {\bf MLB}: magnetic linear birefringence

\noindent {\bf MLD}: magnetic linear dichroism

\noindent {\bf MRAM}: Magnetic random access memory

\noindent {\bf SET}: Single electron transistor

\noindent {\bf SHE}: Spin Hall effect

\noindent {\bf SOT}: Spin orbit torque

\noindent {\bf STT}: Spin transfer torque

\noindent {\bf SWR}: Spin-wave resonance 

\noindent {\bf TAMR}: Tunneling anisotropic magnetoresistance

\noindent {\bf TAMT}: Tunneling anisotropic magneto-thermopower

\noindent {\bf TBA}: Tight-binding approximation

\noindent {\bf TMR}: Tunneling magnetoresistance

\noindent {\bf TMT}: Tunneling magneto-thermopower

\noindent {\bf UCF}: Universal conductance fluctuations 

\noindent {\bf WB}: Walker breakdown 

\noindent {\bf WL}: Weak localization

\section*{Acknowledgments} 
We acknowledge support from the ERC Advanced Grant No. 268066, from the Praemium Academiae of the Academy of Sciences of the Czech Republic,  from the Ministry of Education of the Czech Republic Grant No. LM2011026, and from the Czech Science Foundation Grant No. 14-37427G.

\end{document}